\documentclass[useAMS,usenatbib]{mn2e}
\usepackage{aas_macros}
\usepackage{graphicx}
\usepackage{natbib}
\usepackage{amssymb}
\usepackage{amsmath}

\bibpunct{(}{)}{;}{a}{}{,}

\usepackage[normalem]{ulem}
\usepackage[usenames,dvips]{color}

\def\del#1{{}}

\newcommand{\dps}{\displaystyle}
\newcommand{\dd}{\mathrm{d}}

\newcommand{\ltsima}{$\; \buildrel < \over \sim \;$}
\newcommand{\lsim}{\lower.5ex\hbox{\ltsima}}
\newcommand{\gtsima}{$\; \buildrel > \over \sim \;$}
\newcommand{\gsim}{\lower.5ex\hbox{\gtsima}}
\newcommand{\eps}{\varepsilon}
\newcommand {\apgt} {\ {\raise-.5ex\hbox{$\buildrel>\over\sim$}}\ }
\newcommand {\aplt} {\ {\raise-.5ex\hbox{$\buildrel<\over\sim$}}\ } 

\renewcommand{\vec}{\mathbfit}
\newcommand{\vel}{\upsilon}
\newcommand{\vvel}{\bupsilon}

\newcommand{\hm}{h^{-1}\,\rmn{M}_\odot}
\newcommand{\hg}{h^{-1}\,\rmn{Gyr}}
\newcommand{\kms}{\rmn{km~s}^{-1}}
\newcommand{\kpc}{h^{-1}\,\rmn{kpc}}

\title[Galactic winds driven by cosmic-ray streaming]{Galactic winds driven by cosmic-ray streaming}

\author[M. Uhlig et al.]{M.~Uhlig,$^{1,2}$\thanks{e-mail: maxu@nld.ds.mpg.de (MU);
    christoph.pfrommer@h-its.org (CP)} C.~Pfrommer,$^{3}$\footnotemark[1]
  M.~Sharma,$^{4}$ B.~B.~Nath,$^{4,1}$ T.~A.~En{\ss}lin,$^{1}$ V.~Springel$^{3,5}$\vspace*{0.1cm}\\
  $^{1}$Max-Planck-Institut f\"ur Astrophysik, Karl-Schwarzschild-Str. 1, D-85741 Garching bei M\"unchen, Germany\\
  $^{2}$Max-Planck-Institut f\"ur Dynamik und Selbstorganisation (MPIDS), D-37077 G\"ottingen, Germany\\
  $^{3}$Heidelberg Institute for Theoretical Studies, Schloss-Wolfsbrunnenweg 35, D-69118 Heidelberg, Germany\\
  $^{4}$Raman Research Institute, Sadashivanagar, Bangalore 560 080, India\\
  $^{5}$Zentrum f\"ur Astronomie der Universit\"at Heidelberg, Astronomisches
  Recheninstitut, M\"{o}nchhofstr. 12-14, 69120 Heidelberg, Germany}

\begin{document}
\pagerange{\pageref{firstpage}--\pageref{lastpage}} \pubyear{2008}
\maketitle
\label{firstpage}

\begin{abstract}
  Galactic winds are observed in many spiral galaxies with sizes from dwarfs up
  to the Milky Way, and they sometimes carry a mass in excess of that of newly
  formed stars by up to a factor of ten. Multiple driving processes of such
  winds have been proposed, including thermal pressure due to supernova-heating,
  UV radiation pressure on dust grains, or cosmic ray (CR) pressure. We here
  study wind formation due to CR physics using a numerical model that accounts
  for CR acceleration by supernovae, CR thermalization by Coulomb and hadronic
  interactions, and advective CR transport. In addition, we introduce a novel
  implementation of CR streaming relative to the rest frame of the
  gas. Streaming CRs excite Alfv\'en waves on which they scatter, thereby
  limiting the CR's effective bulk velocity. We find that CR streaming drives
  powerful and sustained winds in galaxies with virial masses $M_{200}\lesssim
  10^{11}\,\rmn{M}_\odot$. In dwarf galaxies ($M_{200}\sim 10^{9}\,
  \rmn{M}_\odot$) the winds reach a mass loading factor of $\sim 5$, expel
  $\sim60\%$ of the initial baryonic mass contained inside the halo's virial
  radius and suppress the star formation rate by a factor of $\sim5$. In dwarfs,
  the winds are spherically symmetric while in larger galaxies the outflows
  transition to bi-conical morphologies that are aligned with the disc's angular
  momentum axis. We show that damping of Alfv\'en waves excited by streaming CRs
  provides a means of heating the outflows to temperatures that scale with the
  square of the escape speed, $k T \propto \vel_\rmn{esc}^2$. In larger haloes
  ($M_{200}\gtrsim 10^{11}\,\rmn{M}_\odot$), CR streaming is able to drive
  fountain flows that excite turbulence, providing another means of heating the
  halo gas. For halo masses $M_{200}\gtrsim 10^{10}\,\rmn{M}_\odot$, we predict
  an observable level of H$\alpha$ and X-ray emission from the heated halo
  gas. We conclude that CR-driven winds should be crucial in suppressing and
  regulating the first epoch of galaxy formation, expelling a large fraction of
  baryons, and -- by extension -- aid in shaping the faint end of the galaxy
  luminosity function. They should then also be responsible for much of the
  metal enrichment of the intergalactic medium.
\end{abstract}


\section{Introduction}
\label{sec:intro}

Large-scale galactic outflows constitute an important process in the evolution of
galaxies. They influence the gas content and therefore the star formation
rate in galaxies. Moreover, outflows from galaxies also affect the
surrounding medium by depositing mass, metals and energy \citep[see][for a
recent review]{Veilleux2005}.

The earliest works on galactic outflows were motivated by the observations of
the starburst galaxy M82 \citep{LyndsSandage1963, Burbidge1964}, and the
detection of broad optical emission lines in some elliptical galaxies
\citep{Osterbrock1960} that implied winds \citep{Burke1968}. It was suggested
that galactic winds cause ellipticals to shed their interstellar medium
\citep[ISM;][]{Mathews1971, Johnson1971}.  Galactic outflows have since been
frequently observed in both local and high redshift galaxies. Multi-wavelength
studies for local galaxies have demonstrated the ubiquity of the phenomenon of
outflows \citep[e.g.,][]{Heesen2009}, while spectroscopic studies of Lyman
Break Galaxies at $z\sim3$ have also revealed the presence of winds
\citep[e.g.,][]{Adelberger2003, Shapley2003}. Large mass-loading factors
(defined as mass loss rate divided by star formation rate, SFR) with values of a
few are estimated for local galaxies and massive star-forming regions at $z\sim2
- 3$ \citep{Martin1999, Heckman2000, Sato2009, Steidel2010, Coil2011}. The use
of classic ionization diagnostic diagrams allows to cleanly separate winds that
are driven by either starbursts or active galactic nuclei
\citep[AGN;][]{Sharp2010}: while AGN photoionization always dominates in the
wind filaments, this is not the case in starburst galaxies where shock
ionization dominates, suggesting that the latter is an impulsive
event. Moreover, after the onset of star formation, there is a substantial delay
before a starburst wind develops \citep{Sharp2010}. This may suggest that
different physical processes, exhibiting different timescales, are responsible
for driving winds in starbursts and AGNs, and reinforces the importance of
studying different physical mechanisms that could drive winds.

Theoretically, it has proven to be challenging to launch those powerful winds
with mass-loading factors of a few from the galactic disc. The first models of
outflows invoke the heating of the ISM by repeated supernovae \citep{Larson1974,
  DekelSilk1986}, in which thermal pressure of the hot gas drives the outflow.
Early simulation studies \citep{MacLow1999} that placed supernova explosions in
simple models for dwarf galaxies predicted comparatively little mass loss.
Recently, simulations have begun to approach the necessary resolution to
explicitly simulate the formation of a multiphase medium with a dense, cold
molecular phase that harbours star formation in full three-dimensional
models. The collective action of stellar winds and supernovae drives hot
chimneys and superbubbles that start to launch galactic outflows
\citep{Ceverino2009}, which may be important for low-redshift galaxies
\citep{Hopkins2011b}. However, these thermal pressure-driven winds suffer from
their tendency to destroy the cool gas through evaporation which limits the
mass-loading factor and calls for alternative mechanisms \citep{NathSilk2009,
  Murray2011}. Observed correlations between wind speed and rotation speed of
galaxies, as well as the SFR, may be easier to explain with radiation pressure
driving on dust grains embedded in outflows \citep{Martin2005, Murray2005,
  NathSilk2009}. This appears to be one promising mechanism of explaining
outflows in high-redshift massive starbursts \citep{Hopkins2011b}.

CRs can also drive a large-scale outflow if the coupling between high energy
particles and thermal gas is strong enough \citep{Ipavich1975,
  Breitschwerdt1991, Zirakashvili1996, Ptuskin1997}. This is expected given the
near equipartition of energy in the Milky Way disc between magnetic fields, CRs,
and the thermal pressure \citep{ZweibelHeiles1997, Beck2001, Cox2005}. Fast
streaming CRs along the magnetic field excite Alfv\'en waves through the
`streaming instability' \citep{KulsrudPearce1969}. Scattering off this wave
field limits the CRs' bulk speed. These waves are then damped, effectively
transferring CR energy and momentum to the thermal gas; hence CRs exert a
pressure on the thermal gas by means of scattering off Alfv\'en waves.

\citet{Ipavich1975} studied this process assuming a spherical geometry and that the waves were
completely damped away. Later, \citet{Breitschwerdt1991} considered a disc
geometry, and calculated the effect of both small and large damping. They found
solutions of the outflow equations with realistic ($\sim1$ M$_{\odot}$ yr$^{-1}$)
mass loss rates from Milky Way-type galaxies.  Such winds can explain the small
gradient in $\gamma$-ray emission as a function of galacto-centric radius
\citep{Breitschwerdt2002}. Similarly, an outflow from the Milky Way driven by
CRs explains the observed synchrotron emission as well as the diffuse soft X-ray
emission towards the Galactic bulge region much better than a static atmosphere
model \citep{Everett2008, Everett2010}. This match is particularly striking in
the hard X-ray band because of the CR Alfv\'en-wave heating that partially
counteracts the adiabatic cooling of the thermal gas and makes a strong case in
favour of CR physics being a necessary ingredient in understanding galactic
winds. CR-driven outflows, in which the CR fluid provides an additional source
of pressure on thermal gas, may eject substantial amounts of gas from spherically
symmetric galaxies with a mass outflow rate per unit SFR of order $0.2 - 0.5$
for massive galaxies \citep{Samui2010}. Hence, CR pressure in starburst galaxies
would provide a negative feedback to star formation and eventually limit the
luminosity \citep{Socrates2008}.

These general ideas and analytical arguments have been supported by detailed
numerical simulations employing an implementation of CR physics in the {\sc
  Gadget} code \citep{Jubelgas2008, Wadepuhl2011}. According to these
simulations, CRs that are accelerated by supernova (SN) shocks can exert enough
pressure to significantly influence star formation, particularly in low mass
galaxies so that the observed faint-end of the satellite luminosity function in
the Milky Way can be successfully reproduced. These simulations assumed the CR
fluid to be perfectly coupled to the thermal gas and did not allow for CR
streaming in the rest frame of the gas. In this paper, we improve upon these
earlier studies and investigate the effect of CR streaming in hydrodynamical
simulations of galaxy formation. We will show that CR streaming is potentially
the dominant energy transport process powering outflows from dwarf galaxies.  In
Section~\ref{sec:CR-streaming}, we introduce the concept of CR streaming and
work out a spherically symmetric wind model driven by CR streaming in
Section~\ref{sec:analystics}.  In Section~\ref{sec:numerics}, we introduce our
simulation setting. In Section~\ref{sec:results}, we study how CR streaming can
drive outflows and assess their specific properties including mass loss rates of
galaxy haloes as well as the CR streaming-induced heating of the halo gas. We
conclude in Section~\ref{sec:conclusions}.  Supporting material on the numerical
implementation, code and resolution tests are shown in the Appendices
\ref{sec:AppCRstreaming} and \ref{sec: Tests}.

\section{Cosmic ray streaming}
\label{sec:CR-streaming}

CR proton populations have several properties which could help galaxies to
launch strong winds. First of all, supernovae, being one of the main energy
sources of the ISM, are believed to convert a significant fraction ($10\%-60\%$)
of their released kinetic energy into CRs via diffusive shock acceleration in
SN remnants \citep{Kang2005, Helder2009}. The CR population forms a very
light fluid in the galaxy, with a significant pressure and the tendency to
buoyantly escape more rapidly than any other of the ISM components from the
galactic disc. Since the CR fluid is coupled via magnetic fields to the thermal
ISM components, the CR pressure gradient helps to lift thermal gas out of 
galactic discs.

Second, the energy loss time-scales of trans- and ultra-relativistic protons due
to Coulomb and hadronic interactions with the thermal ISM are long. The CR loss
times are longer than typical residence times of these particles in the denser
galactic discs, where the target densities are highest (and therefore loss times
are shortest), and they are substantially longer than the cooling time of the
thermal gas.

Third, nearly all the energy lost by the CR population, be it via
particle-particle interactions, via volume work, or via collective plasma
effects (as discussed below) is delivered to the thermal plasma. Dissipating CR
energy into the thermal plasma increases the total pressure of the composite
fluid due to the harder equation of state of the thermal plasma. It can be
easily verified that the increase of thermal pressure is larger than the
decrease in CR pressure by evaluating the amount of energy transferred per
volume element, $\Delta E/V = P/(\gamma-1) = P_\rmn{cr} / (\gamma_\rmn{cr}
-1)$. The pressure can double in the limiting case of  a fluid initially
composed of an ultrarelativistic CR population with $\gamma_\rmn{cr}=4/3$ which
dissipated all its energy. More importantly, any energy injected from CRs into
the thermal plasma above the galactic disc is protected against the severe
radiative losses any thermal gas element would have suffered within the dense
and high-pressure environment within the galactic disc.

Thus, CRs could play an important role as a relatively loss-less vehicle for the
transfer of SN energy into the wind regions. They can energize the wind with a
higher efficiency than the thermal gas, which radiatively looses a large
fraction of the energy injected into it within the galactic discs. The mobility
of CRs to travel rapidly along already opened magnetic field lines and their
ability to open ISM magnetic field lines via the Parker instability
\citep{Parker1966} render them ideal for this function.

The dominant CR transport process along such opened field lines should be
streaming as we will explain now. This process appears naturally provided there
exists a gradient of the CR number density along a field line. Since the
individual CR particles travel with nearly the speed of light on spiral
trajectories guided by the magnetic field direction, a gradient of CRs at a
location leads immediately to a bulk motion of particles from the dense to the
less dense region. This bulk motion is mainly limited by resonant scattering of
CR particles on plasma waves. The scattering isotropizes the CRs' pitch angles,
and thereby reduces the CR bulk speed.

The amplitude of plasma waves which are able to resonate with CRs of a certain
energy therefore determines the rate at which the streaming CR population is
re-isotropized, and thus the limiting bulk speed. The plasma waves are certainly
seeded by the unavoidable turbulent cascade of kinetic energy from large to
small spatial scales in the violent ISM. They are further amplified by plasma
instabilities, which are powered by the free energy of the anisotropic CR
population.  Thus the CRs, while streaming through the plasma, excite the waves
which limit this streaming. The growth of the waves itself is limited by plasma
physical damping processes. The energy of the CRs transferred to the waves is
finally thermalized and heats the plasma.

The delicate question is at which level does this interplay of CR streaming due
to a CR gradient, plasma-wave excitation due to the CR streaming, and
wave-damping processes saturate? What is the resulting effective bulk velocity
achieved by the CRs?

A unique answer to these questions is difficult. In a low-beta plasma, where the
Alfv\'en speed exceeds the sound speed by a large factor, most works on this
topic seem to agree that the effective streaming speed should be of the order of
the Alfv\'en velocity
\citep[e.g.,][]{Wentzel1968,KulsrudPearce1969,KulsrudCesarsky1971,Kulsrud2005}.
In a high-beta plasma, where the Alfv\'en speed is strongly subsonic, the
Alfv\'en speed can not be the limiting velocity for CRs. This should become
clear by the following Gedankenexperiment. If magnetic fields become weaker and
weaker in an otherwise unaltered plasma, the Alfv\'en velocity would approach
zero and any CR streaming, which would be limited to this speed, would
vanish. However, weaker magnetic fields imply weaker coupling of the CRs to the
thermal plasma and therefore intuitively one would expect a larger, or at least
constant transport velocity that is independent of the magnetic field strength
in this limit.

Consequently, a number of authors have worked on the problem of CR streaming in
high-beta plasmas \citep{Holman1979, Acherberg1981, FeliceKulsrud2001,
  YanLazarian2008}, and a summary of these works can be found in
\citet{Ensslin2011}.  The consensus among most of these authors seems to be
that the limiting velocity is of the order of the sound speed in this
situation. It can be several times this speed, or less, depending on details of
the pre-existing wave level, thermal and CR energy density, magnetic topology and
so forth.

Here, we adopt the pragmatic approach of \citet{Ensslin2011} to parametrize our
lack of knowledge on the precise dependence of the magnitude of the CR streaming
speed $\vel_{\mathrm{st}}$ on the plasma parameters and assume that it is
proportional to the local sound speed $c_{\mathrm{s}}$,
\begin{equation}
\vel_{\mathrm{st}}=\lambda_{\mathrm{}}\, c_{\mathrm{s}},\label{eq:streaming speed}
\end{equation}
with a proportionality constant $\lambda_{\mathrm{}}\ge1$.

An important aspect of this work is the energy transfer of the CRs to the
thermal plasma. This energy transfer is able to repower the thermal wind
continuously, in particular at large heights above the galactic disc. The CR
stream has a significantly higher velocity than the thermal wind, since the CRs'
bulk speed with respect to the thermal gas is $\vel_{\mathrm{st}}\sim
c_{\mathrm{s}}$.  Thus the dominant energy flow of a galactic wind region might
be carried by the mobile CR fluid, and not by the much slower thermal gas. The
transfer of energy from CRs to the gas is mediated by plasma waves.  To the CR
population, the process of wave excitation resembles adiabatic work done during
the expansion of the CR fluid with respect to the rest frame of magnetic
irregularities of interacting plasma waves. Those are generated by streaming CRs
and provide efficient scattering partners which are pushed away from the regions
with high CR pressure. If the plasma wave motion would be reversed, they would
compress back the expanded CRs and thereby return exactly the transferred
energy. However, since the CRs are driving these waves, this usually does not
happen under the circumstances we are interested in. Only the wave damping
itself is a dissipative process, which thermalizes this energy and therefore
re-powers the plasma.

For these reasons, the CR energy loss term in the wind equations due to CR
streaming formally resembles an adiabatic loss term, which would be reversible
if the streaming velocity field converged. Just because the relevant streaming
velocity is always diverging away from the region with high CR pressure, the
corresponding energy change of the CR population is always negative and the
thermal plasma is always heated by CR streaming.

\section{Analytical estimates}
\label{sec:analystics}

We begin with analytical estimates of the outflows excited by the heating from
CR streaming. For simplicity, we assume a steady state situation and assume that
CRs first diffuse out to a height above the disc, comparable to the scale
height. Assuming spherical symmetry, we can write the constant gas mass flux per
unit solid angle, $q=\rho\vel r^{2}$, where $\rho$ is the gas mass density,
$\vel$ is the gas velocity, and $r$ is the radial distance.  The mass density of
CRs, $\rho_\rmn{cr}$, is fixed by the constant CR flux per unit solid angle in
the steady state: $q_\rmn{cr}=\rho_\rmn{cr} \vel_\rmn{cr}r^{2}$, where
$\vel_\rmn{cr} =\vel+c_{\mathrm{s}}$ is the CR speed (for $\lambda\sim1$ in
equation~\eqref{eq:streaming speed}).

We will first calculate the terminal wind speed using the Bernoulli
theorem, assuming that there are streamlines along which gas can travel
from the disc to a large distance. In the steady state and for spherical
symmetry, the energy equation for a compressible fluid is given by
\begin{equation}
\frac{1}{r^{2}}\frac{\partial}{\partial r}\left[\rho\vel r^{2}
\left(\epsilon+\frac{\vel^{2}}{2}+\frac{P}{\rho}\right)\right]=
\vvel\cdot\vec{F}+\nabla\cdot\vec{Q}\,,\label{eq:lr}
\end{equation}
where $\vec{F}$ and $\vec{Q}$ represent the external force and energy flux
respectively, $\epsilon$ is the specific internal energy (thermodynamic energy
per unit mass), and $P$ is the pressure. Here we have a two-component fluid
composed of gas and CRs. Using an adiabatic index of $\gamma=5/3$ for the
gas, and writing $c_\rmn{s}^{2}=\gamma P/\rho$, we have the following gas
energy equation,
\begin{equation}
\frac{q}{r^{2}}\frac{\partial}{\partial r}\left(\frac{\vel^{2}}{2}+3\frac{c_\rmn{s}^{2}}{2}+\Phi\right)=
\Lambda_\rmn{cr-heating},\label{eq:gas}
\end{equation}
where $\Phi$ is the gravitational potential and $\Lambda_\rmn{cr-heating}$ is
the heating term due to damping of CR-excited Alfv\'en waves. We neglect
radiative cooling for the tenuous gas in the wind in this analytical treatment.
Since $\rho_\rmn{cr}\ll\rho$, we can safely neglect the bulk kinetic energy of
CRs as well as their gravitational attraction. Thus, we have (using
$\epsilon=P_\rmn{cr}/3$ as an approximation for ultra-high-energy particles),
\begin{equation}
\frac{q_\rmn{cr}}{r^{2}}\frac{\partial}{\partial r}\left(4\frac{P_\rmn{cr}}{\rho_\rmn{cr}}\right)=
-\Lambda_\rmn{cr-heating}.\label{eq:cre}
\end{equation}
Here, the negative sign of $\Lambda_\rmn{cr-heating}$ indicates the loss of CR
energy due to wave excitation.  Adding these two equations and integrating, we
get the following equation for total energy,
\begin{equation}
q\left(\frac{\vel^{2}}{2}+3\frac{c_\rmn{s}^{2}}{2}+\Phi\right)+4q_\rmn{cr}\frac{P_\rmn{cr}}{\rho_\rmn{cr}}=C,
\end{equation}
where $C$ is a constant. Equating the values at the base and the end of a
streamline, we have
\begin{equation}
\frac{\vel_{\infty}^{2}}{2}+\left(3\frac{c_{\rmn{s},\infty}^{2}}{2}+\frac{4q_\rmn{cr}P_{\rmn{cr},\infty}}{q\rho_{\rmn{cr},\infty}}\right)+\Phi_{\infty}=\frac{\vel_\rmn{b}^{2}}{2}+3\frac{c_\rmn{s,b}^{2}}{2}+\frac{4q_\rmn{cr}P_\rmn{cr,b}}{q\rho_\rmn{cr,b}}+\Phi_\rmn{b}\,.
\end{equation}
The sum of the terms inside the parenthesis is $3/2$ times the square of the effective
combined sound speed of gas and CRs at infinity, and is negligible for an
adiabatic flow. We are therefore left with the following expression,
\begin{equation}
\label{eq:v_infty}
\vel_{\infty}^{2}=\vel_\rmn{b}^{2}+3c_\rmn{s,b}^{2}+\frac{8q_\rmn{cr}P_\rmn{cr,b}}{q\rho_\rmn{cr,b}}-2\,\Delta\Phi\,,
\end{equation}
where $\Delta\Phi=\Phi_{\infty}-\Phi_\rmn{b}$.

We will use the gravitational potential due to a baryonic component and the dark
matter halo, given by the Navarro-Frenk-White (NFW) profile
\citeyearpar{Navarro1996}.  Since $\Phi_\infty=0$, we obtain
\begin{equation}
\Delta\Phi=\frac{GM_\rmn{bar}}{r_\rmn{b}}+\frac{GM_{200}}{\ln(1+c)-c/(1+c)} \frac{\ln(1+r_\rmn{b}/r_\rmn{s})}{r_\rmn{b}}.
\end{equation}
Here, $M_\rmn{bar}$ is the total baryonic mass and $M_{200}$ is the halo mass of
a sphere enclosing a mean density that is 200 times the critical density of the
universe, $r_\rmn{s}$ is the characteristic radius of the NFW profile, in which
the density profile is given by
$\rho\propto(r/r_\rmn{s})^{-1}(1+r/r_\rmn{s})^{-2}$.  Assuming the initial gas
speed $\vel_\rmn{b}\sim c_\rmn{s,b}$, the gas sound speed at the base, we have
an expression for the wind speed at large distance which depends on the CR
parameters and the gas sound speed at the base. The condition for an unbound
flow or the existence of the wind is given by,
\begin{equation}
4c_\rmn{s,b}^{2}+\frac{8q_\rmn{cr}P_\rmn{cr,b}}{q\rho_\rmn{cr,b}}\,>2\Delta\Phi\,.
\end{equation}
Multiplying equation~\eqref{eq:v_infty} with the mass density, we arrive at the energy
budget of the problem, 
\begin{equation}
\eps_\rmn{wind} \equiv \frac{\rho\vel_{\infty}^{2}}{2}=
\eps_\rmn{gain} - \eps_\rmn{loss} = 2\rho c_\rmn{s,b}^{2}+
\frac{4\rho q_\rmn{cr}P_\rmn{cr,b}}{\rho_\rmn{cr,b} q}-\rho\Delta\Phi\,,
\end{equation}
which essentially states that the kinetic energy of the wind is the difference
between the total energy gain and the loss of energy against gravity. In other
words, we have,
\begin{equation}
\frac{\eps_\rmn{wind}}{\eps_\rmn{gain}}=
1-\frac{q\Delta\Phi}{2qc_\rmn{s,b}^{2}+\frac{\dps 4q_\rmn{cr}P_\rmn{cr,b}}{\dps\rho_\rmn{cr,b}}}\,.
\end{equation}
Now, using the value of CR mass flux at the base
$q_\rmn{cr}^{}=\rho_\rmn{cr}^{}r_\rmn{b}^{2}\,(\vel_\rmn{b}+c_\rmn{s,b})\sim2\rho_\rmn{cr}c_\rmn{s,b}r_\rmn{b}^{2}$,
we have
\begin{eqnarray}
\vel_{\infty}^{2} & = & 4c_\rmn{s,b}^{2}+\frac{16c_\rmn{s,b}r_\rmn{b}^{2}P_\rmn{cr,b}}{q}-2\Delta\Phi\,,
\label{eq:v_infty2}\\
\frac{\eps_\rmn{wind}}{\eps_\rmn{gain}}
 & = & 1-\frac{q\Delta\Phi}{2qc_\rmn{s,b}^{2}+8c_\rmn{s,b}r_\rmn{b}^{2}P_\rmn{cr,b}}\,.
\end{eqnarray}

In the next step, we want to assess how the wind speed and mass-loading factor
scale with the halo mass and how this compares to the simulations that we will
introduce in Section~\ref{sec:numerics}. To this end, we have to specify the
wind parameters which should reflect realistic ISM values and also match the
late-time behaviour of our simulations with an approximate steady state. We use
$c_\rmn{s,b}=10\,\rmn{km~s}^{-1}$, corresponding to an ISM temperature of
10$^4$~K and assume $P_\rmn{cr,b}=10^{-12}$ erg cm$^{-3}$. We adopt a mass flux
(per unit solid angle) of $q=0.01/(4 \pi)$ M$_{\odot}$ yr$^{-1}$, which is
typical in our simulations that show an almost spherical outflow.

The scaling of the base height $r_\rmn{b}$ exhibits two regimes separated by a
characteristic circular velocity of $\vel_\rmn{c,\,crit} \simeq
120\,\rmn{km~s}^{-1}$. At $\vel_\rmn{c} \lesssim \vel_\rmn{c,\,crit}$, the
vertical scale height increases roughly as $r_\rmn{b}\propto \vel_\rmn{c}$
according to two-dimensional fits of edge-on exponential discs in the
$K_\rmn{s}$-band \citep{Dalcanton2004}.\footnote{Here we assume that the base
  height of the wind scales with the vertical scale height of the disc.} At
$\vel_\rmn{c} \gtrsim \vel_\rmn{c,\,crit}$, the vertical scale height appears to
be independent of $\vel_\rmn{c}$.

This behaviour can be understood theoretically by the following line of
arguments. In shallow gravitational potentials, as found in dwarfs, the vertical
scale height should scale with the system size. Hence, we assume the base height
$r_\rmn{b}$ to scale linearly with the disc size $r_\rmn{d}$ (which is in turn a
function of halo mass, according to the model by \citealt{Mo1998}).  When the
gravitational potential of the halo is deep enough, the vertical scale height is
set by the effective equation of state of the ISM. For an effective equation of
state of $P_\rmn{eff} \propto \rho^2$, the solution to the hydrostatic equation
yields a vertical scale height that is independent of surface mass density
\citep[see, e.g.,][]{Springel2000}. It turns out, that the sub-resolution model
employed by our simulations has such a stiff equation of state just above the
star formation threshold \citep[see Fig.~2 of][]{Springel2003}. Hence, the
vertical scale height should be set by radiative cooling (which breaks the
scale-invariance of gravity seen in smaller systems) in haloes heavy enough to
support a thin disc, i.e., for $M \gtrsim 10^{11} M_\odot$ corresponding to
$\vel_\rmn{c}\gtrsim \vel_\rmn{c,\,crit}$. This justifies a constant vertical
scale height for these large haloes. For our main models, we adopt the following
simple prescription for the base height, $r_\rmn{b}=0.5\, r_\rmn{d}$ for
$r_\rmn{b}<r_{\rmn{b,\,max}}$. We note that these considerations are reflected
by our simulations where we find identical trends of $r_\rmn{b}$ with halo mass.

\begin{figure}
\centerline{\includegraphics[width=\linewidth]{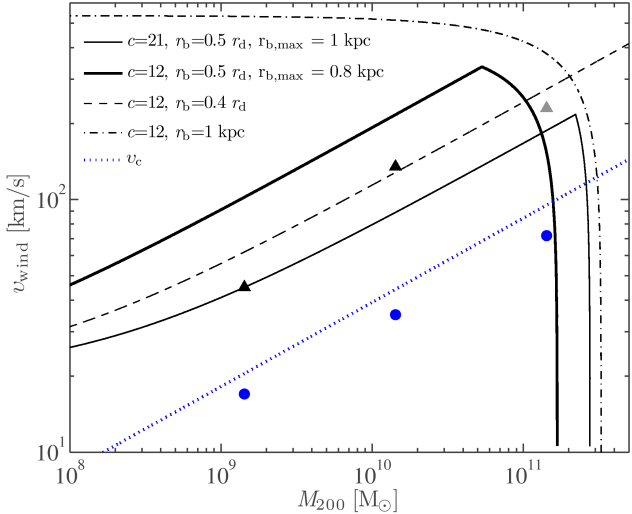}}
\caption{Wind speed as a function of halo mass (equation \eqref{eq:v_infty2})
  for four models where we vary the input parameters, namely the halo
  concentration $c$, the constant of proportionality connecting the base height
  $r_\rmn{b}$ and disc scale $r_\rmn{d}$, and the maximum base height
  $r_{\rmn{b,\,max}}$. The blue dotted curve shows the maximum rotation speed as
  a function of halo mass.  The triangles are the results of the simulation:
  $\vel_\rmn{wind}\simeq(45,~135,~230)$ km s$^{-1}$ for
  $M_{200}=(10^9,~10^{10},~10^{11})\, h^{-1}$ M$_{\odot}$, while the wind in the
  last case (grey triangle) was only sustained for $t<1.5\,\hg$. Filled circles
  show the circular velocity measured in the simulations at the virial radius.}
\label{fig: analytical1}
\end{figure}

\begin{figure}
\centerline{\includegraphics[width=\linewidth]{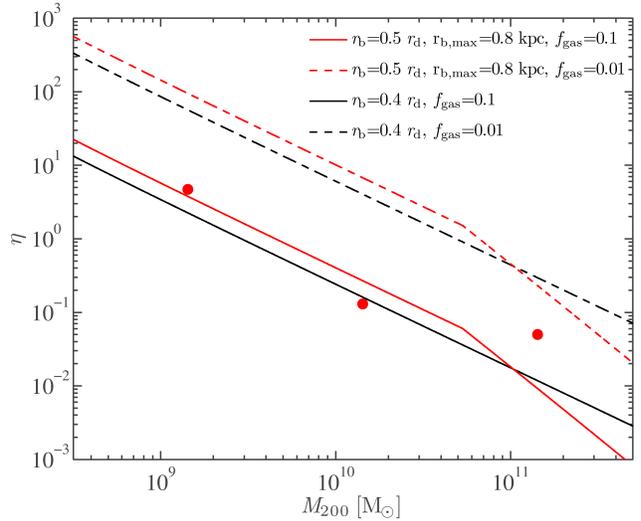}}
\caption{Mass loading factor $\eta=\dot{M}/{\rm SFR}$ (equations \eqref{eq:sfr}
  and \eqref{eq:Mdot}) as a function of halo mass for two values of the gas
  fractions, $f_\rmn{gas} = 0.1$ and 0.01 (solid and dashed). The two different
  models both have a halo concentration $c=12$, as in Fig.~\ref{fig:
    analytical1}.  The analytic models compare well with our simulated values
  for the ``peak mass loading'' ($\eta_\rmn{peak} \equiv \dot{M}_\rmn{max}/{\rm
    SFR}_\rmn{max}$, filled circles).}
\label{fig: analytical2}
\end{figure}

In Fig.~\ref{fig: analytical1}, we show the wind speed at a large distance as a
function of halo mass. This is compared with the rotation speed of the disc
implied by the halo mass, adopting the model of \citet{Mo1998}. We note that in
this model, the disc mass (which we equate with $M_\rmn{bar}$) is a factor $\sim
0.05$ of the halo mass. In the low-mass regime, the wind speed increases with
halo mass (in accordance with the simulations) up to wind velocities of
$\vel_\rmn{wind}\equiv\vel_\infty\sim(200-350)\,\rmn{km~s}^{-1}$, depending on
the adopted parameters. At halo masses around $M_{200}\sim
10^{11}\,\rmn{M}_\odot$, the CR-driven wind is not any more powerful enough to
overcome the increasing potential difference of the halo and
ceases. It is informative to contrast this behaviour of our main models (solid lines)
with two limiting cases where we either adopt a constant $r_\rmn{b}$
(dash-dotted) or do not impose a cutoff value $r_{\rmn{b,\,max}}$ to the
$r_\rmn{b} - r_\rmn{d}$ scaling (dashed).

The behaviour of our main models is in reassuring agreement with our
simulations where the wind speed increases with halo mass up to a
critical halo mass of $M_{200}\sim 10^{11}\,\rmn{M}_\odot$. At this
mass scale, we see an onset of a wind that stalls after $< 1.5\,\hg$,
coinciding with the time of the formation of the thin disc. Here the
increasing stellar mass deepens the central potential to the point
where the CR-driving is not any more powerful enough to support a
wind; instead we end up with a fountain flow. The normalization of the
wind speed and the transition mass varies with changes in the base
height, $r_\rmn{b}$, and the halo concentration, i.e., the effective
potential difference from the starting point of the wind to
infinity. Interestingly, in the model with the high concentration $c=21$
in Fig.~\ref{fig: analytical1}, the wind speed starts to flatten
towards lower halo masses. This is because for these small haloes the
(constant) sound speed starts to dominate over the terms accounting
for the loss of energy against gravity and the energy gain due to CR
pressure in equation~\eqref{eq:v_infty2}.

To estimate the mass loading of the wind, we first estimate the disc surface
density $\Sigma_\rmn{d}=M_\rmn{d}/(\pi r_\rmn{d}^2)$, where the disc mass is
$M_\rmn{d} =0.05 M_\rmn{200}$. The gas surface density can be written as
$\Sigma_\rmn{gas} = f_\rmn{gas}\Sigma_\rmn{d}$, where $f_\rmn{gas}$ is the gas
mass fraction.  To determine the SFR per unit area, $\Sigma_\rmn{SFR}$, we use
the Schmidt-Kennicutt relation \citep{Kennicutt1998},
\begin{equation}
\frac{\Sigma_\rmn{SFR}}{M_\odot \,\rmn{yr}^{-1}\, \rmn{kpc}^{-2}} 
= 2.5\times 10^{-4}\,\left(\frac{\Sigma_\rmn{gas}}{M_\odot \,\rmn{yr}^{-1}\, \rmn{pc}^{-2}}\right)^{1.4}.
\end{equation}
Hence the SFR of the disc is obtained as
\begin{equation}
\rmn{SFR} = \pi r_\rmn{d}^2 \Sigma_\rmn{SFR}
=\pi r_\rmn{d}^2 \left[ 2.5\times 10^{-4}
\left(\frac{f_\rmn{gas} \Sigma_\rmn{d}}{\rmn{M}_\odot\, \rmn{yr}^{-1}\, \rmn{pc}^{-2}}\right)^{1.4}\right] .
\label{eq:sfr}
\end{equation}

We can estimate the value of the mass flux per unit solid angle, $q$, for
escaping winds, by requiring that $\vel_\rmn{wind} = \vel_\rmn{esc} =
3\vel_\rmn{c}$ in the wind equation (equation~\eqref{eq:v_infty2}) and by inverting it to
get $\dot{M}=4\pi q$ (for a spherical outflow):
\begin{equation}
\dot{M} = \frac{2^6 \pi c_{\rmn{s,b}} r_{\rmn{b}}^2 P_{\rmn{cr,b}}}{9 v_{\rmn{c}}^2 + 2 \Delta \phi - 4 c_{s,b}^2}.
\label{eq:Mdot}
\end{equation}
This is reasonable since the escape speed for a NFW halo is $\le 3
\vel_c$ \citep{Sharma-Nath2012}.  We can then divide this by the
corresponding SFR of equation~\eqref{eq:sfr} to obtain the mass
loading factor $\eta=\dot{M}/{\rm SFR}$.

In Fig.~\ref{fig: analytical2}, we compare the estimated mass loading factor
with the corresponding values obtained in the simulation. We find that our
simple theoretical estimate nicely describes the mass loading observed in
simulations, which scales as $\eta \propto M_{200}^{-2/3} \propto
\vel_{c}^{-2}$. Hence we conclude that the mass loading in CR-driven winds
scales with the halo mass in exactly the way that is needed to explain the
low-mass end of luminosity function according to semi-analytical models of
galaxy formation \citep{Bower2011} and phenomenological simulation models
\citep{Puchwein2012}. CR driven winds are thus also very promising candidates to
help explaining the ``missing satellites problem'' in the Milky Way.

\section{Numerical simulations}
\label{sec:numerics}

\subsection{Simulated physics}
 
We now turn to numerical simulations of galaxy formation that include
CR physics.  Our simulations were carried out with an updated and
extended version of the distributed-memory parallel TreeSPH code {\sc
  Gadget-2} \citep{Springel2005}. Gravitational forces were computed
using a tree algorithm.  Hydrodynamic forces are computed with a
variant of the smoothed particle hydrodynamics (SPH) algorithm that
conserves energy and entropy where appropriate, i.e., outside of
shocked regions \citep{Springel2002}.

Radiative cooling was computed assuming an optically thin gas of
primordial composition (mass-fraction of $X = 0.76$ for hydrogen and
$1-X = 0.24$ for helium) in collisional ionization equilibrium,
following \citet{Katz1996}. We account for star formation according to
a sub-resolution model by \citet{Springel2003} which assumes that star
forming regions establishes a self-regulated regime, described by an
effective equation of state. The self-regulation arises due to the
interplay of cold, dense clouds that continuously form stars and
supernova feedback that evaporates these clouds, creating a hot
ambient medium. The rate of star formation is adjusted such as to
reproduce the observed Schmidt-Kennicut law. As far as the important
parameters of this model are concerned, we choose $t_{0}^{\star}=3.0$
Gyr, thereby setting the characteristic time-scale of star formation,
and a density threshold of star formation of
$\rho_{\mathrm{th}}\simeq0.1\,\mathrm{cm^{-3}}$ (in units of hydrogen
atoms per unit volume). This choice is motivated by \citet{Dubois2008}
who studied wind formation using similar initial conditions and a
similar description of star formation. Their work differs from ours in
that they considered superbubbles as the mechanism that drives
outflows and they observed the strongest winds for the star formation
time-scale stated above. Since CRs as well as superbubbles are powered
by supernova explosions, adopting their time-scale should also give
the most favourable results in our case.

We model CR physics with a formalism that follows the most important CR
injection and loss processes self-consistently while accounting for the CR
pressure in the equations of motion \citep{Pfrommer2006, Ensslin2007,
  Jubelgas2008}. In our methodology, the non-thermal CR population of each
gaseous fluid element is approximated by a simple power law spectrum in particle
momentum, characterized by an amplitude, a low-momentum cut-off, and a fixed
slope $\alpha = 2.5$. Adiabatic CR transport processes such as compression and
rarefaction, and a number of physical source and sink terms, which modify the CR
pressure of each particle, are included. As the most important source of
galactic CRs, we consider CR acceleration by supernova remnants with a spectral
index for the freshly accelerated CR population of $\alpha_{\mathrm{SN}}=2.4$
and an acceleration efficiency, i.e., the fraction of supernova feedback energy
directed into the CR population, ranging in between $\zeta_{\mathrm{SN}}=0.1$ and
0.3 (where the latter value is our standard choice). As sinks of CRs, we
consider thermalization by Coulomb interactions and catastrophic losses by
hadronic interactions.

In addition to the advective CR transport, we have implemented CR
streaming relative to the rest frame of the gas. The details of the
numerical implementation in {\sc Gadget-2} are described in
Appendix~\ref{sec:AppCRstreaming}. We set the streaming speed equal to the
local sound speed (i.e., $\lambda=1$ in equation \eqref{eq:streaming
  speed}).

\subsection{Initial conditions and settings}

\begin{table*}
\begin{tabular}{c|c|c|c|c|c|c|c|c|c|}
\hline 
$M_{200}$ & $R_{200}$ & Resolution & $m_{\mathrm{DM}}$ & $m_{\mathrm{gas}}$ & $\epsilon_{\mathrm{grav}}^{\mathrm{DM}}$  & $\epsilon_{\mathrm{grav}}^{\mathrm{gas}}$  & $\varepsilon$ & NGB & \tabularnewline
$\left[\hm\right]$ & $\left[\kpc\right]$ & (DM/gas)  & $[h^{-1}\,\mathrm{M}_{\odot}]$ & $[h^{-1}\,\mathrm{M}_{\odot}]$ & $\left[\kpc\right]$ & $\left[\kpc\right]$ & \phantom{\Large M} &  & \tabularnewline
\hline 
\hline 
$10^{9}$ & $17$ & $(2/1)\times1\mathrm{e}5$ & $4.3\times10^{3}$ & $1.3\times10^{3}$ & $0.05$ & $0.034$ & $0.004$ & 64 & \tabularnewline
\hline 
$10^{10}$ & $35$ & $(2/1)\times1\mathrm{e}5$ & $4.3\times10^{4}$ & $1.3\times10^{4}$ & $0.106$ & $0.072$ & $0.004$ & 64 & \tabularnewline
\hline 
$10^{11}$ & $75$ & $(2/1)\times1\mathrm{e}5$ & $4.3\times10^{5}$ & $1.3\times10^{5}$ & $0.226$ & $0.153$ & $0.01$ & 64 & \tabularnewline
\hline 
\end{tabular}
\caption{Simulation parameters adopted in this study. Here, $M_{200}$ and $R_{200}$ denote
  the virial mass and radius, $m_{\mathrm{DM}}$ and $m_{\mathrm{gas}}$ denote the particle
  masses used to sample dark matter and baryonic gas,
  $\epsilon_{\mathrm{grav}}^{\mathrm{DM}}$ and
  $\epsilon_{\mathrm{grav}}^{\mathrm{gas}}$ are the gravitational softening
  lengths for the two components, $\varepsilon$ regulates the size of the
  streaming time step (see equation \eqref{eq:timestep}) and NGB is the 
  number of SPH smoothing neighbours used in this study.}
\label{SimParams}
\end{table*}
In our simulations, we consider disc galaxy formation in isolated dark matter
haloes, according to the standard picture \citep{Fall1980}. Initially, the dark
matter potential well is filled with a hydrostatic, slowly rotating, baryonic
gas in equilibrium with the dark matter. Right after the start of the
simulation, radiative cooling sets in, removing thermal pressure support of
the gas atmosphere which will then collapse, subject to gravity, to form a
rotationally supported galactic disc in the centre of our halo. The gas in the
disc becomes colder and denser until it starts to form stars at a rate given by
our sub-resolution model. This picture of galaxy formation in isolation is
clearly oversimplified, since galaxy formation is a highly hierarchical process,
shaped by frequent mergers of different dark matter haloes. However, isolated
haloes enable us to study the impact of CRs in a comparatively clean and
well-controlled environment, so that our simulations should be able to reveal
the significance of CR streaming for driving outflows and to make some
qualitative predictions on the size of the effects.

Both dark matter and baryonic gas are set up in such a way that they follow the
same NFW density profile. Furthermore, the two components are initially in
equilibrium, i.e., the gas atmosphere will be stable in the absence of radiative
cooling when evolved over a Hubble time. In addition, the halo carries an
initial amount of angular momentum described by the spin parameter
$\lambda_\rmn{spin}=0.05$, and with a radial distribution in agreement with
results from full cosmological simulations \citet{Bullock2001}. We adopt a
concentration parameter of the NFW profile of $c=12$ in all our runs. The
initial mass fraction of baryons is kept fixed in all cases at its universal
value, $\Omega_{\mathrm{b}}/ \Omega_{\mathrm{m}}=0.133$, to ease comparison to
the results of \citet{Jubelgas2008}.  Our simulations represent the baryons with
$10^{5}$ SPH-particles and the dark matter with twice as many particles, which
produces numerically converged results (see Appendix \ref{sec:
  resolution}). Further, we use $64$ SPH smoothing neighbours in all our
simulations. An overview of the most important numerical parameters adopted in
our simulations is given in Table \ref{SimParams}.

\section{Results}
\label{sec:results}

\subsection{Launching of galactic winds}
\label{sec:launching}

\begin{figure*}
\begin{tabular}{cc}
{\large {\em CR streaming and advection}} & {\large {\em CR advection-only}} \\
\includegraphics[clip,scale=0.43]{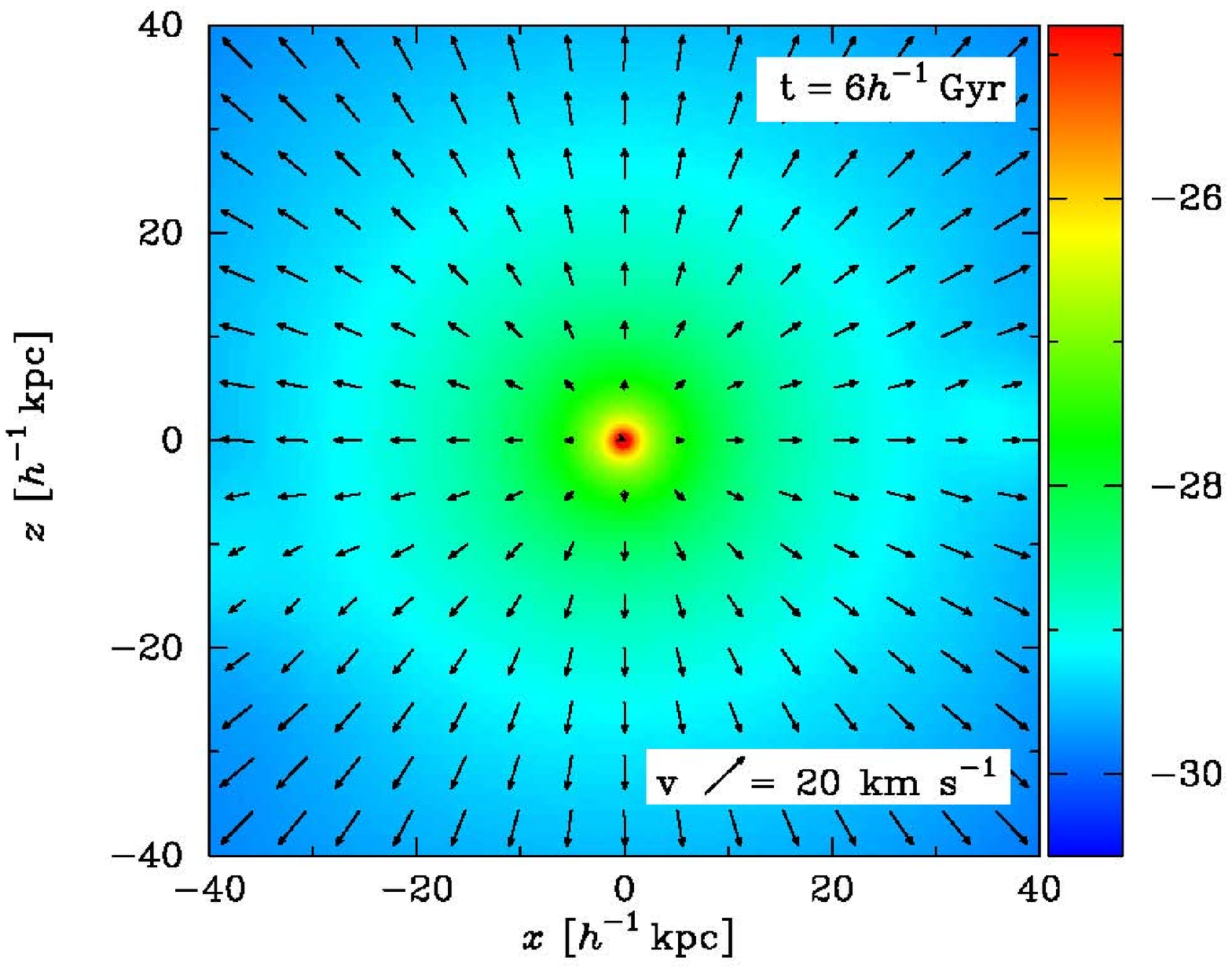} & 
\includegraphics[clip,scale=0.43]{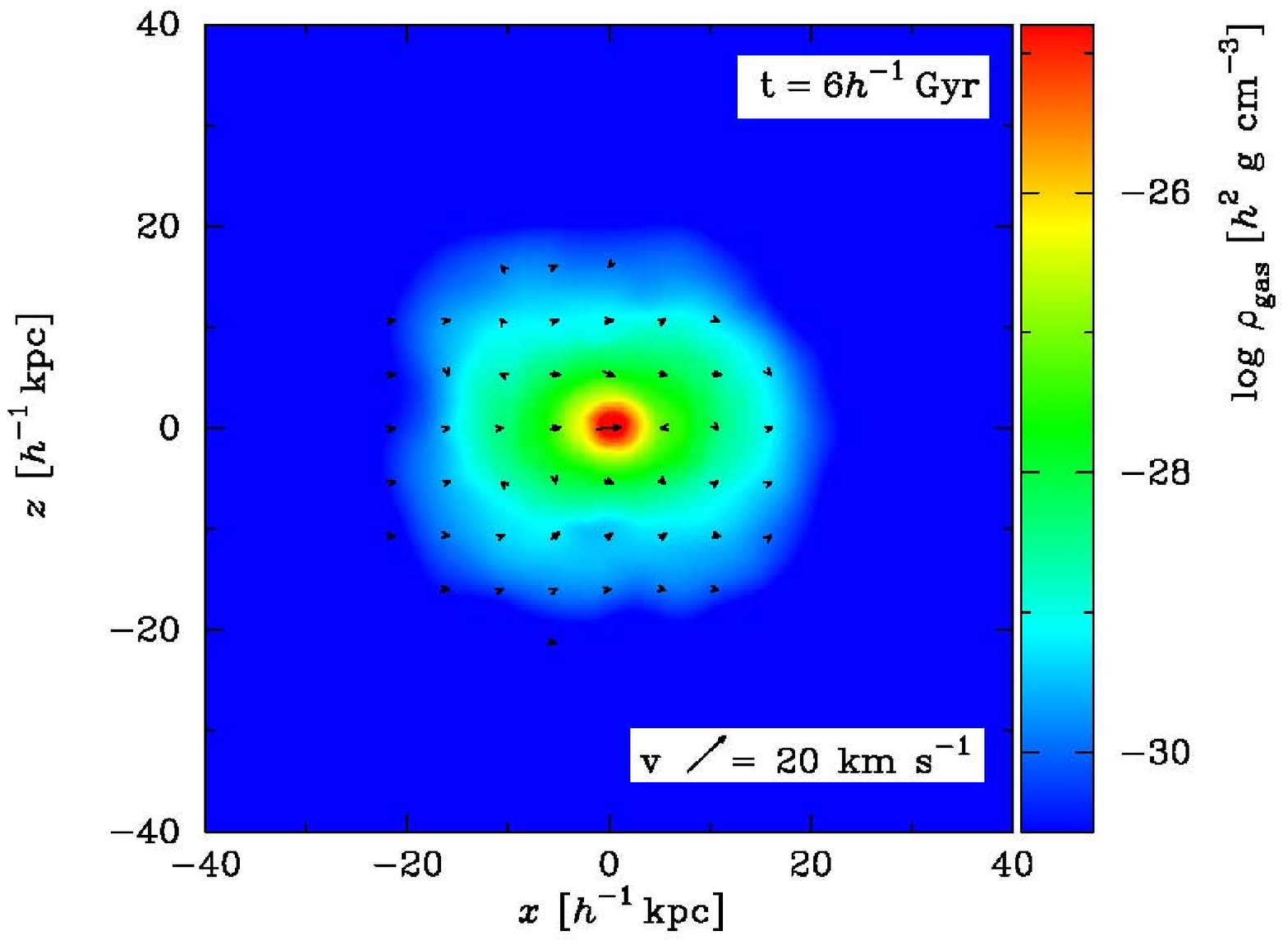}\\
\includegraphics[clip,scale=0.43]{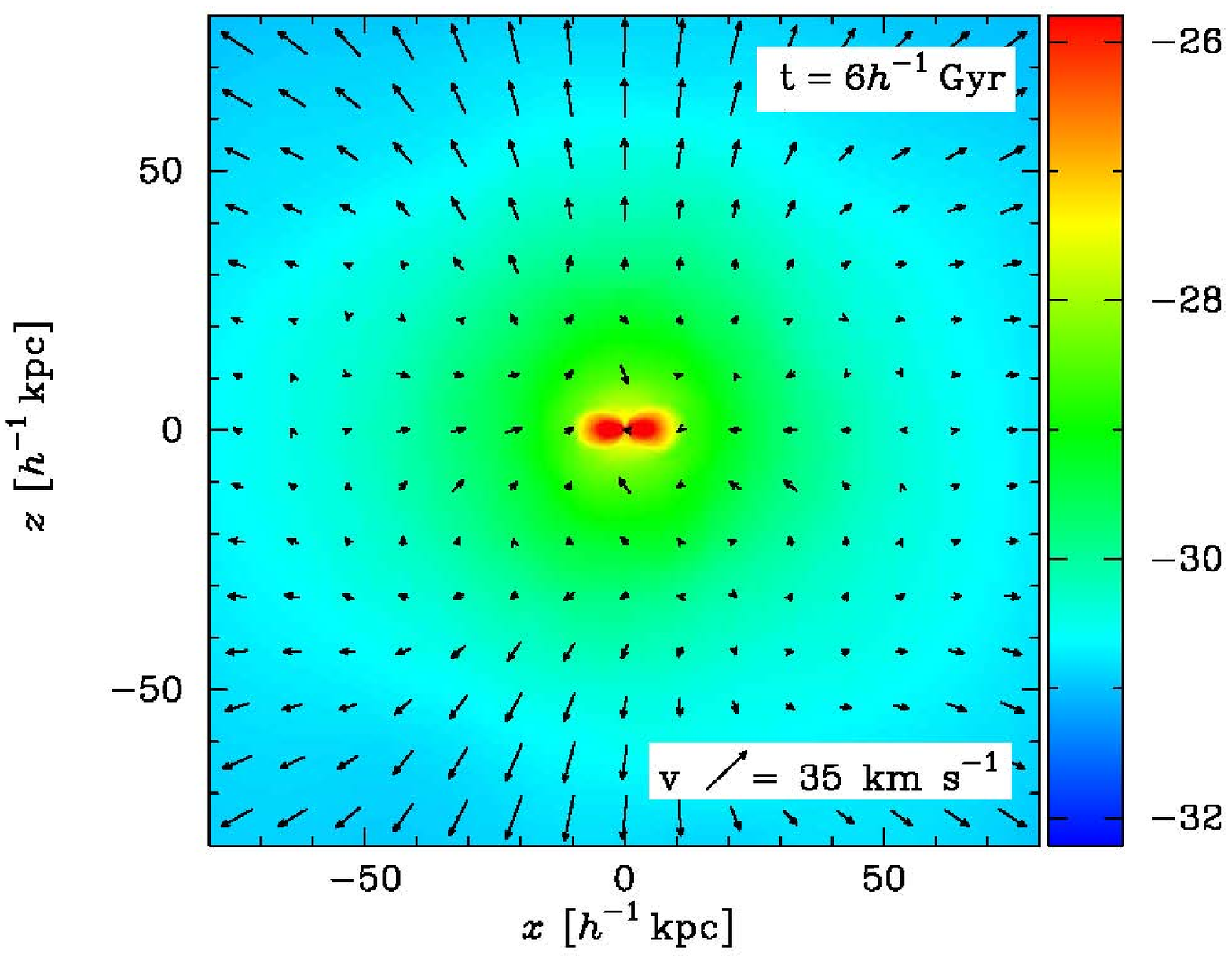} & 
\includegraphics[clip,scale=0.43]{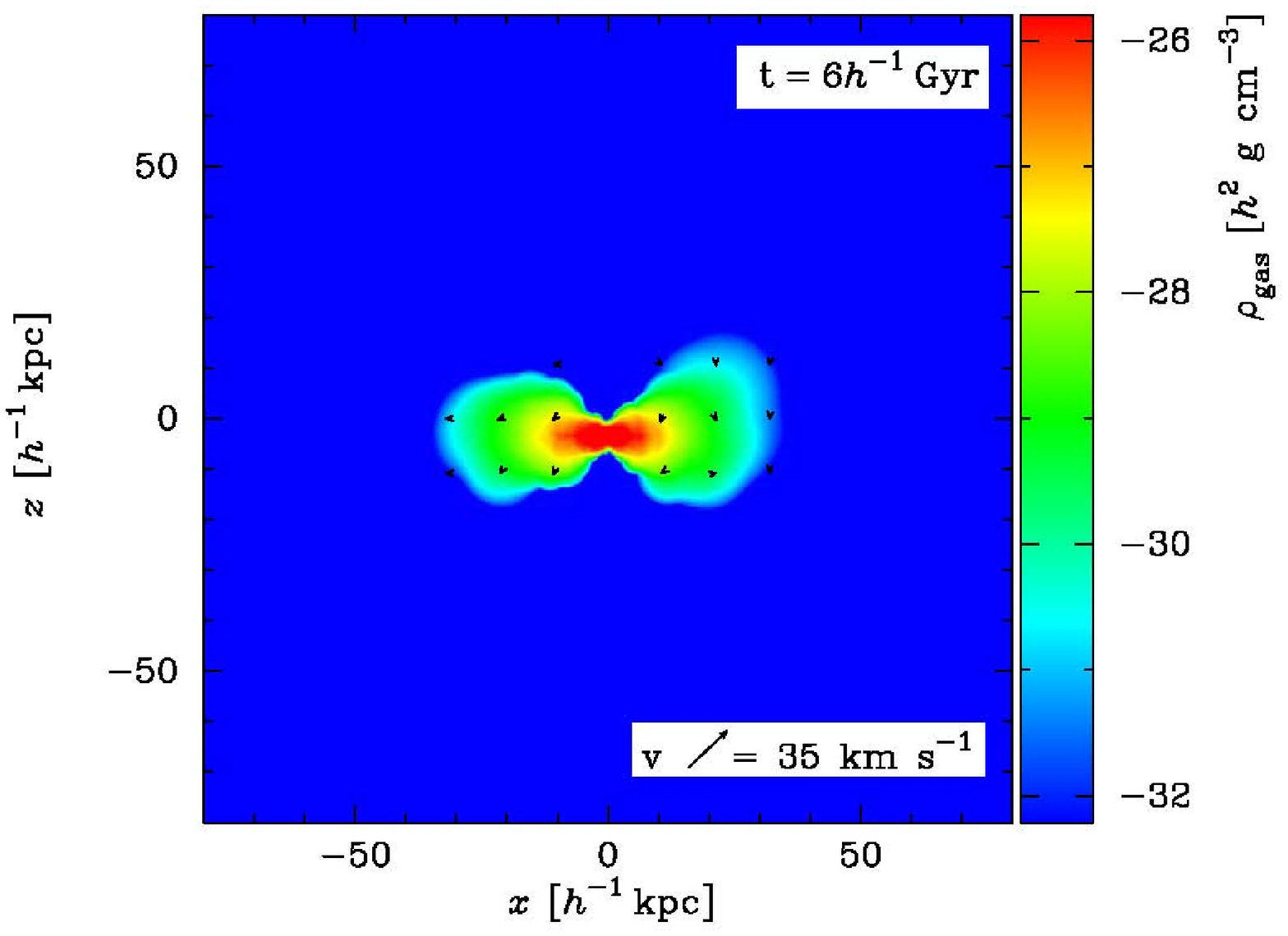}
\end{tabular}
\caption{Maps of the baryonic gas density and the gas velocity field (indicated
  by arrows) for a halo of total mass $10^{9}\,\hm$ (top panels) and $10^{10}\,\hm$
  (bottom panels) at $t=6\,\hg$. The maps show an edge-on slice through the
  midplane of the galaxy that forms in the centre.  The left panels correspond
  to the simulations with CR streaming and advection, while the right panels
  show the simulations with purely advective CR transport. While slow
  (thermal) gas motions dominate the velocity field in the advection-only run,
  the simulation employing CR streaming and advection shows a powerful
  outflow.}\label{fig: rho-maps}
\end{figure*}

Since outflows cannot occur in simulations without CR feedback (or
radiation pressure), at least for the specific star formation model
that we use, we will in the following compare simulations with only
advective CR transport to simulations with both, advective and
streaming transport of CRs.  Since the advection-only case was shown
not to drive outflows \citep{Jubelgas2008}, this numerical setup
enables us to assess the question whether CR streaming is able to
launch galactic winds. In Figure~\ref{fig: rho-maps}, we show maps of
the baryonic gas density in haloes of total mass $10^{9}\,\hm$ (top
panels) and $10^{10}\,\hm$ (bottom panels), seen in an edge-on slice
through the midplane of the galaxy that forms in the centre. Those
maps are supplemented by velocity vectors of the gas and are shown at
$t=6\,\hg$ after the start of the simulation when the initial burst of
star formation has ended and the galaxies have settled approximately
into steady state. For reference, the virial radii of the two haloes
are $R_{200}\approx17\,\kpc$ and $R_{200}\approx 35\,\kpc$,
respectively.

When comparing the simulations with CR streaming and advection (left)
to the advection-only case (right), it becomes clear that the
additional mobility of the streaming CRs enables the driving of a
powerful outflow.  This manifests itself in a shallower density
profile in the CR-streaming case, suggestive of a substantial mass
transport beyond the virial radius.  The outward pointing gas
velocities with maximal values of
$\vel_{\mathrm{max}}\gtrsim\vel_{\rmn{esc}}$ demonstrate the presence
of a wind driven by streaming CRs that extends well beyond $3
R_{200}$, allowing the gas to escape from the gravitational influence
of the halo. At the largest scales after $\sim 3$~Gyrs, we obtain
converged values for the wind velocities of $\vel\sim 45 \,\kms$ and
$135 \,\kms$ for our two haloes ($10^{9}\,\hm$ and $10^{10}\,\hm$),
respectively. (Note however that owing to our simplified initial
conditions, our simulations cannot be taken as a realistic picture for
radii larger than a few $R_{200}$ where the wind will encounter the
anisotropic mass distribution of the cosmic web.)

Comparing the CR-streaming simulations of the different halo masses,
we observe a quasi-spherical outflow in the dwarf halo with
$10^{9}\,\hm$. In contrast, there is a bi-conical, hour-glass shaped
wind in the $10^{10}\,\hm$ halo within $R_{200}$ that starts to become
more spherical on larger scales. This different wind morphology can be
traced back to the density distribution of the star-forming ISM which
determines the initial conditions at the base of the outflow. While
the shallow potential of the dwarf halo is not able to confine the ISM
into a thin disc and instead has an ISM with a quasi-spherical
morphology, the larger halo shows a well-confined galactic disc. Such
a configuration has a vertical stratification of CR pressure which
determines the initial direction of the CR streaming and focuses the
forming wind into azimuthal symmetry.

We now detail the physics of the launching mechanism of the wind. Initially, the
CR pressure distribution peaks at the star forming regions in the galactic disc
which accelerate them in SN remnant shocks. This causes a steep CR pressure gradient
that implies the onset of CR streaming out of the galaxy. A fluid element above
the disc which is in approximate hydrostatic equilibrium with the surrounding
fluid elements receives the (instreaming) CR flux and becomes overpressured. CR
loss processes (wave heating as a result of CR streaming instabilities or
particle-particle interactions) transform a part of the CR pressure to the
thermal plasma.  Since the thermal plasma has a harder equation of state, it
gains more in terms of pressure than the CRs have lost by this transfer (see
Section~\ref{sec:CR-streaming}). This increases the total pressure of this fluid
element furthermore. This excess pressure causes a weak shock that accelerates
the gas in the stratified atmosphere. Momentum conservation ensures that this
weak shock moves in the same (outwards) direction as the original CR-streaming
flow. The rarefaction wave, moving in the opposite direction, dilutes the gas in
the tail and causes a buoyant rearrangement of the flow.  We see that the
streaming CRs are crucial in this picture for launching the outflow.

\begin{figure}
\centerline{\includegraphics[width=\linewidth]{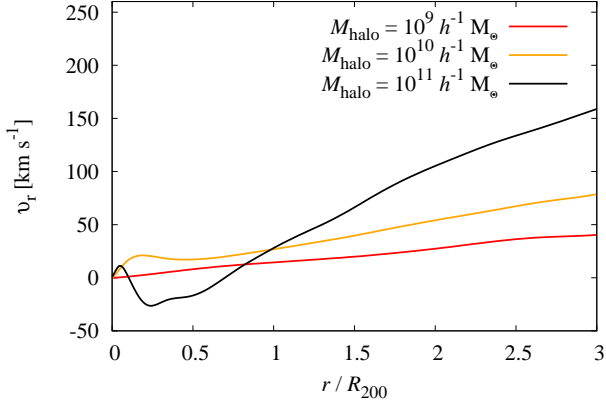}}
\caption{Radial velocity of the gas in a cone of opening angle $20^\circ$,
  aligned with the angular momentum axis of the disc for our three haloes of
  masses $10^{9}\,\hm$, $10^{10}\,\hm$, and $10^{11}\,\hm$ at $2.5\,\hg$. In the
  two lower-mass haloes (as well as for $r>0.7\,R_{200}$ in the $10^{11}\,\hm$
  halo), we see an accelerating wind over the entire radial range which is due
  to a continuous CR momentum and energy deposition during the ascent of the
  wind in the gravitational potential. For $r<0.7\,R_{200}$ in the
  $10^{11}\,\hm$ halo, the wind stalls and falls back onto the disc (shown by
  the negative velocity values) which is the characteristics of a fountain
  flow.}
\label{fig: velocity}
\end{figure}

A central question for winds is the evolution of their outflow velocity along
the streamlines and whether they are energy or momentum driven. In the first
case, it is known to be difficult to obtain the observed large mass loading
factors. Additionally, if the (uncertain) clumping factor in the wind is larger
than unity, the wind would radiate away more energy per unit time than in the
smooth case which could potentially stall it. In Fig.~\ref{fig: velocity} we
show the radial velocity of the gas in a cone of opening angle $20^\circ$,
aligned with the angular momentum axis of the disc. In the two lower-mass haloes,
we see an accelerating wind over the entire radial range. This is due to the
particular property of CR streaming-driven winds, which represents neither a
classic energy-driven nor a momentum-driven wind in a sense that they are not
preloaded with those conserved quantities. Instead, they are continuously
re-loaded with energy and momentum during their ascent in the gravitational
potential through the Alfv\'en-wave heating term in the energy equation and the
$\nabla P_\rmn{cr}$ term in the momentum equation.

While there was initially a wind launched in the $10^{11}\,\hm$ halo that
propagates on scales $r>0.7\,R_{200}$ (at $2.5\,\hg$), it apparently stalled for
$r<0.7\,R_{200}$ and started to fall back as a fountain flow onto the galactic
disc (see Fig.~\ref{fig: velocity}). We observe the same behaviour for later
times in the $10^{10}\,\hm$ halo.  The reason for this is the increasing stellar
mass and the formation of a thin galactic disc which causes the wind to be
launched deeper in the gravitational potential (implying a small $r_b$) and to
loose more binding energy during its ascent in the potential.\footnote{In both
  cases, the thermal pressure distribution is monotonically decreasing for
  increasing $r$ so that there is no pressure barrier building up in the halo
  due to CR Alfv\'en-wave heating that could potentially stall the wind.}
Another effect that can partly affect the sustainability of winds in larger
haloes are the increased CR losses due to Alfv\'en-wave heating in the dense ISM
which is characterized by steep CR pressure gradients. For our current
description of the ISM with an effective equation of state, we may overestimate
the CR losses as CRs are kept for a longer time in the dense phase of the
ISM. In reality, they may be able to escape more quickly into the warm and more
dilute phase of the ISM that is characterized by a smoother CR pressure gradient
so that less CR energy is lost at the initial phases of the wind
launching. Without improving our modelling of the ISM or the CR transport within
the ISM the resulting wind velocities could thus be underestimated.  We will
return to this point in Section~\ref{sec: mass loss} and estimate its effect on
the cumulative mass loss in a $10^{9}\,\hm$ halo.

\begin{figure*}
\begin{tabular}{cc}
{\large {\em $10^{9}\,\hm$ halo}} & {\large {\em $10^{10}\,\hm$ halo}} \\
\includegraphics[clip,scale=0.43]{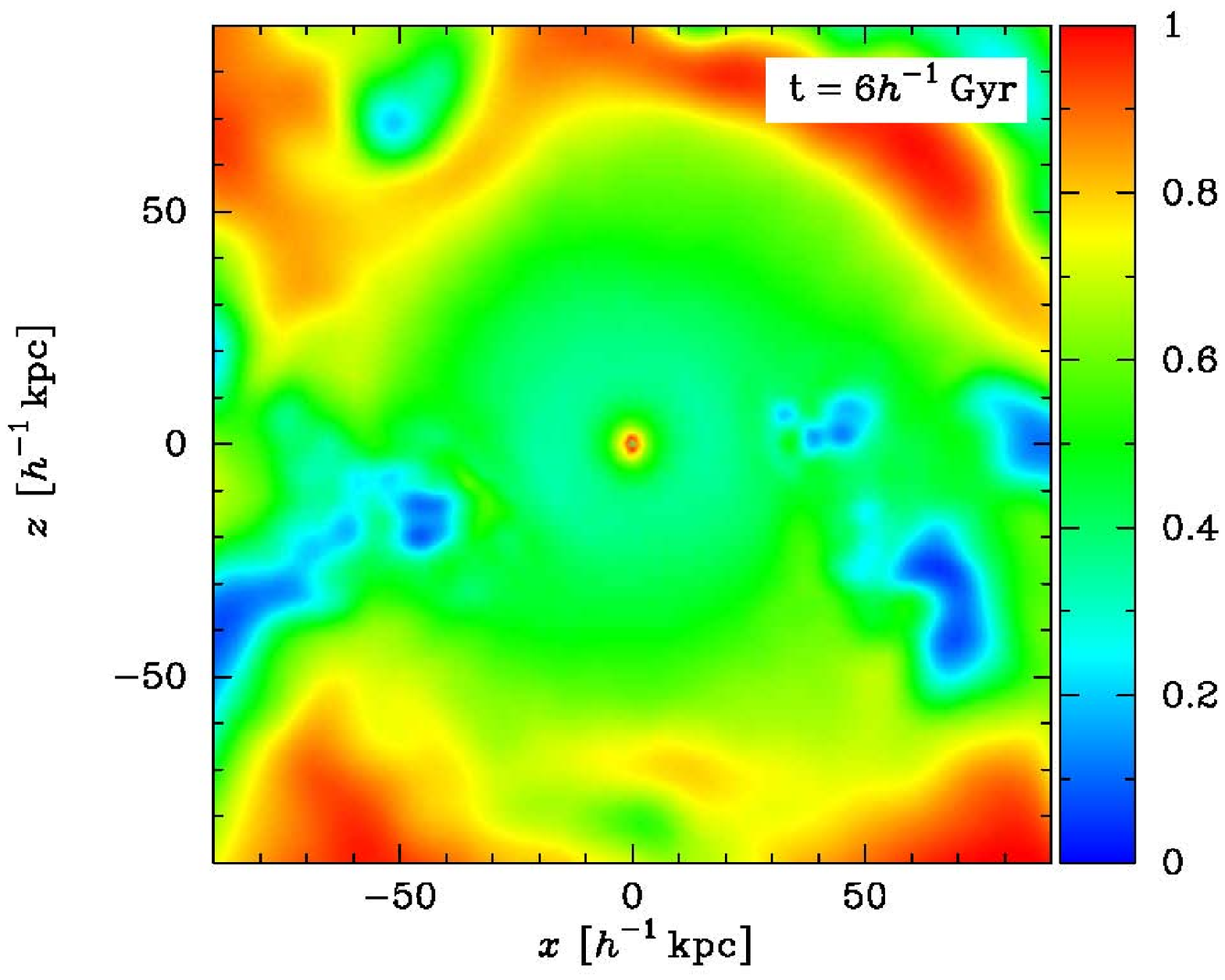} & 
\includegraphics[clip,scale=0.43]{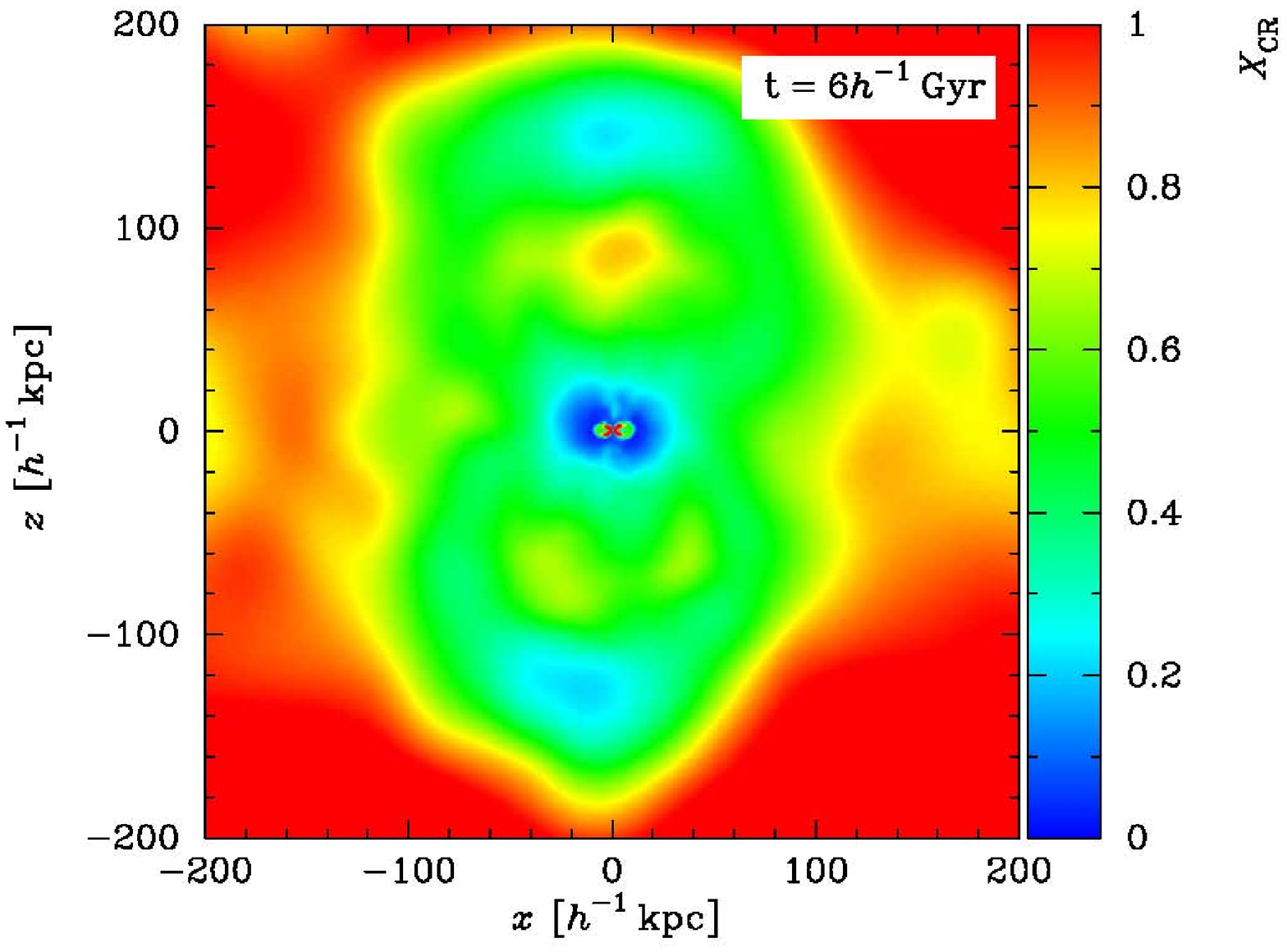}
\end{tabular}
\caption{CR-to-thermal pressure ratio, $X_\rmn{cr} = P_\rmn{cr} / P$, in an
  edge-on slice through the galactic disc. Shown is the simulation with CR
  streaming and advection in haloes of $10^{9}\,\hm$ (left) and $10^{10}\,\hm$
  (right) at time $t=6\,\hg$. The CR-to-thermal pressure ratio is less than $50\%$ in
  the vicinity of the centre, because of loss processes that effectively
  transfer CR energy into the thermal reservoir, but becomes dominant at larger
  heights due to the softer adiabatic index of CRs.}
\label{fig: XCR}
\end{figure*}

To highlight the launching mechanism of the wind, we study the CR-to-thermal
pressure ratio, $X_\rmn{cr} = P_\rmn{cr} / P$, in an edge-on slice through the
galactic disc in Fig. \ref{fig: XCR}. Generally, we find an increasing CR
pressure fraction, $X_\rmn{cr}$, at larger radii. As the wind propagates, the
composite gas of CRs and thermal plasma experiences adiabatic expansion so that
the pressure drops less quickly than the thermal pressure due to the composite's
softer equation of state. Thus, the CR-to-thermal pressure ratio rises to
$X_\rmn{cr}\gtrsim 1$ at large radii.  Apparently, this effect wins over CR
energy losses due to CR Alfv\'en-wave heating during the ascent of the wind in
the halo potential.

The $X_\rmn{cr}$ map of the $10^{9}\,\hm$ halo shows a homogeneous morphology for
$r<R_{200}$ with values around $X_\rmn{cr} \simeq 0.5$. At larger radii, the
morphology of the $X_\rmn{cr}$ map becomes patchier. It is interesting that CR
streaming is unable to smooth this inhomogeneous CR pressure distribution. This
is because of the large wind velocities which reach values of $\vel\sim50\,\kms$,
whereas the sound speed (equal to the streaming speed in our model) is only
around $c_{\mathrm{s}}\sim(5-10)\,\kms$ or less. Thus, advection dominates the
transport on these scales which is not expected to result in a smooth
distribution of CR pressure and apparently does not significantly mix. The
$X_\rmn{cr}$ map of the $10^{10}\,\hm$ halo shows the hour-glass morphology and
supports the picture of a bi-conical outflow driven by CR streaming. Clearly
visible are the torus-shaped regions in blue on scales below the virial radius
which indicate a low relative CR pressure. This is due to a converging vortex
flow towards the midplane and a necessary consequence of the bipolar
outflow. Here, the CR pressure component is disfavoured in comparison to the
thermal component upon adiabatic compression.

\subsection{Mass loss due to galactic winds}
\label{sec: mass loss}

\begin{figure*}
\begin{tabular}{cc}
\includegraphics[scale=0.59]{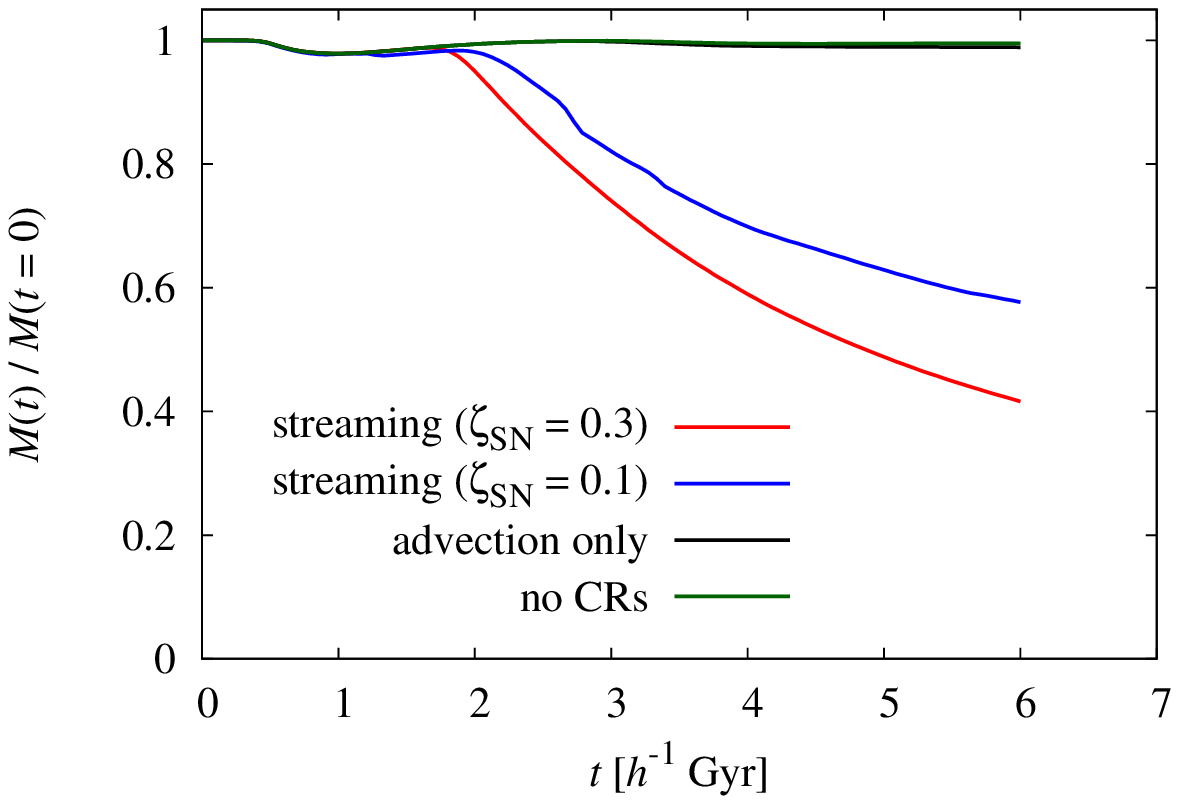} & 
\includegraphics[scale=0.59]{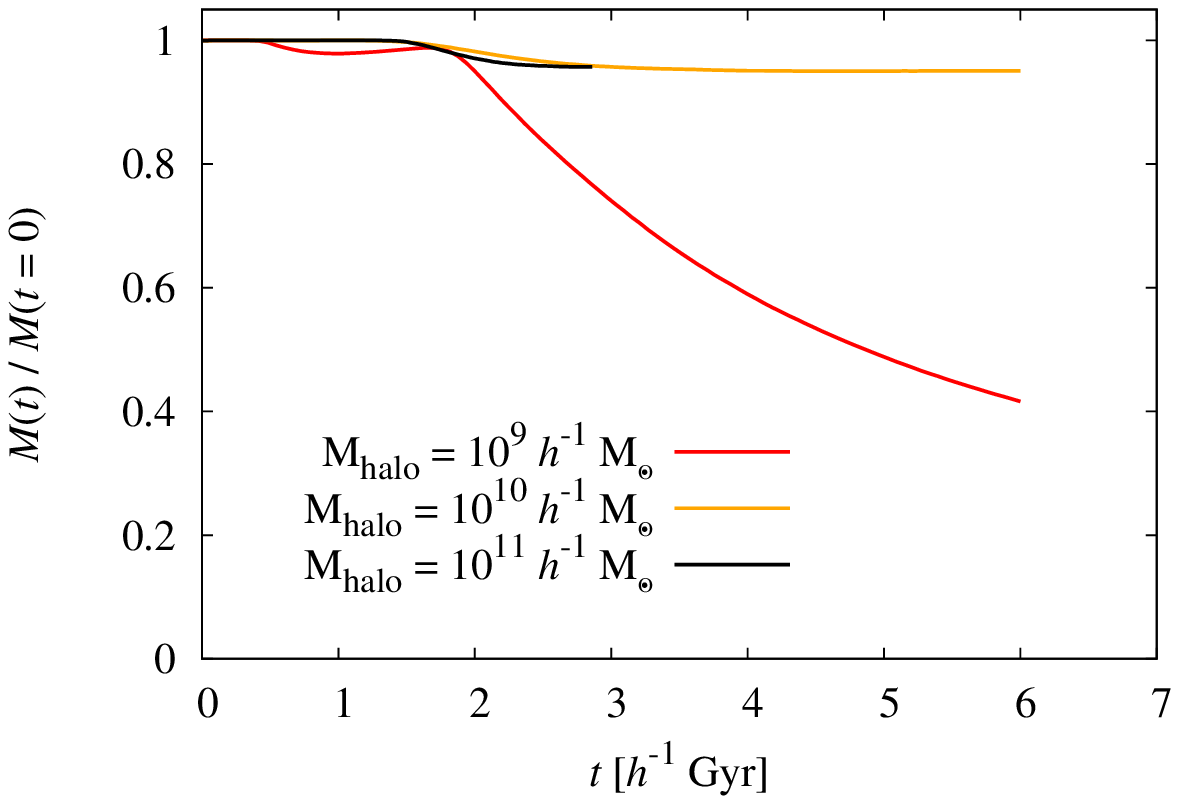}
\end{tabular}
\caption{Time-evolution of the baryonic mass (in form of gas and stars) enclosed
  within the virial radius, $M(t)$, normalized to the initial mass at the
  beginning of the simulations. Left: we compare $M(t)$ for four different
  scenarios (as indicated in the legends) in a halo of total mass $10^{9}\,\hm$.
  Right: we show $M(t)$ for three different halo masses, $10^{9}\,\hm$,
  $10^{10}\,\hm$, and $10^{11}\,\hm$, always adopting the CR streaming and advection
  scenario with $\zeta_{\mathrm{SN}}=0.3$.}
\label{fig: mass loss}
\end{figure*}

\begin{figure*}
\begin{tabular}{cc}
{\large {\em $10^{9}\,\hm$ halo}} & {\large {\em $10^{10}\,\hm$ halo}} \\
\includegraphics[scale=0.59]{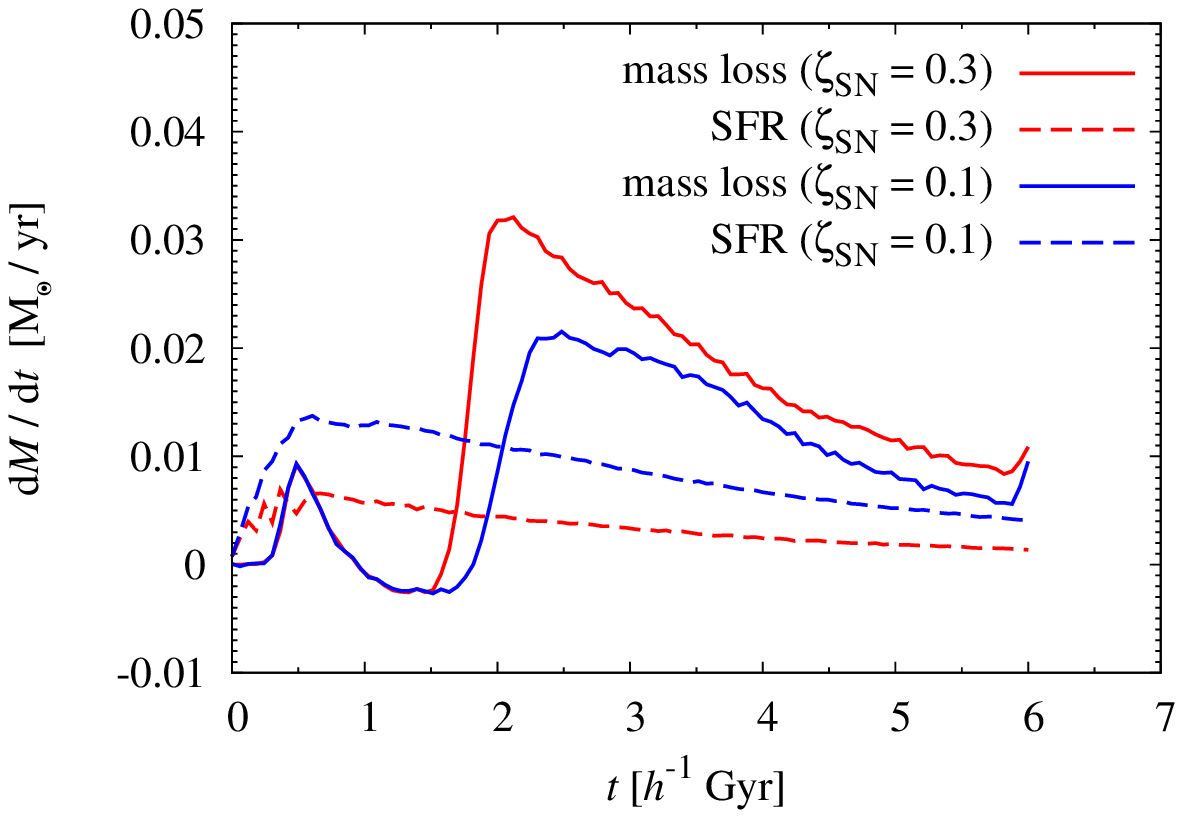} & 
\includegraphics[scale=0.59]{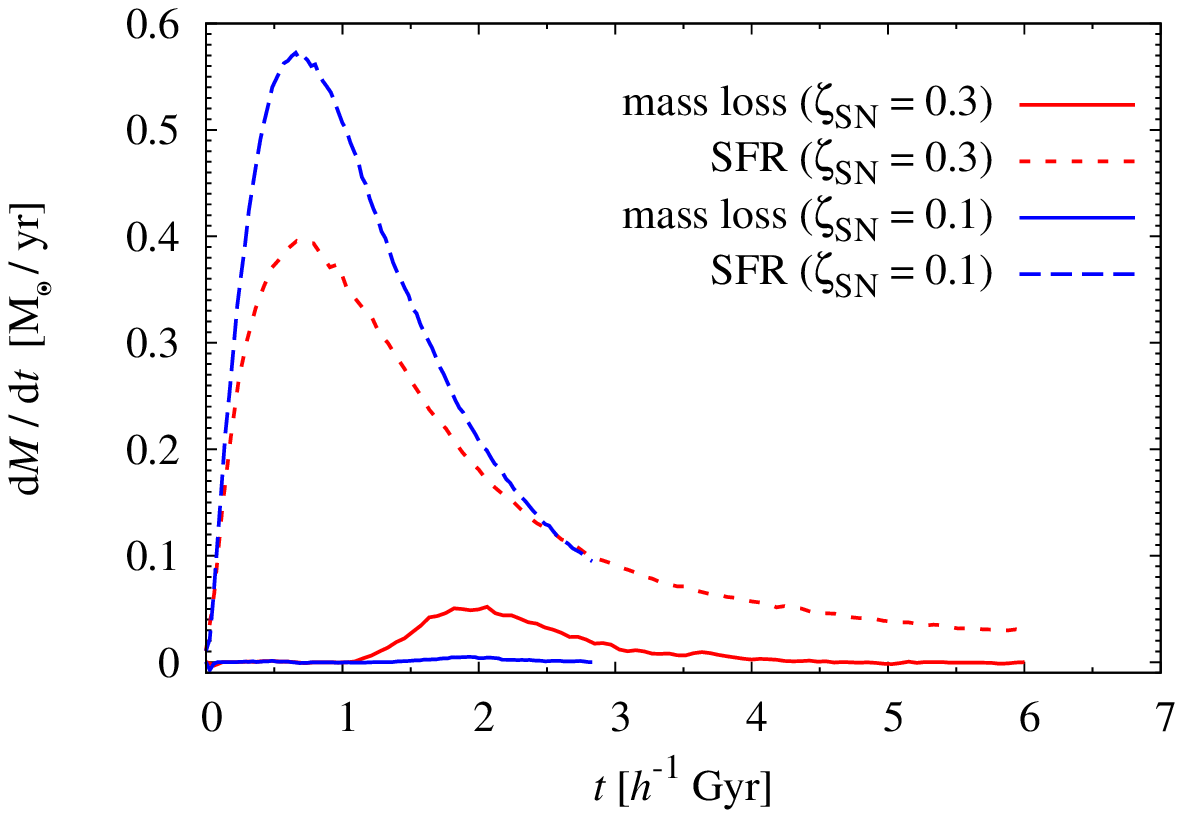}
\end{tabular}
\caption{Mass loss rates and SFRs as a function of time for the CR streaming and
  advection case, computed within the virial radius in haloes of total mass
  $10^{9}\,\hm$ (left) and $10^{10}\,\hm$ (right). The mass loss rate was
  calculated for two different acceleration efficiencies,
  $\zeta_{\mathrm{SN}}=0.1$ and 0.3.}
\label{fig: mass loss rates}
\end{figure*}

After studying the launching mechanism of galactic winds through CR streaming,
we will now quantify the associated mass loss and the mass loading of the
wind. In Fig. \ref{fig: mass loss}, we show the fractional mass as a function of
time for a reference simulation without CR feedback, for a simulation with
advective CR transport-only, and for two runs that additionally include CR
streaming and differ only in the adopted CR acceleration efficiencies of
$\zeta_{\mathrm{SN}}=0.1$ and 0.3. We define fractional mass as the baryonic
mass in form of stars and gas contained within a certain radius at a given time,
normalized to the total gas mass within this radius at the beginning of the
simulation.

The reference run without CR feedback exhibits no mass loss and so does the
advection-only run, in accordance with the findings in
Section~\ref{sec:launching}. The fact that there is no wind in the former is
connected to the implementation of stellar feedback that we use, which does not
allow for outflows without additional adjustments. In contrast, we detected a
strong wind in our simulation with CR streaming which results in a
considerable mass loss (see Fig. \ref{fig: mass loss}). Depending on the assumed
CR acceleration efficiency, a fraction of $(40-60)$\% of the original gas mass
contained within the virial radius of the $10^{9}\,\hm$ halo at the start of the
simulation is expelled until $t\sim6\,\hg$. The non-zero slope of the mass loss
history indicates a moderate continuation of the mass loss if we increased the
run time of our simulation.

The mass loss history shows a strong dependence on halo mass with a fractional
mass loss of only $\sim 5$\% from the virial radius in the case of the $10^{10}$
and $10^{11}\,\hm$ haloes (see right panel of Fig.~\ref{fig: mass loss}). Thus,
these outflows are weaker in comparison to that in the dwarf halo of
$M_{200}=10^{9}\,\hm$, and do not result in a severe mass loss. Since
the mass loss history has saturated in the two high-mass haloes, further mass
loss is not expected for increasing simulation time.  Apparently, the
gravitational attraction is too strong for the CRs to launch a strong wind in
these cases, but certainly drives powerful fountain flows.

However, the mass loss history in Fig. \ref{fig: mass loss} demonstrates that
the outflow is not able to develop until $t\sim1.8\,\hg$ which is due to the ram
pressure from inflowing gas that the wind needs to overcome first. Ram pressure
can therefore significantly delay wind formation and is probably more important
in this respect than the shallow gravitational potential of this $10^{9}\,\hm$
halo. This result is in agreement with previous studies of outflows
\citep{Fujita2004,Dubois2008} and may be partly due to our simplified initial
conditions. If most of the gas mass is accreted onto a galaxy along filaments
that cover a small solid angle, the ram pressure in directions other than the
filaments would be substantially reduced, possibly allowing for an earlier onset
of a wind after the first burst of star formation.

It is interesting to compare the mass loss rate in the wind with the SFR. These
are expected to be roughly proportional to each other, as simple theoretical
predictions and observations suggest. Thus, in Fig. \ref{fig: mass loss rates},
we show the time evolution of the SFR and the mass loss rate within the virial
radius for two CR streaming simulations that differ only in the acceleration
efficiency of CRs. The influence of ram pressure is apparent, which delays a
significant escape of gas until $t\sim 1.8\,\hg$.\footnote{In both simulations,
  there is already a small amount of gas leaving the halo around
  $t\sim0.7\,\hg$. This is not due to a wind but rather caused by the build-up
  of thermal and CR pressure in the centre of the galaxy owing to star
  formation, that reaches its maximum at the same time. This will slowly push a
  small amount of gas over the boundary of the halo. However, the negative mass
  loss rate immediately thereafter indicates that this gas flow does not
  propagate far, but returns soon.}  Once the CR-driven wind overcomes the ram
pressure, the mass loss rate increases sharply.

\begin{figure*}
\begin{tabular}{cc}
{\large {\em $10^{9}\,\hm$ halo}} & {\large {\em $10^{10}\,\hm$ halo}} \\
\includegraphics[scale=0.59]{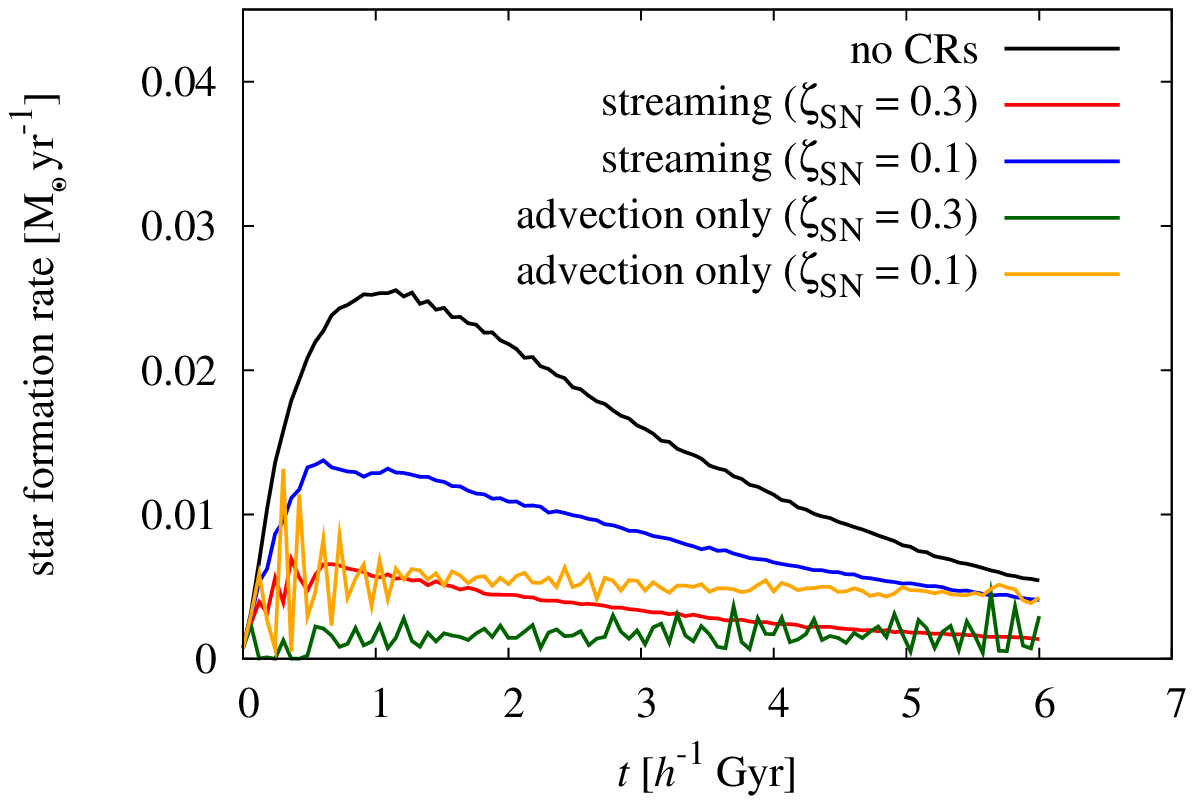} & 
\includegraphics[scale=0.59]{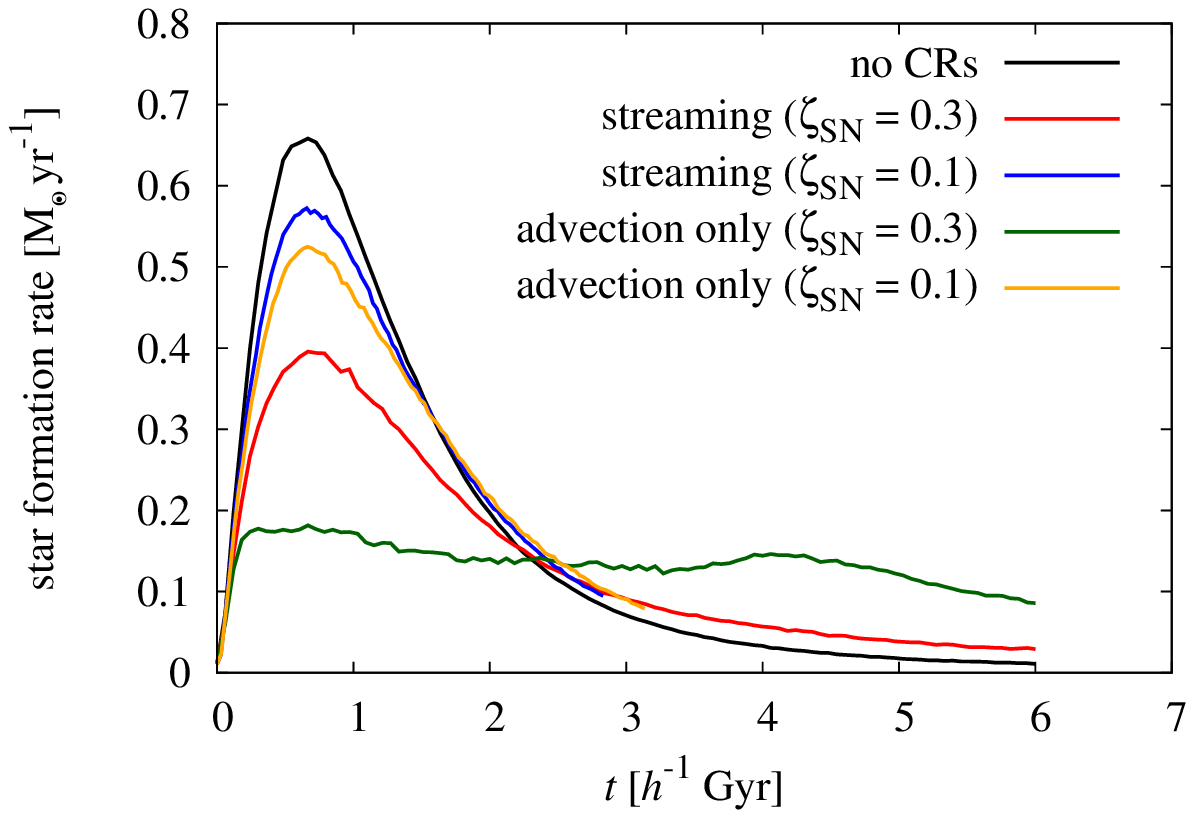} \\
\includegraphics[scale=0.59]{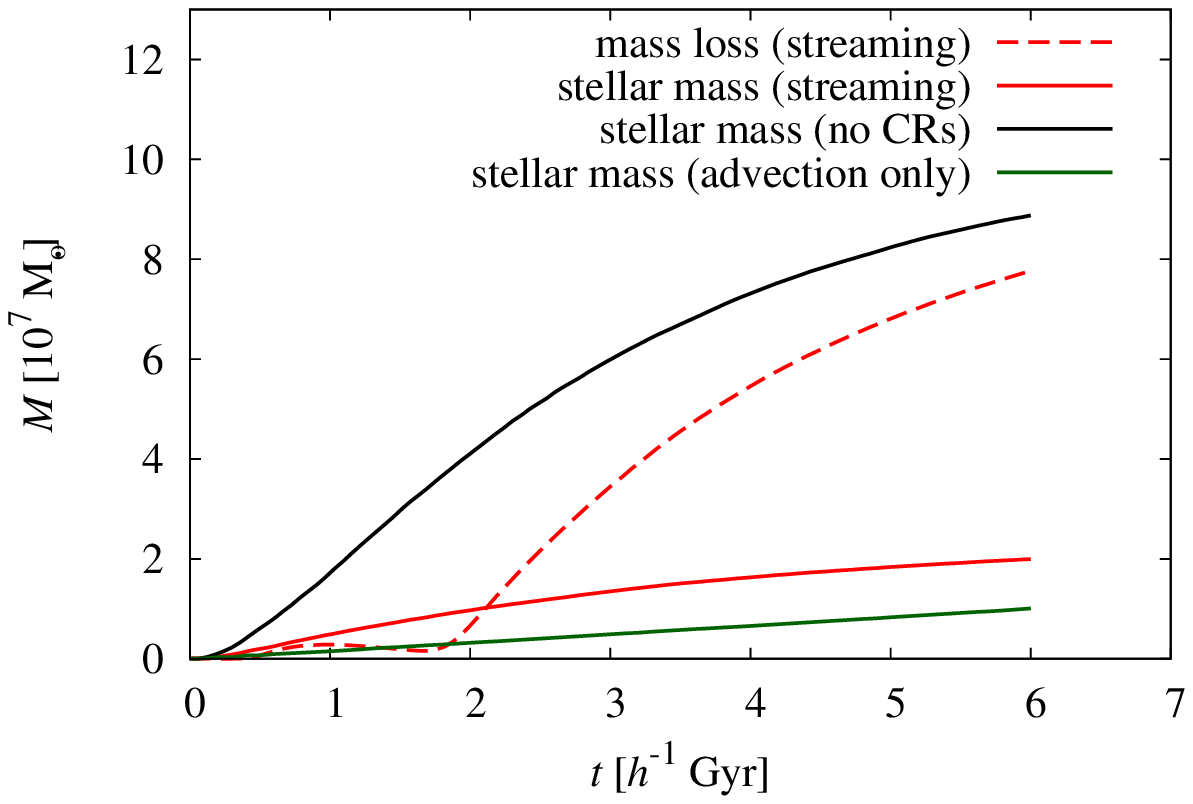} &
\includegraphics[scale=0.59]{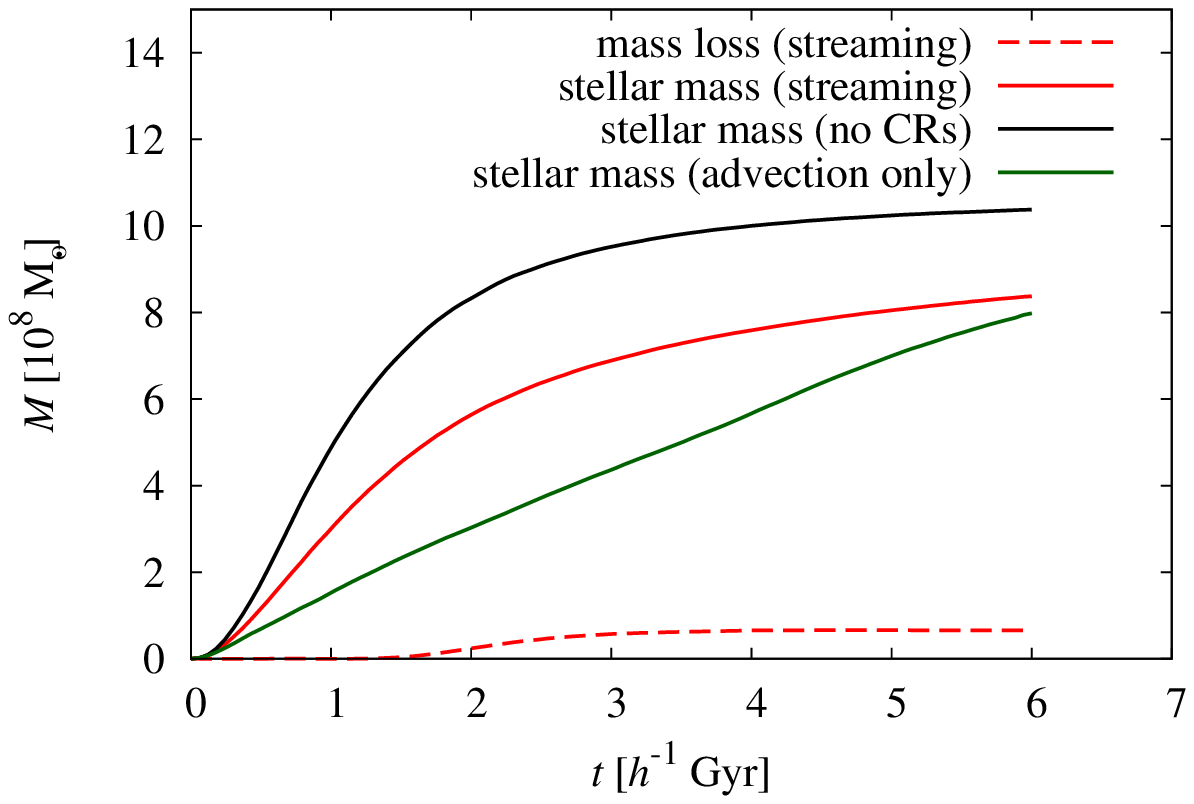}
\end{tabular}
\caption{Top panels: SFR as a function of time in a halo of total mass
  $10^{9}\,\hm$ (left) and $10^{10}\,\hm$ (right). We compare five different
  simulations: a reference simulation without CRs (black line) and four runs
  with CR feedback, differing only in the acceleration efficiency and the
  transport scheme, i.e., advective transport-only (orange line:
  $\zeta_{\mathrm{SN}}=0.1$, green line: $\zeta_{\mathrm{SN}}=0.3$) and CR
  streaming and advection (blue line: $\zeta_{\mathrm{SN}}=0.1$, red line:
  $\zeta_{\mathrm{SN}}=0.3$), respectively. CR feedback leads to a strong
  suppression of the SFR in comparison to the reference simulation. Bottom
  panels: Time-evolution of the total mass lost in the wind out of the virial
  radius $R_{200}$ (dashed red line), and of the total mass in stars (solid red
  line) for the CR streaming run. These are compared to the total mass in stars
  in the advection-only run (green line) and in the reference simulation without
  CRs (black line). Both runs with CRs adopt $\zeta_{\mathrm{SN}}=0.3$. In the
  $10^{9}\,\hm$ halo, the amount of mass removed from the halo in the CR
  streaming run is considerably larger than the stellar mass that forms in this
  simulation, in agreement with the large mass loading of the outflow. Moreover,
  the total mass in stars is significantly less than in the reference
  simulation.}
\label{fig: sfr}
\end{figure*}

In case of the high acceleration-efficiency run for the dwarf haloes of
$10^{9}\,\hm$, the mass loss rate is about $5$ times the SFR and about $1.5$
times the SFR for the lower acceleration efficiency (we always consider the
``peak mass loading'', i.e., we divide the respective maxima of the mass loss
and star formation histories). For dwarfs, the mass loading of these winds is
therefore considerable. In the end of our high acceleration-efficiency
simulation, we find that the wind has expelled a gas mass fraction of
$\simeq60\%$ while $\simeq15\%$ of the initial gas mass ended up in stars (which
was considerably smaller in comparison to the stellar mass fraction of
$\simeq70\%$ for our run without CR feedback, see Fig.~\ref{fig: sfr}). In
contrast, the wind in the $10^{10}\,\hm$ halo only has a small mass loading of
about $0.13$ for a high acceleration efficiency. While only $\simeq5\%$ of the
initial gas mass was expelled by the wind in this halo, the stellar mass
fraction dropped from $\simeq80\%$ down to $\simeq60\%$ due to feedback by CR
streaming as well as by advected CRs (Fig.~\ref{fig: sfr}). Apart from that, in
the low acceleration-efficiency CR streaming run there is no mass loss out
of the virial radius.

The energy transfer from the CRs to the thermal plasma following the excitation
and damping of hydromagnetic waves within the dense ISM, that has a
comparatively short cooling time, causes a large fraction of this energy to be
lost to cooling radiation. If CRs are injected into the warm phase of the ISM
from where they would drive a wind, then our simplified treatment of the
multiphase structure of the ISM, which does not explicitly treat the warm and hot
gas in superbubbles, underestimates the effect of CR streaming and hence the
associated mass loss rates. To estimate the importance of this effect, we switch
off the conversion of CR to thermal pressure in our CR streaming implementation
and find a fraction of $90$\% of the original gas mass contained within the
virial radius of the $10^{9}\,\hm$ halo is already expelled by $t\sim4\,\hg$. For
the reminder of this work, we always include CR-wave heating but bear in mind
that the results may be too conservative as a result of too efficient CR energy
losses to radiation in the dense, cool phase of the ISM.

\subsection{Suppression of star formation rates}

To study how CR feedback influences the star formation process, we
plot the star formation history in Fig.~\ref{fig: sfr}. We vary the
simulated physics and compare a model without CR feedback, one with
advective CR transport only, and one that additionally includes CR
streaming. For each of our CR models, we employ two CR acceleration
efficiencies of $\zeta_{\mathrm{SN}}=0.1$ and 0.3,
respectively. Obviously, if feedback by CRs is taken into account, the
SFR is suppressed in comparison to the reference simulation without
CRs.  In case of the CR streaming model with a high acceleration
efficiency, the SFR is suppressed by about $80\%$ ($40\%$) at the peak
of the star formation history for our halo with $10^{9}\,\hm$
($10^{10}\,\hm$). In our model that only accounts for advective CR
transport, this suppression is even increased and amounts to more than
$90\%$ ($70\%$) for our halo with $10^{9}\,\hm$ ($10^{10}\,\hm$); in
accordance with the findings of \citet{Jubelgas2008}.  As laid out
there in detail, the suppression of the SFR by CR feedback is less
effective in higher mass haloes which attain higher central gas
densities due to their deeper potential wells. This is because
adiabatically compressing a composite of CRs and thermal gas
disfavours the CR pressure relative to the thermal pressure due to the
softer equation of state of CRs. Hence, for the higher central
densities present in larger haloes the relative importance of the CR
pressure as well as their modulating effect on the SFR decreases.

There are two reasons why the inclusion of CR streaming suppresses the SFR in
comparison to the reference model without CRs.  First, CR pressure that quickly
builds up as stars form and eventually explode will hinder the gas from
collapsing to densities much larger than the threshold of star formation. This
is due to the inefficient cooling processes of CRs as opposed to the thermal gas
that cools comparatively quickly. Consequently, the SFR, that scales with the
gas surface density as $\Sigma_\rmn{SFR}\propto \Sigma_\rmn{gas}^{1.4}$, will be
reduced. As soon as CRs are removed by cooling processes, collapse can start
again and the SFR increases.  The overall effect is a suppression of star
formation as well as an oscillatory behaviour of the SFR, that can be appreciated
in Fig.~\ref{fig: sfr}. If CR transport via streaming is included, these
oscillations basically vanish because freshly accelerated CRs continuously stream
away from the star forming regions, allowing the gas to reach higher densities
and consequently enhancing the SFR in comparison to the advection-only run.

Second, the wind launched by CR streaming prevents a large fraction of the
infalling gas from accreting onto the galactic disc. Hence, the amount of gas
accumulated in the galactic disc is reduced. This lowers the available amount of
gas that may be turned into stars as well as the surface gravity. The resulting
smaller central gas densities imply a lower SFR.

The runs adopting a low CR acceleration efficiency of $\zeta_{\mathrm{SN}}=0.1$ for
the CRs show less suppression of the SFR in comparison to their corresponding
high-efficiency counterparts. This is due to the lower contribution of CR
pressure to the total pressure in the star forming regions, weakening the
influence of CRs on the process of star formation, as outlined
above. Additionally, the mass loading of the winds is also reduced in this case.
Nevertheless, the suppression is still significant and amounts to $\sim40\%$
($\sim80\%$) at the maximum of the star formation history in the CR streaming
model (advection-only model) for the $10^{9}\,\hm$ halo.

The bottom panels of Fig.~\ref{fig: sfr} show the time-evolution of the total
amount of mass lost from the halo's virial radius for our CR streaming
simulation with $\zeta_{\mathrm{SN}}=0.3$ as well as the total amount of mass
formed in stars at a given time in this simulation. We contrast this to the
cumulative stellar mass for a model without CRs and our model with advective CR
transport-only. This shows again that the mass loading of the outflow in the CR
streaming case is large, with the total mass loss being about $4$ times as large
as the mass in stars at the end of the simulation. Furthermore, the final
stellar mass at the end of the CR feedback runs is reduced by $\sim(80-90)\%$
relative to the final stellar mass in the reference simulation (considering the
$10^{9}\,\hm$ halo). This is in good agreement with the suppression factor of
the SFR at the peak of star formation history. As expected, the reduction is
strongest in the model that exhibits only advective CR transport which results
in the least amount of stars formed at the end of the simulation.

In contrast, the baryonic mass lost in the galactic wind of the larger
halo of $10^{10}\,\hm$ is only $\sim8$\% of the total mass in stars
at $t\sim6\,\hg$, reflecting the small mass loading of the wind (see
Section~\ref{sec: mass loss}). In this larger halo, the final stellar
mass at the end of both CR feedback runs is only reduced by $\sim20\%$
relative to the final stellar mass in the reference simulation. The
large reduction of the SFR at the peak of $\sim70\%$ in the
advection-only run is therefore not seen in the final stellar
mass. This is because of the different star formation history which
shows more continuous rather than a bursty behaviour of the CR
streaming models for this halo.

\subsection{How CR streaming heats the halo gas}

\begin{figure*}
\begin{tabular}{cc}
\includegraphics[clip,scale=0.43]{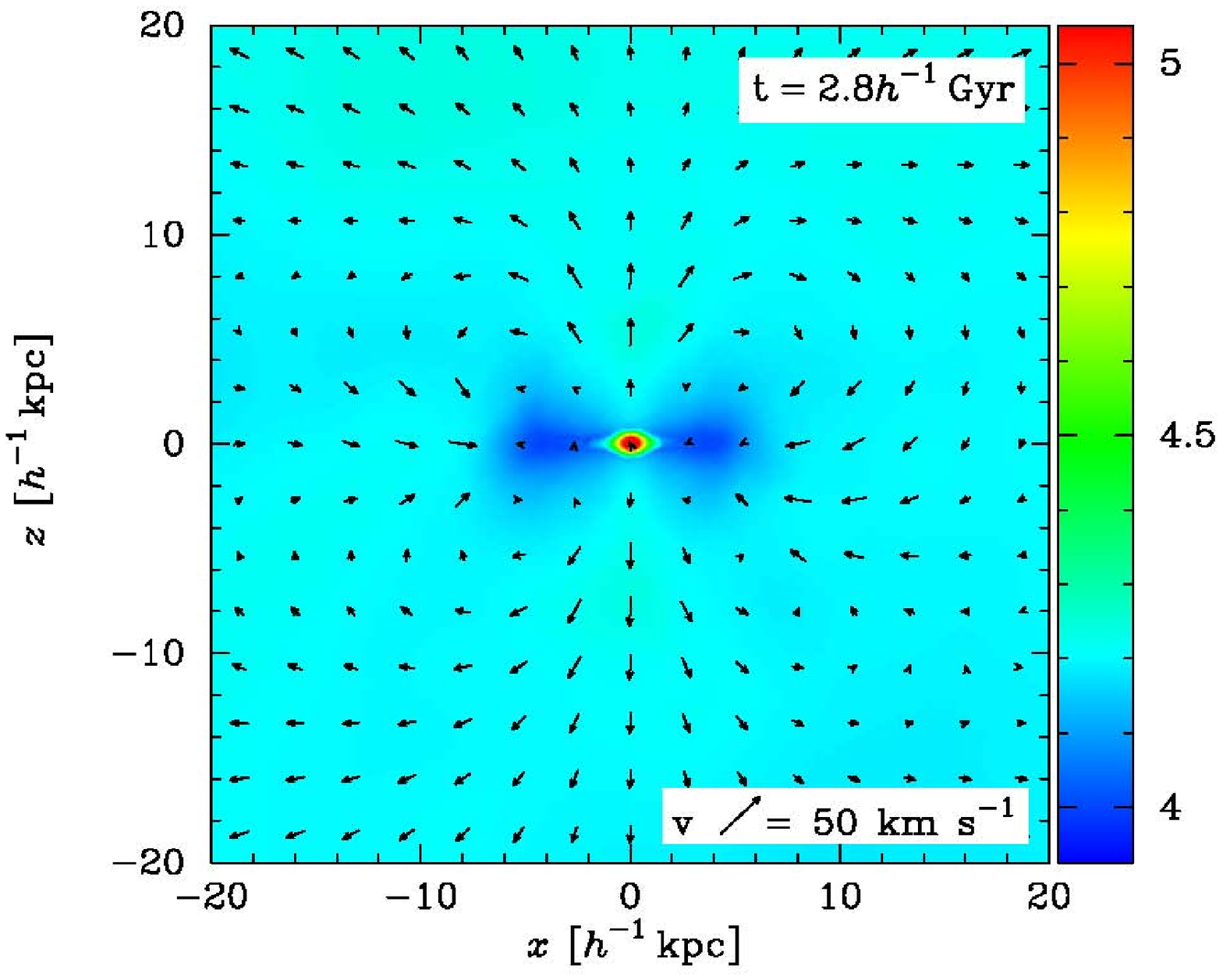} & 
\includegraphics[clip,scale=0.43]{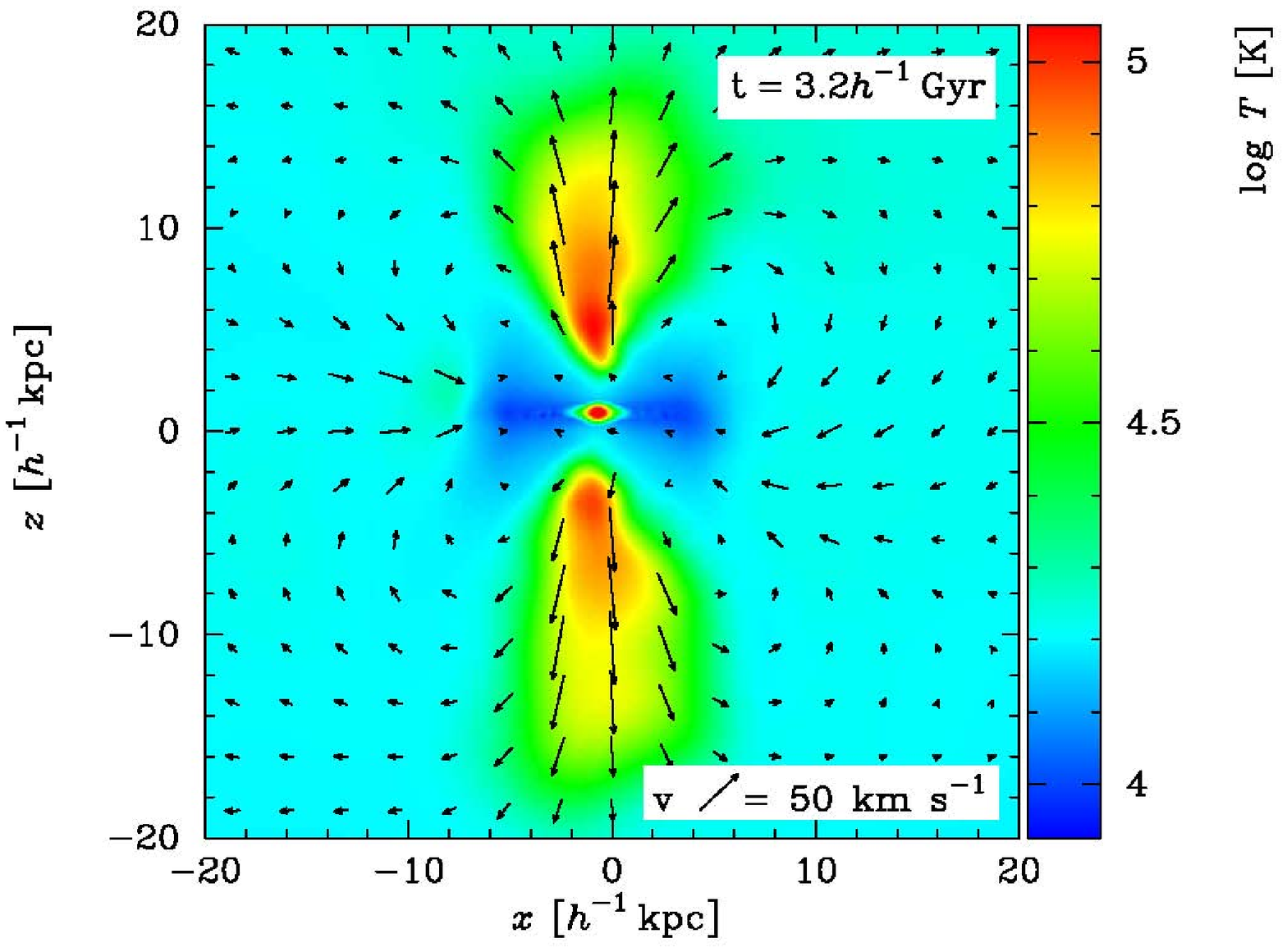}\\
\includegraphics[clip,scale=0.43]{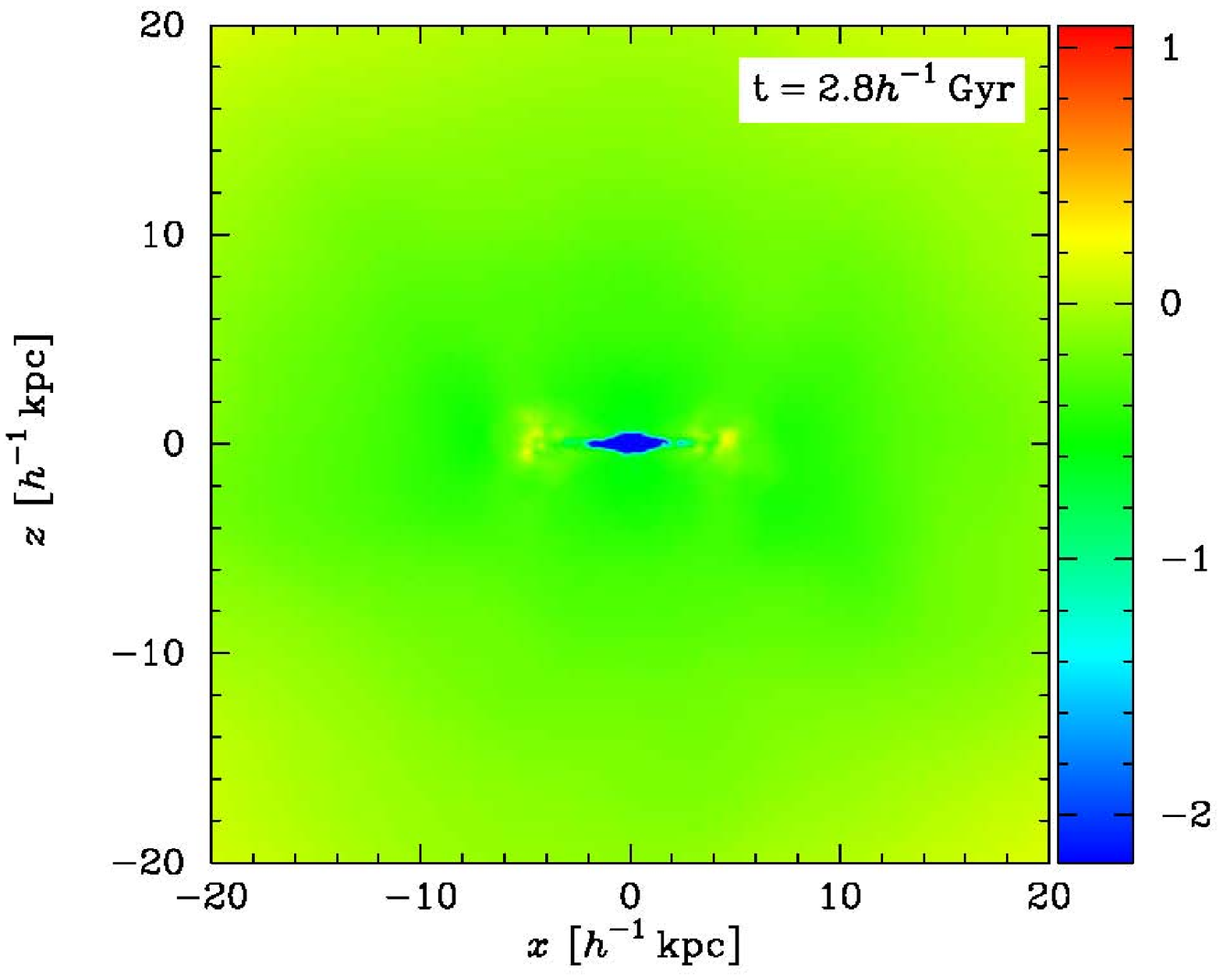} & 
\includegraphics[clip,scale=0.43]{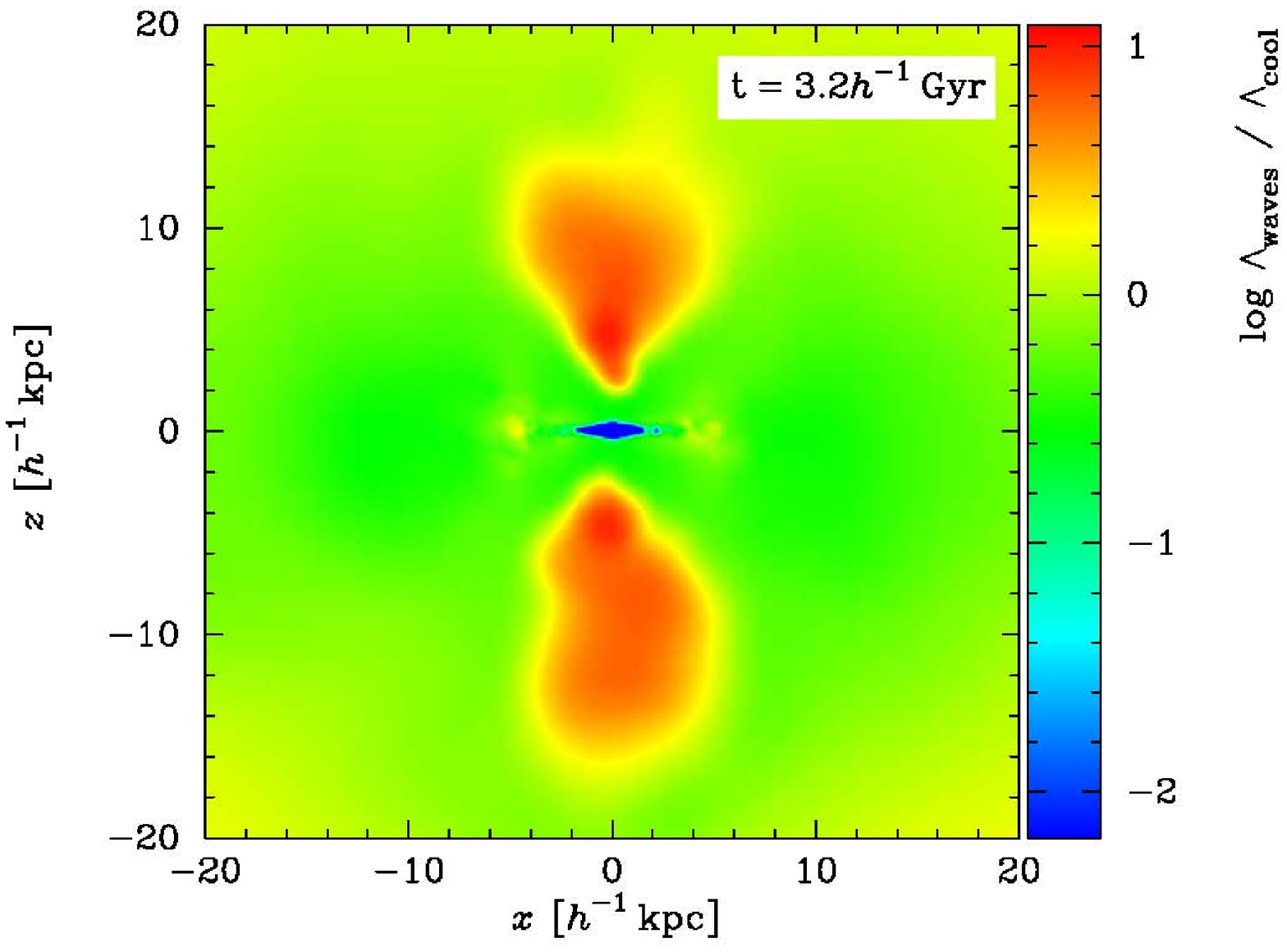}\\
\includegraphics[clip,scale=0.43]{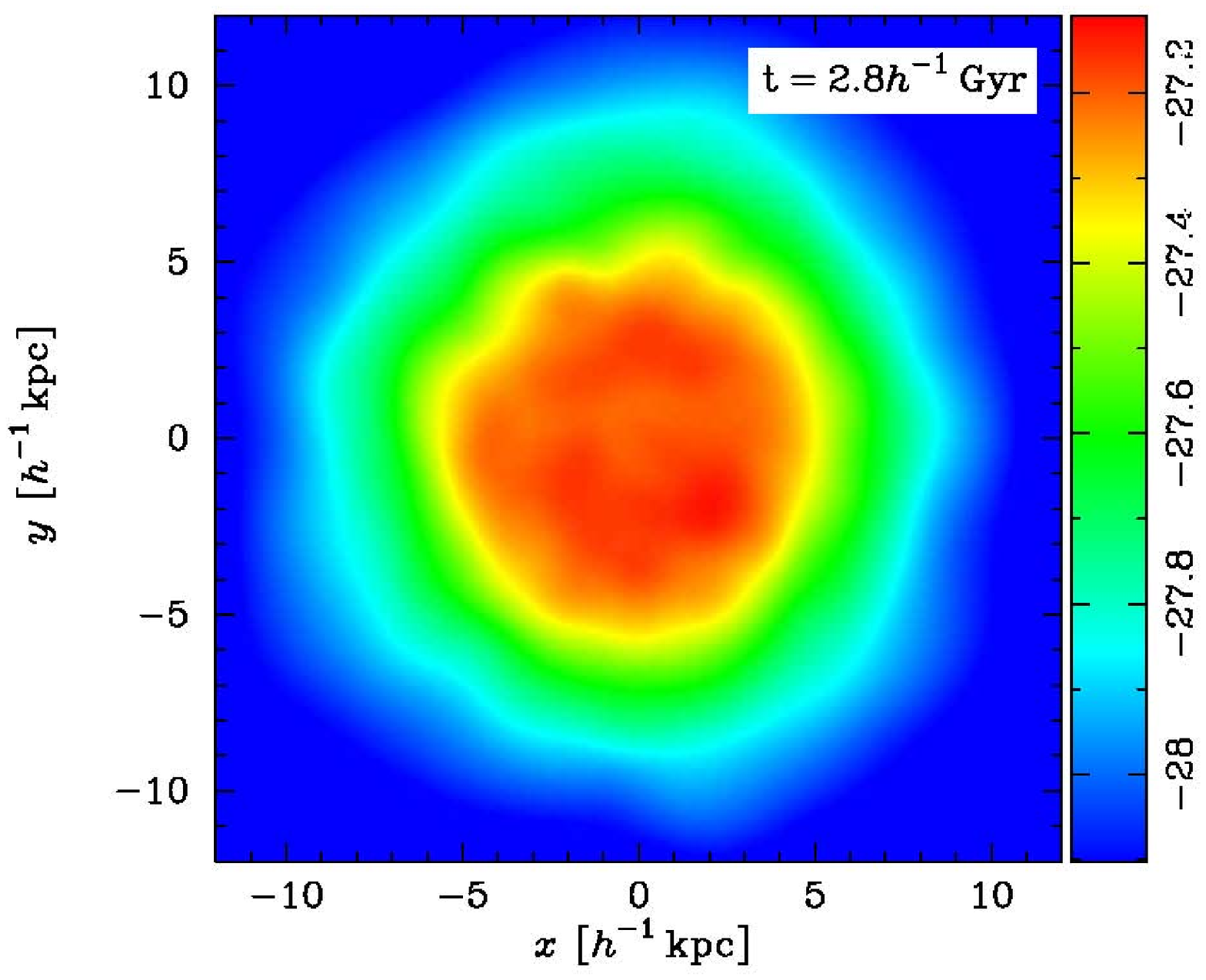} & 
\includegraphics[clip,scale=0.43]{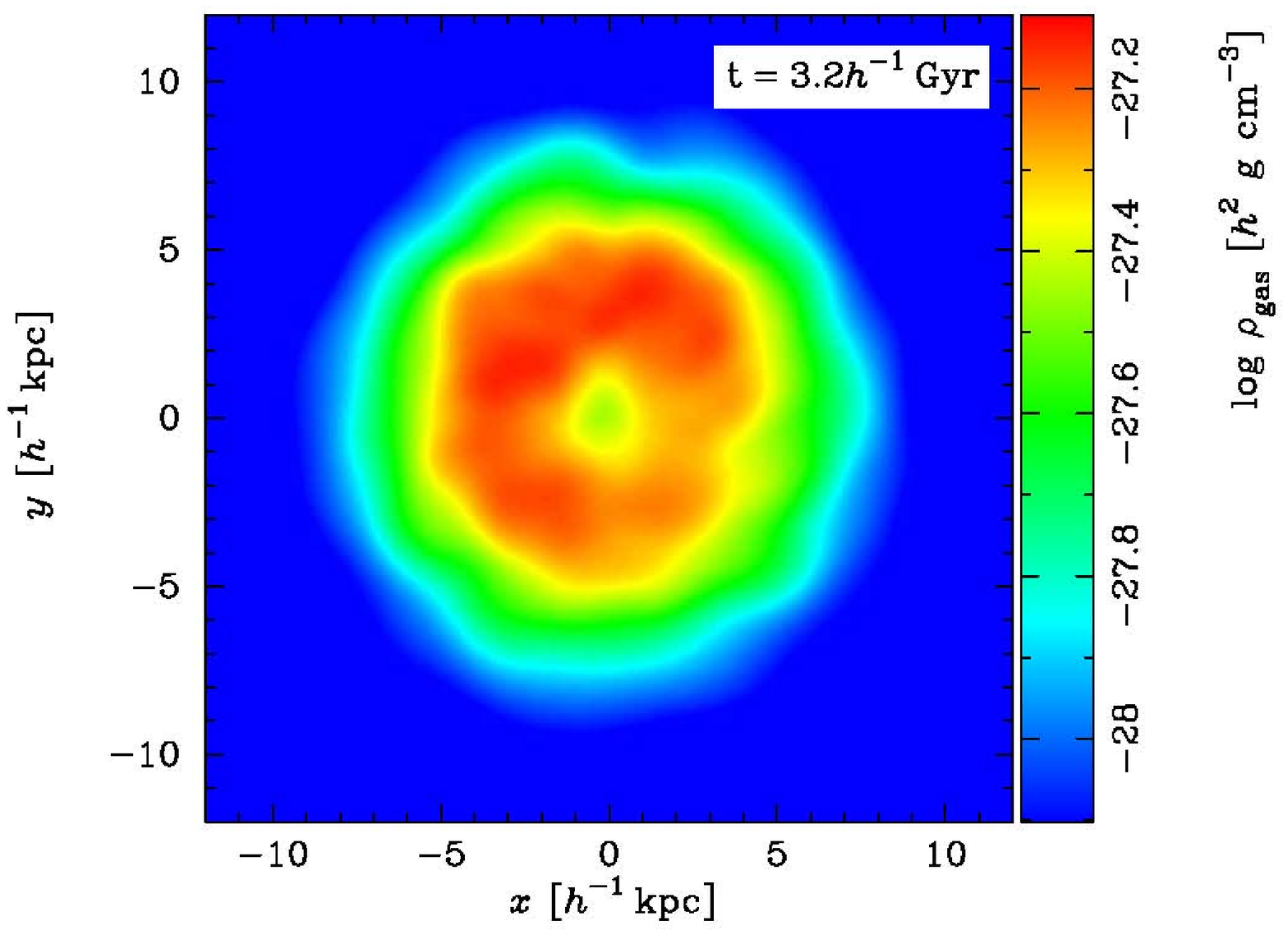}
\end{tabular}
\caption{Heating mechanism of the halo gas in a
  halo of total mass $10^{10}\,\hm$ for the CR streaming model. Top panels:
  Time evolution of the gas temperature and the gas velocity field in an edge-on
  slice through the galactic disc. Results are shown at two different epochs in
  the simulation, at $t=2.8\,\hg$ (left panels) and $t=3.2\,\hg$ (right panels).
  The maximum velocities encountered in these runs are 26 $\kms$ (left panels)
  and 70 $\kms$ (right panels). Collimated cavities of hot gas form below and
  above the disc and propagate further up in the course of time. Middle panels:
  Corresponding plots for the time-evolution of the ratio of the wave heating
  rate due to the damping of self-excited waves to the gas cooling rate,
  $\Lambda_{\mathrm{waves}}/\Lambda_{\mathrm{cool}}$, demonstrating that wave
  heating is responsible for heating the cavities.  Bottom panels: Gas density
  in a cutting plane parallel to the disc at a height of $z=4.5\,\kpc$ (to assess the
  density distribution in the hot cavities). At $t=3.2\,\hg$, an under-dense
  channel forms in the centre where the cooling rate is consequently lowered so
  that wave heating can dominate over the reduced cooling rate.}
\label{fig: temperature}
\end{figure*}

\begin{figure*}
\begin{tabular}{ccc}
\hspace{-1.5em}{\large {\em $10^{9}\,\hm$ halo}} & 
\hspace{-3em}  {\large {\em $10^{10}\,\hm$ halo}} & 
\hspace{-3em}  {\large {\em $10^{11}\,\hm$ halo}} \\
\hspace{-1.5em}\includegraphics[clip,width=0.35\textwidth]{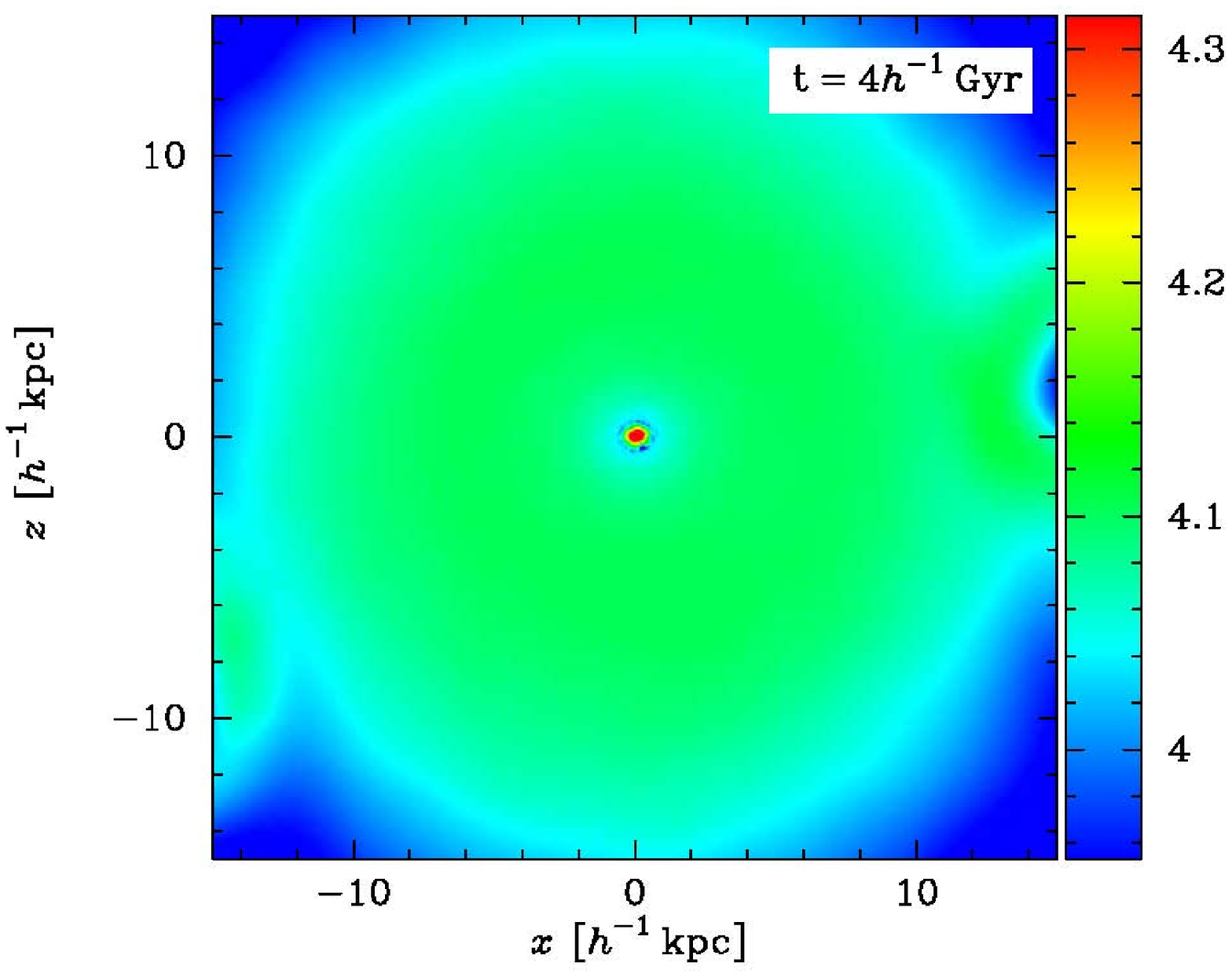} &
\hspace{-3em}  \includegraphics[clip,width=0.35\textwidth]{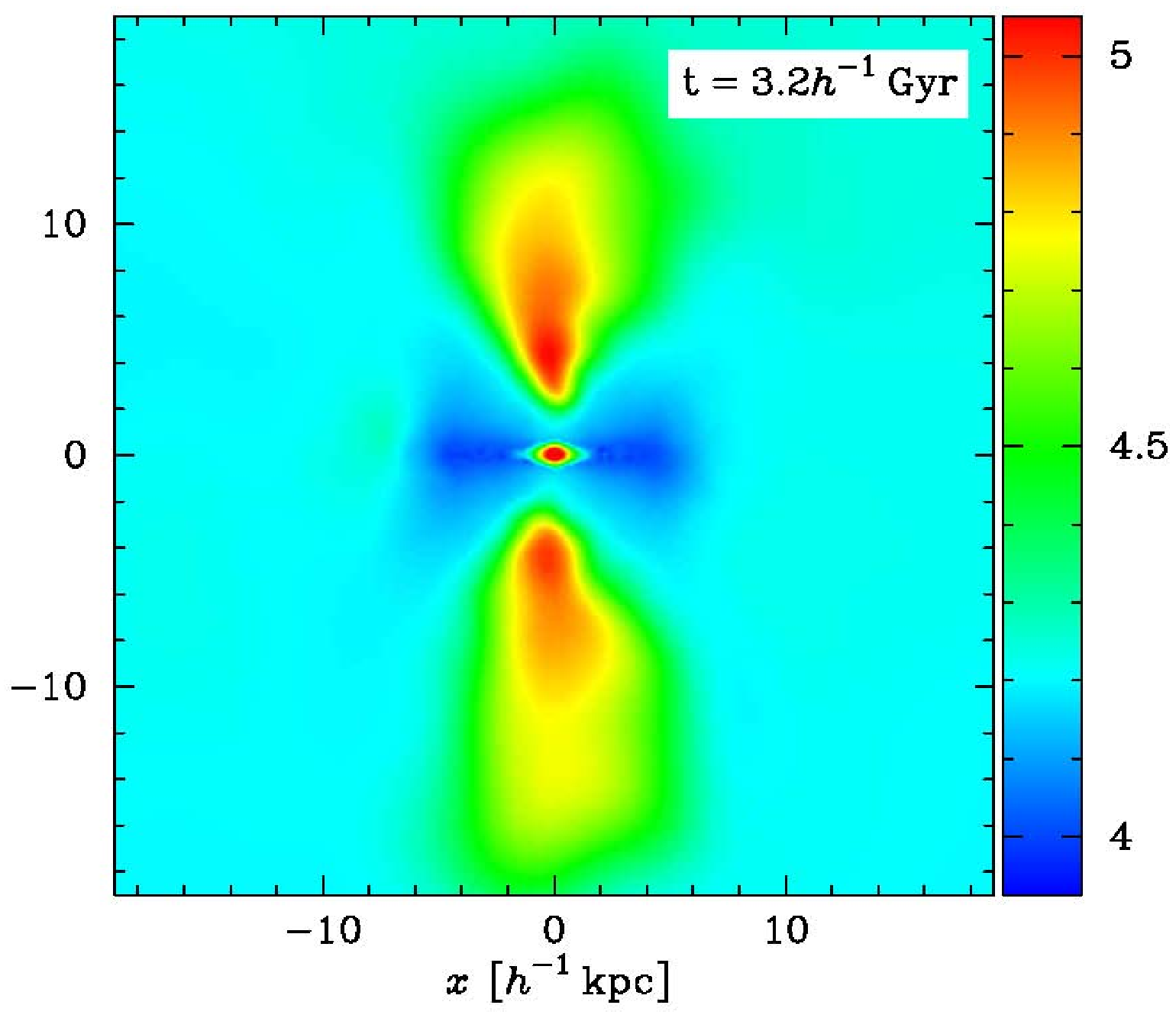} &
\hspace{-3em}  \includegraphics[clip,width=0.35\textwidth]{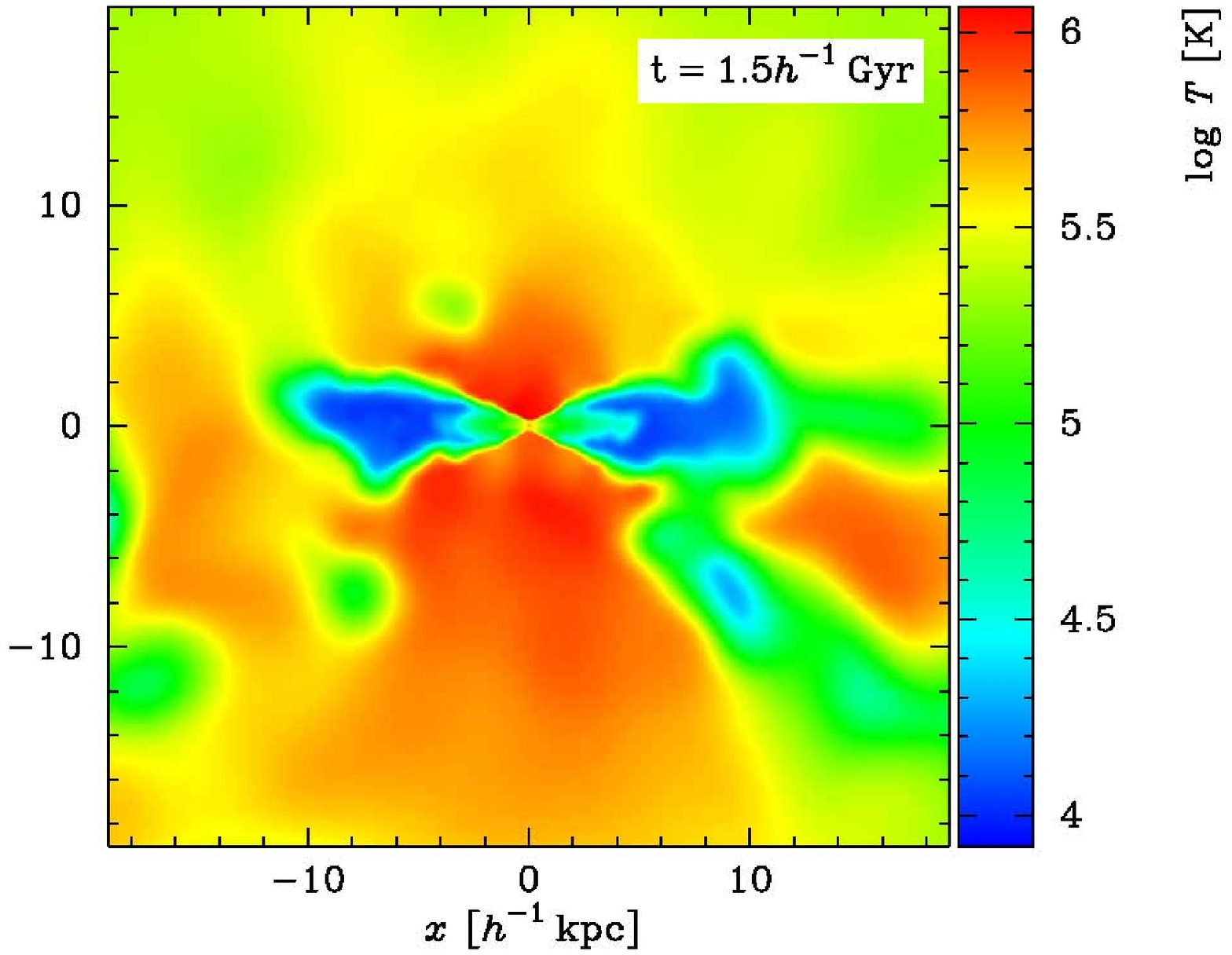}
\end{tabular}
\caption{Temperature distribution at the time of maximum CR Alfv\'en-wave
  heating in an edge-on slice through the galactic disc. We compare three
  different haloes of mass $10^{9}\,\hm$, $10^{10}\,\hm$, and $10^{11}\,\hm$
  (left to right) in our CR streaming model with an acceleration efficiency of
  $\zeta_{\mathrm{SN}}=0.3$. Note that the resulting halo temperatures roughly
  scale as $k T \propto \vel_\rmn{wind}^2 \sim \vel_\rmn{esc}^2$. The
  temperature structure resembles that of the wind, which implies that with
  increasing halo size, the morphology of the hot patches becomes more
  conical. The broadening of the hot regions in our $10^{11}\,\hm$ halo is
  associated with the inability of CR streaming to drive a sustained wind that
  escapes from such a halo. Hence, the kinetic energy of the fountain flow drives
  turbulence which dissipates energy and thereby heats larger regions of the
  halo gas.}
\label{fig: T-comparison}
\end{figure*}

In this section, we assess the role of the wave heating that is intimately
connected to the physics of CR streaming. To this end, we only concentrate on
our CR streaming model that successfully launches winds.  In Fig.~\ref{fig:
  temperature} we plot the gas temperature in an edge-on slice at two different
epochs, $t=2.8\,\hg$ and $t=3.2\,\hg$, about $1\,\hg$ after the outflow started.
At $t=3.2\,\hg$ we can see collimated flows through chimneys of hot gas,
extending up to $\left|z\right|\sim20\kpc$ above and below the disc.  In these
chimneys, the gas temperatures reach values greater than $T\sim10^{5}$~K. In the
centre of the maps, we can see that the disc is also heated up to
$T\sim10^{5}$~K owing to supernova feedback.\footnote{We are using a
  sub-resolution model of the ISM and star formation \citep{Springel2003}. It
  assumes that star formation establishes a self-regulated regime of the ISM,
  which arises due to the interplay of cold, dense clouds that continuously form
  stars and supernova feedback that evaporates these clouds, creating a hot
  ambient medium. The sub-resolution model describes the two phases by an
  effective equation of state that is characterized by a mean mass weighted
  temperature which we plot in Fig.~\ref{fig: temperature}.} These hot regions
are surrounded by a torus-shaped region, filled with comparatively cool gas at
$T\sim10^{4}$~K.

The only conceivable process which could provide a source of heating several
kiloparsecs above the disc, where the gas density is low, must be the damping of
CR-excited waves. In Sec.~\ref{sec:CR-streaming} we have seen that the
hydromagnetic waves are created by streaming CRs in the background plasma. Those
waves are damped on short time-scales, thereby transferring energy to the
plasma. This heating process is in operation as long as CRs are streaming
through the rest frame of the gas, i.e., as long as there is a gradient in the
CR pressure. Thus, unlike Coulomb or hadronic losses, which scale with gas
density, wave heating is also important in low-density regions. In order to
demonstrate that this process is responsible for the hot gas chimneys, we show
an edge-on slice through the galaxy with the ratio of the radiative cooling rate
of the gas to the wave-heating rate of equation~\eqref{eq:wave heating} due to the
damping of self-excited waves,
$\Lambda_{\mathrm{waves}}/\Lambda_{\mathrm{cool}}$ (middle panels of
Fig.~\ref{fig: temperature}).

At $t=2.8\,\hg$, wave heating is nowhere able to overcome the radiative cooling
of the gas. This is in particular true for the galactic disc, where the density
is so large that the cooling dominates the heating by more than two orders of
magnitudes.  At $t=3.2\,\hg$, however, we can see the structure of the hot gas
chimneys again, traced by the evidence that the wave heating dominates the
cooling there by about a factor of ten, showing that heating via wave damping is
indeed the process that creates the hot cavities in our simulations.

To better illustrate the reasons for the dominance of wave heating over gas
cooling above and below the disc, we take a slice parallel to the disc that cuts
the outflowing gas stream at a height of $z=4.5\,\kpc$ above the disc, which is
near the base of the upper chimney. We plot the gas density in this slice in the
bottom panels of Fig.~\ref{fig: temperature}. At $t=2.8\,\hg$, when the cavities
have not yet formed, we see that the outflow covers a circular area in the plane
and features a denser core in the centre, with decreasing density towards the
outskirts. This indicates that the outflow is very collimated. Later, at
$t=3.2\,\hg$, an under-dense hole has appeared in the very centre of the
core. Thus, the gas cooling rate will drop there, so that the CR-wave heating
can heat up the gas.

What is the reason for this under-dense channel in the centre of the outflowing
material? Interestingly, the hot chimneys first occur about $1\,\hg$ after the
outflow started. The likely reason for this is that the CR-driven wind becomes more
and more collimated as time progresses, owing to the disc that forms in the
centre of the simulation box.  When the wind first occurs, the disc formation is
not yet finished, so that it still has an approximately spherical shape. At
these early times, CRs can stream at a large angle with respect to the $z$-axis
without encountering too much disc material, resulting in an outflow with a wide
opening angle. Later, when the disc has formed, the inertia of the dense and
cool star forming gas prohibits a long pathway of the CRs through the disc and
forces an outflow with a smaller opening angle. When the outflow is collimated
enough, it will then quickly dig a diluted channel in the old ejecta above the
disc with a correspondingly lower cooling rate.

This argument is also in agreement with the fact that we do not observe chimneys
of hot gas for our lower mass halo. The weaker gravity of those dwarf haloes does
not support a thin disc forming at the centre, but a rather spherical density
distribution which is unable to collimate the wind. This is highlighted in
Fig.~\ref{fig: T-comparison} which shows temperature maps for haloes with
$10^{9}\,\hm$, $10^{10}\,\hm$, and $10^{11}\,\hm$. These show a clear dependence
of the wave heating on the halo mass. The resulting halo temperatures roughly
scale as $k T \propto \vel_\rmn{wind}^2 \sim \vel_\rmn{esc}^2$ with the largest
halo in our simulations reaching temperatures in excess of $10^6$~K so that the
outflow is expected to emit thermal bremsstrahlung emission. Besides, the larger
SFR obtained for the higher mass halo and correspondingly the higher CR energy
densities could contribute to the preferred creation of the chimneys in the
higher mass halo. We also see in our largest halo that the heated regions are
less collimated in comparison to the intermediate-mass haloes: the wind starts
deeper in the gravitational potential of this halo (see
Section~\ref{sec:analystics}) and looses a good fraction of its kinetic energy
in climbing up the greater potential difference so that it is unable to
counteract the ram pressure of the infalling gas (that reaches also bigger
infall velocities in assuming the larger gravitational binding energy of this
halo). However, the interaction of the outflow with the infalling gas results in
violent turbulence that also dissipates by cascading down in length scale,
thereby additionally heating the halo gas. Potentially, the additional momentum
deposition from radiation pressure may help in circumventing the stalling of the
CR-driven wind seen for this halo.

Another process that certainly enhances the collimation of the wind
into a narrow channel is directly connected to the physics of CR
streaming. Recall that the magnitude of wave heating due to CR-excited
hydromagnetic waves is proportional to the streaming speed which is
equal to the sound speed in our model. Thus, CRs will stream fastest
and deposit most of their energy in the high-temperature gas inside
the channel, thereby further increasing the temperature there and more
importantly the outflow velocity because of enhanced thermal
pressure. This is illustrated by the increasing magnitudes of the
velocity vectors in the outflow (see top panels of Fig. \ref{fig:
  temperature}).  We find maximum velocities of $\sim70\,\kms$ inside
the hot channel, providing the gas with sufficient momentum to
overcome the gravitational attraction of the halo.

Moreover, one can see vortex like structures in the velocity field at the sides
of the chimneys, where gas is stripped off from the flow and returns to the
disc. The vortical motions are presumably created by shear along the sides of
the outflow and they converge towards the base of the cavities, where gas is
then compressed to high densities, facilitating efficient radiative cooling
(similar effects also occur in the case of hot, rising supernova bubbles, see
\citealt{Robinson2004}). This converging gas motion therefore explains the cool
torus structure surrounding the disc.  As time progresses, the vortical motions
will rip the warm chimneys more and more apart so that a warm shell of gas is
created around torus and the galaxy.

\subsection{Maps of H$\alpha$ emission}
\label{sub:Maps-of-Ha emission}

In the previous section we have seen that the wave heating owing to CR
streaming results in characteristic, high-temperature structures. In
these structures, the hydrogen gas is highly ionized and therefore
allows for optical line emission of recombining electrons.  Important
in this context is the H$\alpha$ line (the dominant line of the Balmer
series corresponding to the $3\rightarrow2$ transition in the hydrogen
atom), because hydrogen is the most abundant element in the Universe
and thus ubiquitous in astronomical spectra. H$\alpha$-line emission
from hot gas has been detected in many galaxies that exhibit a wind,
e.g., in the well-known starburst galaxy M82 \citep{Bland1988}.  Thus,
we calculate a map the expected H$\alpha$ emission in our CR streaming
and advection simulation to see how well they compare to these
observations.

The H$\alpha$ emissivity, $\varepsilon\,(\rmn{H}\alpha)$, is given by
\citep{Spitzer}
\begin{equation}
\varepsilon\left(\rmn{H}\alpha\right)=h\nu_{\alpha}\alpha_{\mathrm{eff}}(T)\, n_{\mathrm{e}}n_{\mathrm{p}},
\label{eq:Emissivity}
\end{equation}
where $h\nu_{\alpha}=1.88\,$eV is the energy of a H$\alpha$ photon,
$\alpha_{\mathrm{eff}}(T)$ is the effective recombination coefficient and
$n_{\mathrm{e}}n_{\mathrm{p}}$ denotes the product of the (free) electron number
density and the proton number density, respectively.  The product
$\alpha_{\mathrm{eff}}(T)\, n_{\mathrm{e}}n_{\mathrm{p}}$ gives the number of
H$\alpha$ photons emitted per second and per cubic centimetre. The temperature
dependence of the effective recombination coefficient can be fitted,  
\begin{equation}
\alpha_{\mathrm{eff}}(T)=10^{-13}\,\frac{2.708\, T_{4}^{-0.648}}{1+1.315\,T_{4}^{0.523}}\,\mathrm{cm^{3}s^{-1}},
\label{eq:effective recombination coefficient}
\end{equation}
with $T_{4}\equiv T/10^{4}\,$K \citep[][assuming case B]{Pequignot1991}. Using equation~\eqref{eq:effective recombination
  coefficient} we can then integrate the emissivity, equation~\eqref{eq:Emissivity},
along the line-of-sight. 

\begin{figure}
\includegraphics[clip,scale=0.43]{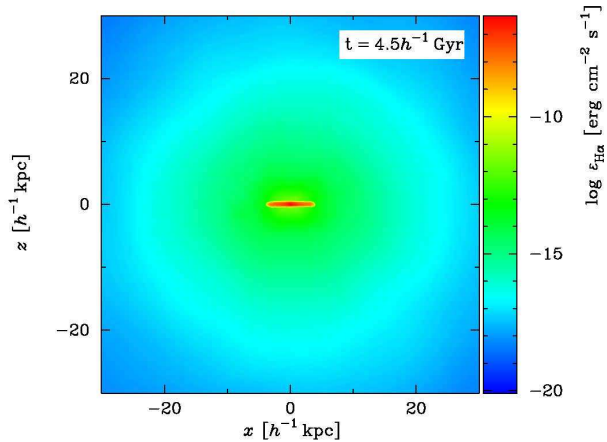}
\caption{H$\alpha$ emission (edge-on slice) in our CR streaming simulation at
  $t=4.5\,\hg$ for a total halo mass of $10^{10}\,\hm$. We see diffuse H$\alpha$
  emission in the halo ($R_{200}\sim35\,\kpc$), which is explained by ionized
  gas, entrained in the wind.}
\label{fig: Halpha Emission}
\end{figure}

The result is shown in Fig.~\ref{fig: Halpha Emission}, where we show
the H$\alpha$ emissivity in an edge-on view of our simulation at
$t=4.5\,\hg$, when the wind is already well-developed. The disc
dominates the surface brightness map due to the supernova feedback
which heats the gas to sufficiently high temperatures so that the gas
becomes ionized. As a result, the product
$n_{\mathrm{e}}n_{\mathrm{p}}$ is large and consequently also the
recombination rate. Interestingly, there is a halo of H$\alpha$
emission with a characteristic surface brightness of
$\sim10^{-13}\,\mathrm{erg\, cm}^{-2}\, \rmn{s}^{-1}$ that extends to
scales of $\sim20\,\kpc$.  Such a diffuse emission might be
detectable, given reported detections at comparably low levels
\citep{Young1996}. This emission is caused by the outflow that
transports ionized gas into the halo and beyond.  Especially inside
the cavities and in their surroundings, where the streaming CRs heat
the gas via wave excitation, the gas has a high degree of ionization,
ranging from $\sim70\%$ to up to $100\%$ in some
places. Unfortunately, the overall gas density in the chimneys is very
low, so that their clear structure that we saw in the temperature maps
is not visible in H$\alpha$.  There is in addition even lower surface
brightness diffuse emission with an elliptical morphology as a result
of the galactic wind, extending up to $\left|z\right|\sim 80\,\kpc$.

\subsection{Limitations of the model}
\label{sec:limits}

We need to caution that some of our conclusions depend on certain assumptions
used in this work and may not be fully generic to the physics of winds driven by
CR streaming.  (1) The sub-resolution model for star formation that we adopt
here is arguably not well-suited for studying outflows. In particular, the
scheme is designed to produce stars in a quiescent mode, i.e., star formation in
general only takes place at an average rate. In reality, however, winds are
typically observed for starburst galaxies \citep[cf.][and references
therein]{Veilleux2005}, which exhibit high rates of star formation and therefore
inject a larger amount of thermal and CR pressure. Apart from that, the star
formation scheme does not allow for an explicit treatment of supernova remnants
and of the hot, over-pressurized gas bubbles they contain. These bubbles are
expected to rise buoyantly in the disc potential of their host galaxy which
should open channels along which the wind can escape much easier.  (2) The use
of such a sub-resolution model with an effective equation of state of the ISM
may overestimate the energy transfer from the CRs to the thermal plasma
following the excitation and damping of hydromagnetic waves within the dense
ISM, that has a comparatively short cooling time, causing a large fraction of
this energy to be lost to cooling radiation (as shown in Section~\ref{sec: mass
  loss}). This overestimate of CR streaming losses may be particularly important
in larger haloes with $M_{200}\gtrsim 10^{10}\,\rmn{M}_\odot$ where larger ISM
overdensities are reached due to the larger gravity.  When explicitly modelling
these two phases, it has been shown that CRs do not effectively couple to gas
within cool clouds since they do not exert forces inside of cool clouds
\citep{Everett2011}. Hence, if CRs were injected into the warm phase of the ISM
(for an improved ISM modelling in the simulations), CR streaming could more
efficiently drive winds, possibly producing larger mass loading factors for
haloes with $M_{200}\gtrsim 10^{10}\,\rmn{M}_\odot$.  (3) To isolate the physics
of CR streaming, we have here used an extremely simplified model for a forming
and evolving galaxy. Our initial conditions assume spherical symmetry which
causes the gas accretion and the associated ram pressure to be initially
spherically symmetric. However, a galaxy that forms in a cosmological
environment accretes the gas in part along filaments that have a much smaller
angular covering factor, hence implying a ram pressure that varies with angle
and could potentially modify the resulting wind morphology. This effect is
particularly important for collimated winds in larger galaxy haloes with
$M_{200}\gtrsim 10^{10}\,\rmn{M}_\odot$. Those should feel a substantially
reduced amount of ram pressure in the wind direction since filaments are
generally not aligned with the disc's angular momentum axis. (4) In our
approach, we use a simplified treatment of the CR streaming speed and equate it
to the local sound speed. However, a realistic value of $\vel_\rmn{st}$ should
depend on the pre-existing wave level, thermal and CR energy density, magnetic
topology, and the damping mechanism of the Alfv\'en waves, to name a few.  (5)
We do not follow MHD; in particular, we assume the existence of locally open
field lines along which CRs can escape into the halo.  While only 10\% of the
galactic SN remnants will create flux tubes that reach heights
$\gtrsim10\,\rmn{kpc}$ above the disc, SN remnant lobes are expected to overlap,
suggesting that at every location there is sufficient supply of open field lines
\citep{Breitschwerdt1993}. This lets this assumption appear plausible (see also
Appendix~\ref{sec:AppCRstreaming}). (6) We neglect diffusive shock acceleration
of CRs in winds from high-mass Wolf-Rayet stars. The recent detection of
TeV-gamma rays from the young stellar cluster Westerlund 2 is compatible with
the hypothesis of pion-decay emission of relativistic nuclei interacting with a
close-by molecular cloud complex \citep{HESS2011}. If confirmed, that would
imply that CR feedback could already be an important mechanism in dispersing
molecular clouds when the most massive stars turn on, long before the first star
became a supernova, making it possibly a competing mechanism to radiation
pressure in launching outbursts.

\section{Discussion and Conclusions}
\label{sec:conclusions}

\subsection{Properties of winds driven by CR streaming}

Using numerical simulations and analytical arguments, we show in this paper that
CR feedback is able to power large-scale galactic outflows. These are not only
able to escape the galactic disc, but also its host dark matter halo. In
particular, we demonstrate that CR streaming relative to the rest frame of the
gas is necessary to drive winds through CRs. Solely considering advective CR
transport, i.e., employing the flux-freezing approximation would not result in
galactic outflows. In this advective-only case, the CR pressure does volume work
on the gaseous discs (in the low-mass haloes), causing them to become more dilute
which increases the thermal cooling time. As a result, the SFR declines
strongly. After some time, the CR pressure is dissipated such that the gas can
collapse again, triggering another ``cycle'' of star formation and causing an
oscillatory mode of star formation with substantial suppression of the stellar
mass formed by a factor of 8 for a halo of mass $M_{200}\lesssim
10^{9}\,\rmn{M}_\odot$ \citep{Jubelgas2008}. In this case, most of the CR energy
is eventually transferred to the thermal gas, which radiates it away. In
contrast, allowing for CR streaming in the rest frame of the gas down their
pressure gradient enables CRs to quickly escape their acceleration sites in the
dense, star forming regions, and to move into the CR-dilute, low density gas
above the disc (or in between the cold phase of the ISM). There, the CR cooling
rates (due to Coulomb or hadronic interactions) are significantly reduced since
all these processes scale with the gas density.  Thus, a big fraction of their
initial energy can be used to power the wind. Moreover, launching a wind from a
low-density gas is considerably easier than from the dense phase, because the
smaller column density of the material that needs to be removed. Star formation
is still suppressed in comparison to the case without CR feedback, but less in
comparison to the advection-only case. Remarkably, the star formation rate is
much smoother and does not show the oscillatory behaviour discussed above.

We find that CR streaming powers winds most effectively in small dwarf haloes
where the wind speed exceeds the escape velocity for host halo masses
$M_{200}\lesssim 10^{11}\,\rmn{M}_\odot$. Below this mass threshold, we find
that the wind velocity increases as a function of host halo mass. We can trace
this behaviour back to the potential difference the wind has to climb until it
can escape, and which is effectively set by the scale height of the disc. On
dwarf scales, the disc scale height is increasing with circular velocity so that
the system is approximately scale invariant. In larger galaxies with
$M_{200}\gtrsim 10^{11}\,\rmn{M}_\odot$, the disc height is set by the effective
equation of state of the ISM and roughly constant with halo mass. This breaks
scale invariance, causes an increasing potential difference for larger haloes,
and renders CR streaming not to be powerful enough to be solely responsible for
galactic winds. While CR-driven winds are approximately spherical in dwarf haloes
($M_{200} \simeq 10^{9}\,\rmn{M}_\odot$), the denser discs in larger haloes are
able to shield the solid angles subtended by the disc and hence focus the wind
into bi-conical outflows around the angular momentum axis of the disc.

In a halo of mass $M_{200}=10^{9}\,\hm$, the relative baryonic mass loss amounts
to $\sim60\%$ ($\sim40\%$) of the initial gas mass contained inside the halo's
virial radius, assuming a CR acceleration efficiency of
$\zeta_{\mathrm{SN}}=0.3$ ($\zeta_{\mathrm{SN}}=0.1$). Similarly, the mass
loading factor of the wind increases towards low-mass haloes, reaching values as
great as $5$ for a halo of $M_{200}=10^{9}\,\hm$. The reason for these increased
mass losses towards smaller halo masses are the shallower gravitational
potential wells of these systems. However, despite the inability of CR streaming
to drive winds in large galaxies that escape the halo, we nevertheless find
powerful fountain flows. While we expect a smaller impact of CR streaming
towards larger halo masses $M_{200} \gtrsim10^{10}\,\rmn{M}_\odot$, we caution
that our simplified simulation setting may artificially weaken the CR streaming
efficiency of driving winds in large galaxies. First, our employed
sub-resolution model with an effective equation of state of the ISM may
overestimate the CR energy losses in the ISM due to Alfv\'en wave excitation
since larger ISM overdensities are reached in larger galaxy haloes due to their
greater potential depths. Second, a cosmologically forming galaxy accretes the
gas in part along filaments that have a much smaller angular covering
factor. Since filaments are generally not aligned with the disc's angular
momentum axis, this reduces the accretion ram pressure that collimated winds
(aligned with the angular momentum axis) have to exert $p\dd V$-work on.  Future
work that realistically models the ISM in a galaxy forming in a cosmological
setting is required to quantify these effects.

Another unique property of CR-driven winds is the effective heating of the
outflow by dissipating Alfv\'en waves that are excited by streaming CRs. Because
the wind speeds increase as $\vel_\rmn{wind}\propto \vel_\rmn{esc}$, we also see
an increased rate of heating with increasing halo mass that results in
temperatures of the halo gas of $k T \propto \vel_\rmn{wind}^2$. We predict
outflows in haloes with masses $M_{200}\gtrsim 10^{10}\,\rmn{M}_\odot$ to reach
temperatures in excess of $3\times10^5$~K so that the outflow is expected to
emit thermal bremsstrahlung emission.

We would like to emphasize that neither the classic energy-driven nor the
classic momentum-driven wind fully captures the physics of CR streaming-driven
winds.  The classical picture assumes a pre-loading of the wind with energy or
momentum. Instead, the physical mechanism of CR-driven winds implies wave
excitation by the CR streaming instability that transfers both, energy and
momentum to the gas. This manifests itself through the {\em continuous}
Alfv\'en-wave heating term in the energy equation and the $\nabla P_\rmn{cr}$
term in the momentum equation. CR-driven winds are a two-stream phenomenon
composed of the gas flow (carrying momentum) and the fast CR stream
(exerting $p\dd V$-work on the gas) which continuously re-powers the wind
with energy and momentum during its ascent in the gravitational potential.

\subsection{Comparing different physical mechanisms for galactic winds}

\begin{figure}
\centerline{\includegraphics[width=\linewidth]{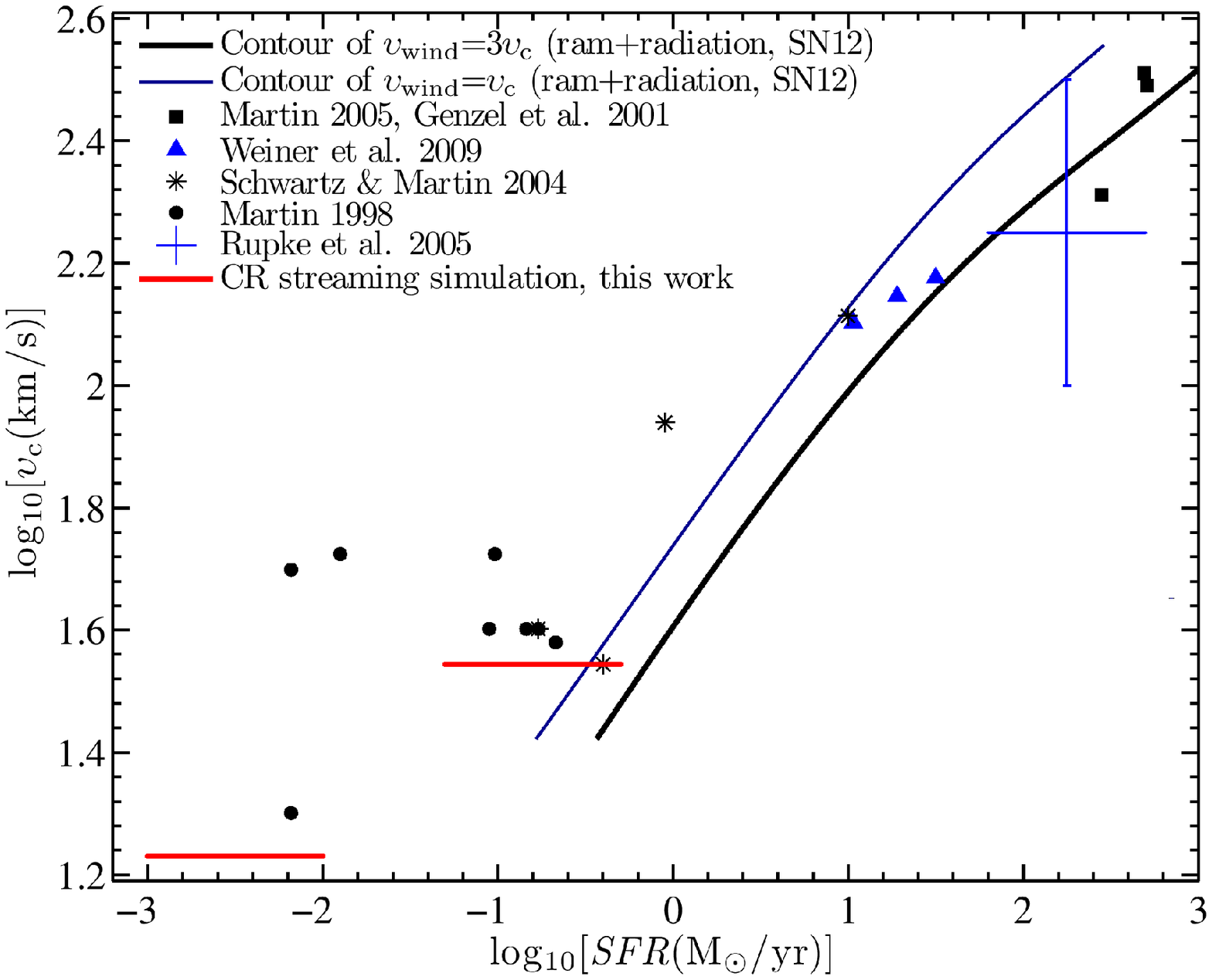}}
\caption{Contours of $\vel_\rmn{wind}$ for cold ($T \sim 10^4$ K) wind models
  driven by a combination of radiation and ram pressure \citep[equation
  (12) of][]{Sharma-Nath2012}. Different symbols indicate observations of winds
  with (partly) considerable mass loadings \citep{Martin1998, Genzel2001,
    Schwartz2004, Rupke2005, Martin2005, Weiner2009}. In these models, the
  region to the lower right of these contour lines allows for powerful winds
  driven by ram and radiation pressure, ejecting the gas from the halo. In the
  region to the upper left, these two processes are not able to drive galactic
  winds, leaving the observational data in this region unexplained. The range of
  SFRs encountered in our CR streaming simulations is shown with red lines and
  indicates the importance of this process in driving winds, especially for
  galaxies showing smaller values of $\vel_\rmn{c}$ and SFRs. }
\label{fig: discussion}
\end{figure}

We now scrutinize different physical mechanisms responsible for driving galactic
winds and put CR streaming -- the mechanism studied here -- into the context of
other plausible processes, namely winds driven by radiation or ram pressure.
Generally, it is believed that galaxies have winds when their star formation per
unit area exceeds a threshold value of $0.1\, \rmn{M}_\odot \,\rmn{yr}^{-1}
\,\rmn{kpc}^{-2}$. In our simulation of CR-driven winds, the SFR densities are
lower than this threshold.  Therefore, this process can address winds with low
SFRs.

In Fig.~\ref{fig: discussion}, we show contours of $\vel_\rmn{wind}$ in the
parameter space spanned by $\vel_\rmn{c}$ and SFR for cold ($T \sim 10^4$ K)
wind models driven by a combination of ram pressure and radiation pressure
\citep{Sharma-Nath2012}.  According to these models, the region of parameter
space to the lower right of these contour lines allows for powerful winds driven
by ram and radiation pressure that are able to eject the gas from the halo,
i.e., it is easier to drive winds for a smaller gravitational potential well
(smaller $\vel_\rmn{c}$) and a greater total power input to the system (larger
SFR). In contrast, in the region to the upper left, these two processes are not
able to drive galactic winds and demonstrate the failure of ram and radiation
pressure driven wind models to explain the wind observations in starburst dwarfs
\citep{Martin1998, Schwartz2004}. Our simulation results for winds driven by CR
streaming are shown with red lines (indicating the range of SFRs) and highlight
the importance of CR streaming in driving winds, especially for galaxies showing
small values of $\vel_\rmn{c}$ and SFRs.

Good examples are the winds observed in the dwarf starburst galaxies NGC 4214
\citep[$\rmn{SFR} = 0.17\,\rmn{M}_\odot\,\rmn{yr}^{-1}$ and
$\vel_{\rmn{c,max}}=40\,\rmn{km\,s}^{-1}$;][]{Walter2001, Martin2005} and NGC
1569 \citep[$\rmn{SFR} = (0.16 - 0.4)\,\rmn{M}_\odot\,\rmn{yr}^{-1}$ and
$\vel_{\rmn{c,max}}=35\,\rmn{km\,s}^{-1}$;][]{Martin2002, Stil2002}. The latter
shows direct evidence for a metal-enriched wind from a dwarf starburst galaxy
with a mass loading factor of 9 that carries nearly all the metals ejected by
the starburst. Both galaxies populate the no-wind region considering only ram
and radiation pressure. These dwarf starbursts are close to our
$10^{10}\,\hm$ halo in the parameter space shown in Fig.~\ref{fig:
  discussion}. The CR-driven bi-conical outflow in this galaxy resembles the
observed winds in NGC 4214 and NGC 1569, consistent with a similar physical origin.

While wind velocities and masses can be reliably determined for neutral phases
(as it is the case for the two examples discussed), and for H$\alpha$-emitting
warm ionized gas (as done in \citealt{Martin1998}), the velocity of the hot gas
($10^6-10^7$~K) is more difficult to determine. However, it is believed to be
higher than that of the neutral gas, with velocities of order its own thermal
sound speed of several hundred to a thousand kilometres per second. How does a
neutral cloud (or some of its material) get entrained into the wind, in
particular for a CR-streaming driven wind? Such a wind is necessarily magnetized
with contributions from open field lines (extending from the disc into the
halo), flux frozen disc fields, as well as self-amplified field by means of CR
streaming. This magnetized super-Alfv\'enic and supersonic wind (with respect to
the gas in the cloud that exhibits comparably low temperatures and sound speeds)
impacts the neutral cloud and forms a contact discontinuity. Independent of the
degree of magnetization, the cloud sweeps up enough wind magnetic field at the
interface to the wind to build up a dynamically important sheath
\citep{Gregori2000, Lyutikov2006, Asai2007, Dursi2008, Pfrommer2010}.

The field strength of this magnetic draping layer is set by a competition
between `ploughing up' and slipping around of field lines, maintaining a
magnetic energy density in steady state that is of order the wind ram pressure
seen by the cloud. This is an inherently three-dimensional problem where the
third dimension is essential in allowing for field lines to slip around the
obstacle. This strongly magnetized layer modifies the dynamics of the cloud,
potentially suppressing hydrodynamic instabilities and mixing \citep{Dursi2007},
and changing the geometry of stripped material. Most importantly, magnetic
draping accelerates the cloud into the direction of the wind through magnetic
tension of draped field lines that are anchored in the hot phase of the wind,
streaming already ahead of the cloud \citep{Dursi2008}. Note that these authors
study the opposite problem of a cloud moving through a magnetized wind (which
can be transformed by means of a Galilean transformation to our problem at hand)
and find that the deceleration through magnetic tension is always more
important, by a factor of $\simeq 3.7$, than for the case of highly turbulent
(Reynolds number $\rmn{Re} \simeq 1000$) hydrodynamic drag. The terminal
velocity that the cloud can reach should be a sizable fraction of the wind
velocity, and may depend on magnetic geometry and the magnetization of the wind.

This emerging picture suggests that different physical processes dominate the
launching and powering of the wind, with a relative contribution that depends on
the SFR and $\vel_\rmn{c}$ \citep{Hopkins2011b, Sharma-Nath2012}. For increasing
values of SFR and $\vel_\rmn{c}$, we seem to have first CR streaming, then ram
pressure, and finally radiation pressure as the dominant agent in powering
winds. We conclude that (1) CR-driven winds are most effective in dwarfs,
ejecting a large fraction ($\sim 60\%$) of metal-enriched baryons into the IGM,
and (2) CR Alfv\'en-wave heating is especially efficient in heating a fountain
flow in high-mass galaxies. While we expect that our results for this physical
scenario are robust irrespective of redshift or environment, we caution that
much more work is needed to firmly confirm this picture, especially given the
limitations in modelling these processes at the present time (see
Section~\ref{sec:limits}).

\subsection{Cosmological and observational implications of CR-driven winds}

In hierarchical structure formation, dwarf galaxies represent the
building blocks of larger galaxies. CR-driven winds appear to be very
efficient in removing baryons from small haloes, hence this process
should be the dominant feedback mechanism during the formation of
these dwarfs, i.e., shortly before and after reionization. We expect
the successive mergers of dwarfs to be less gas rich than without this
feedback mechanism and to form less stars at high redshift. As galaxy
systems grow larger, the ejected gas may be accreted later, but, due
to the modified thermal history and angular momentum, may end up at
different places within the larger galaxies. Hence, it is not
inconceivable that CR-driven winds in small-scale haloes have an
indirect effect up to scales of $L_*$ at the knee of the luminosity
function; but this needs to be carefully studied in future work.

Additionally, we expect our CR-driven winds to influence the early history of
metal enrichment during the epoch of dwarf formation because of their
considerable mass loading factors. This important role of low-mass systems in
the metal enrichment of the IGM is in line with expectations from other authors
\citep{Nath1997,Madau2001,Samui2010}.

What about specific observational predictions of CR-driven winds?  The
bi-conical structure of the hot gas cavities that formed in the $10^{10}\,\hm$
halo has some similarities with H$\alpha$-structures, e.g., in the well-known
starburst galaxy M82 \citep{Bland1988} or with the dwarf starburst galaxy NGC
1569 \citep{Martin2002}. However, the bi-conical structure observed in the
temperature field has no direct resemblance in our simulated H$\alpha$ emission
maps, due to the low density of the gas in these cavities. This is the same
problem that commonly makes observations of winds difficult, which face the
difficulty of detecting the weak emission from a wind that is in general hot and
dilute \citep[see][and references therein]{Veilleux2005}. Our simulations are
not capable of describing the full spectrum of fluid phenomena that are likely
to be associated with these hot cavities. Fluid instabilities of
Kelvin-Helmholtz and Rayleigh-Taylor type \citep{Robinson2004} as well as
thermal instabilities should be important because of the density contrast around
the chimneys and the shear along their sides. In addition, there will be
entrainment of denser clouds from the interstellar medium when the hot wind
pushes on them \citep{MacLow1988}. All these phenomena will influence the
morphology of the cavities and could enhance the density locally, so that their
signature might indeed be visible in H$\alpha$. However, we lack the necessary
resolution to see these effects in our simulations and the SPH method may also
significantly contribute to this limitation, as recent studies found
\citep{Sijacki2011}.  Moreover, magnetic fields that are confined within the hot
chimney gas and rise up along with it are of particular importance. They will
support the chimneys against the hydrodynamical instabilities and shear which
try to tear them apart \citep{Robinson2004, Dursi2007, Ruszkowski2007,
  Dursi2008}. Nevertheless, our simulations predict a low but observable level
of extended diffuse H$\alpha$ emission in haloes of mass $10^{10}\,\hm$.  In
fact, extended H$\alpha$ emission forming an H$\alpha$-halo, that extends
several kiloparsecs away from the disc, has been reported for several
galaxies. The strength and spatial extent of the emission seem to be positively
correlated with the SFR in the galaxy \citep[e.g.,][]{Rand1996}. This highlights
the importance of studying CR-driven winds including a better description of
star formation and stellar feedback.

Some galaxies also exhibit radio haloes with an overall correlation between
H$\alpha$ and radio emission. Such a correlation would naturally arise when CRs
are convected into the halo, as is the case in our simulations.  To discriminate
a pure advective scenario from our model, the spectral age of CR electrons could
potentially be decisive \citep[see][for such a
measurement]{Heesen2009}. Including CR streaming should decrease the spectral
age since streaming electrons escape faster into the halo. However, the
adiabatic expansion history in both scenarios needs to be identical to allow for
a fair comparison. Future work that includes MHD will address those questions
and allow for more definite predictions of the expected radio synchrotron
emission in this scenario.

\section*{Acknowledgments}
We thank Crystal Martin and Peng Oh for useful discussions. We also thank the
referee for a thorough reading of the manuscript and for his constructive
comments. This research was supported in the framework of the DFG Forschergruppe
1254 ``Magnetization of Interstellar and Intergalactic Media: The Prospects of
Low-Frequency Radio Observations''. C.P. gratefully acknowledges financial
support of the Klaus Tschira Foundation and would furthermore like to thank KITP
for their hospitality during the conference and workshop on ``First Light and
Faintest Dwarfs'' (which was supported by the National Science Foundation under
Grant No. NSF PHY05-51164). B.B.N. thanks the Alexander-von-Humboldt foundation
for supporting his visit to the Max-Planck-Institute for Astrophysics
(MPA). B.B.N. also thanks the MPA for the hospitality during his
visit. V.S. acknowledges support by the DFG Research Centre 881 ``The Milky Way
System''.  The rendered plots in this work were produced using the visualization
tool SPLASH by \citet{Price2007}.  The simulations were performed at the
Rechenzentrum Garching (RZG) of the Max Planck Society as well as using
resources of the bwGRiD project\footnote{Member of the German D-Grid initiative,
  funded by the Ministry of Education and Research (Bundesministerium f{\"u}r
  Bildung und Forschung) and the Ministry for Science, Research and Arts
  Baden-Wuerttemberg (Ministerium f{\"u}r Wissenschaft, Forschung und Kunst
  Baden-W{\"u}rttemberg)}.

\bibliography{ref}
\bibliographystyle{mn2e}

\appendix

\section{Implementation of CR streaming}
\label{sec:AppCRstreaming}

Here, we discuss our implementation of CR streaming into the
Lagrangian SPH code
{\sc Gadget-2}. First, we translate the transport equations for CR energy and
number density into Lagrangian form and then cast them into the language of SPH.
Our starting point are the evolution equations for CR energy density and number
density which can be derived from the CR transport equation
\citep[e.g.,][]{Skilling1971, Skilling1975},
\begin{eqnarray}
\frac{\partial\varepsilon_{\rmn{cr}}}{\partial t} \!\!&=&\!\!
(\vvel+\vvel_{\rmn{st}})\cdot\nabla P_{\rmn{cr}}-
\nabla\cdot\left[\left(\vvel+\vvel_{\rmn{st}}\right)\left(\varepsilon_{\rmn{cr}}+P_{\rmn{cr}}\right)\right],
\label{eq:energy equation}\\
\frac{\partial n_{\rmn{cr}}}{\partial t}\!\!&=&\!\!
-\nabla\cdot\left[\left(\vvel+\vvel_{\rmn{st}}\right)n_{\rmn{cr}}\right],
\label{eq:number density equation}
\end{eqnarray}
where $\vvel$ is the gas speed, $\vvel_{\rmn{st}}$ is the streaming speed of CRs
and $P_{\rmn{cr}}$ is the CR pressure. Here and in the reminder of this
derivation, we neglect all CR source terms as well as CR diffusion. We model CR
sources and sinks explicitly in our numerical CR formalism \citep{Pfrommer2006,
  Ensslin2007, Jubelgas2008}. For the limiting case of strong CR-wave scattering
(the Bohm limit), the effective diffusion velocity is usually much slower than
the CR streaming velocity, justifying our approximation. As discussed in
Section~\ref{sec:CR-streaming}, the streaming speed is assumed to be proportional
to the local sound speed, $c_\rmn{s}$, and anti-parallel to a unit vector along the CR
pressure gradient $\nabla P_{\rmn{cr}}$ in our model, i.e.,
\begin{eqnarray}
\vvel_{\mathrm{st}}=-\lambda\, c_{\mathrm{s}}\,
\frac{\nabla P_{\rmn{cr}}}{\left|\nabla P_{\rmn{cr}}\right|},
\label{eq:streaming velocity}
\end{eqnarray}
where $\lambda\geq1$ is a proportionality constant for a plasma where
the thermal pressure dominates over the magnetic pressure. The CR
streaming velocity has been taken to be opposite to the CR pressure
gradient rather than the gradient of the CR number density of a CR
energy interval (which would be the formal criterion for evaluating
the direction of the CR streaming velocity). For power-law momentum
distributions, $n_{\rmn{cr}}\propto P_{\rmn{cr}}$, and hence the
gradients of $n_{\rmn{cr}}$ and $P_{\rmn{cr}}$ point into the same
direction. If this was not the case, i.e., if different energy regimes
dominated $n_{\rmn{cr}}$ and $P_{\rmn{cr}}$, then we would be
interested in the energy regime that dominates the CR pressure (which
is responsible for driving the wind) and hence in the streaming
direction given by the CR pressure gradient.

Since we do not follow MHD in our simulations, we are unable to project
the CR pressure gradient onto the magnetic field lines to determine the formally
correct streaming direction. However, as we will now argue, our approach should
correctly capture the main aspects of CR streaming and the associated galactic
mass loss even without magnetic fields. In hydrostatic equilibrium, the gradient
of the sum of thermal and CR pressure is determined by the gradient of the
gravitational potential,
\begin{eqnarray}
\frac{1}{\rho}\nabla\left( P+P_\rmn{cr}\right) = -\nabla\Phi.
\end{eqnarray}
Above the disc, when wave heating induced by streaming CRs provides substantial
heat input, the CR pressure distribution determines the total pressure
distribution.  For CRs to escape the galactic disc via streaming requires
locally open field lines (that are stretched to large heights
$\gtrsim10\,\rmn{kpc}$ above the disc). While only 10\% of the galactic SN
remnants will create flux tubes that reach those altitudes, SN remnant
lobes are expected to overlap, suggesting that at every location there is
sufficient supply of open field lines \citep{Breitschwerdt1993}. The field
topology at the disc-halo interface and beyond will be determined by the
buoyancy of the CR component that drive a large-scale upward convection and
possibly a galactic mass loss \citep{Parker1966}. Hence, the topology of the
open field line structure is determined by the CR pressure stratification which
itself is shaped by the gravitational potential gradient. This should establish
a field topology consistent with our assumptions, at least when averaged over
sufficiently large scales so that small-scale fluctuations are averaged out.

To derive the Lagrangian form of the CR evolution equations, we define a
Lagrangian time derivative,
\begin{eqnarray}
\frac{\dd}{\dd t}=\frac{\partial}{\partial t}+\vvel\cdot\nabla,\label{eq:convective derivative}
\end{eqnarray}
and introduce the specific CR energy, $\tilde{\eps}_\rmn{cr}$, and CR particle
number, $\tilde{n}_\rmn{cr}$, through
\begin{eqnarray}
\varepsilon_{\rmn{cr}}&=&\tilde{\varepsilon}_{\rmn{cr}}\rho,\\
n_{\rmn{cr}}&=&\tilde{n}_{\rmn{cr}}\rho\,.
\end{eqnarray}
After using the continuity equation for the gas, $\dd\rho/\dd t =
-\rho\nabla\cdot\vvel$, we arrive at the Lagrangian form of the CR
evolution equations,
\begin{eqnarray}
\rho\frac{\dd\tilde{\varepsilon}_{\rmn{cr}}}{\dd t}\!\!\!&=&\!\!\!
\vvel_{\rmn{st}}\cdot\nabla P_{\rmn{cr}}
-P_{\rmn{cr}}\nabla\cdot\vvel
-\nabla\cdot\left[\vvel_{\rmn{st}}\left(\rho\tilde{\varepsilon}_{\rmn{cr}}+P_{\rmn{cr}}\right)\right],
\label{eq:energy equation 3} \\
\rho\frac{\dd\tilde{n}_{\rmn{cr}}}{\dd t}\!\!\!&=&\!\!\!
-\nabla\cdot\left[\vvel_{\rmn{st}}\,\rho\,\tilde{n}_{\rmn{cr}}\right].
\label{eq:number density equation 3}
\end{eqnarray}
The first term on the right-hand side of equation \eqref{eq:energy equation
  3} is the wave heating term due to self-excited waves that get
(almost instantly) damped in the plasma. It results in an energy loss
of CRs and will be addressed at the end of this Appendix. For the
moment, we neglect this term in our further derivation. The second
term on the right-hand side of equation~\eqref{eq:energy equation 3} is the
only term involving the plasma velocity $\vvel$. It expresses energy
changes due to adiabatic compression (or expansion) of CRs when the
gas flow converges (diverges). Since it is already included in our
standard CR implementation, we will not discuss it further  in the
following. The remaining term (as well as the only term on the
right-hand side of equation \eqref{eq:number density equation 3}) describes
CR streaming, i.e., how the CR energy and number changes for an
individual gas mass element due to the CR streaming motion into (or
out of) that particular mass element.  Combining
equation \eqref{eq:streaming velocity} for the CR streaming velocity with
equations \eqref{eq:energy equation 3} and \eqref{eq:number density
  equation 3}, we obtain
\begin{eqnarray}
\rho\frac{\dd\tilde{\varepsilon}_{\rmn{cr}}}{\dd t}&=&
\nabla\cdot\left(\kappa_{\tilde{\varepsilon}}\nabla P_{\rmn{cr}}\right),
\label{eq:energy equation 4}\\
\rho\frac{\dd\tilde{n}_{\rmn{cr}}}{\dd t}&=&
\nabla\cdot\left(\kappa_{\tilde{n}}\nabla P_{\rmn{cr}}\right).\label{eq:number density equation 4}
\end{eqnarray}
These equations take a form that is similar to a diffusion equation with the
formal diffusivities
\begin{eqnarray}
\kappa_{\tilde{\varepsilon}} & = & \lambda\, c_\rmn{s}\,\frac{\rho\tilde{\varepsilon}_{\rmn{cr}}+P_{\rmn{cr}}}{\left|\nabla P_{\rmn{cr}}\right|},\label{eq:effective diffusi1}\\
\kappa_{\tilde{n}} & = & \lambda\, c_\rmn{s}\,\frac{\rho\tilde{n}_{\rmn{cr}}}{\left|\nabla P_{\rmn{cr}}\right|},\label{eq:effective diffusi2}
\end{eqnarray}
for CR energy density and number density, respectively.  Therefore, concerning
the discretized SPH form of equations \eqref{eq:energy equation 4} and
\eqref{eq:number density equation 4}, the same recipe as for thermal conduction
\citep{Cleary1999,Jubelgas2004} and CR diffusion \citep{Jubelgas2008}
can be applied%
\footnote{In Appendix~\ref{sec: Numerical Tests}, we will show in one-dimensional test cases that this indeed leads to reasonable
  results.}. This approach rests on an SPH representation of the Laplacian (see,
e.g., \citealt{Jubelgas2004}) that circumvents second derivatives of the
smoothing kernel, which otherwise would increase numerical noise. Noise is
particularly problematic when an explicit time integration scheme is
used for the diffusion equation, as in our case,
because it can cause overshoots that may eventually reverse the direction of
transport, leading to unphysical results unless a quite small timestep
is used \citep{Jubelgas2004}. 

Despite using this scheme, we cannot avoid the occurrence of second order kernel
derivatives completely, since equations \eqref{eq:effective diffusi1} and \eqref{eq:effective
  diffusi2} contain the magnitude of the CR pressure gradient that also needs to
be SPH-smoothed and therefore already contains a first-order kernel
derivative. Leaving this problem aside for the moment, we obtain the following
evolution equations for the CR quantities of SPH particle $i$,
\begin{eqnarray}
\frac{\dd\tilde{\varepsilon}_{\rmn{cr},i}}{\dd t} \!\!& = &\!\! 2\sum_{j}\frac{m_{j}}{\rho_{i}\rho_{j}}\bar{\kappa}_{\tilde{\varepsilon}}^{ij}\left(\frac{P_{\rmn{cr},j}-P_{\rmn{cr},i}}{|\vec{x}_{ij}|^{2}}\right)\vec{x}_{ij}\cdot\nabla_{i}\overline{W}_{ij},\\
\frac{\dd\tilde{n}_{\rmn{cr},i}}{\dd t} \!\!& = &\!\! 2\sum_{j}\frac{m_{j}}{\rho_{i}\rho_{j}}\bar{\kappa}_{\tilde{n}}^{ij}\left(\frac{P_{\rmn{cr},j}-P_{\rmn{cr},i}}{|\vec{x}_{ij}|^{2}}\right)\vec{x}_{ij}\cdot\nabla_{i}\overline{W}_{ij},
\end{eqnarray}
where the sum extends over all smoothing neighbours $j$ of particle $i$, and
\begin{eqnarray}
\bar{\kappa}_{\tilde{\varepsilon}}^{ij} & = & 2\frac{\kappa_{\tilde{\varepsilon}}^{i}\,\kappa_{\tilde{\varepsilon}}^{j}}{\kappa_{\tilde{\varepsilon}}^{i}+\kappa_{\tilde{\varepsilon}}^{j}},\\
\bar{\kappa}_{\tilde{n}}^{ij} & = & 2\frac{\kappa_{\tilde{n}}^{i}\,\kappa_{\tilde{n}}^{j}}{\kappa_{\tilde{n}}^{i}+\kappa_{\tilde{n}}^{j}},
\end{eqnarray}
represent the harmonic means of the formal diffusivities of
equations \eqref{eq:effective diffusi1} and \eqref{eq:effective diffusi2} (each one
corresponding to the positions of $i$ and $j$),
respectively. Further, $\overline{W}_{ij}$ denotes a symmetrized kernel,
i.e., $\overline{W}_{ij}=0.5\left[W(\vec{r}_{ij},h_{i})+W(\vec{r}_{ij},h_{j})\right]$
and $\vec{x}_{ij}=\vec{x}_{i}-\vec{x}_{j}$ is the vector connecting the
positions of both partners. The CR pressure gradient at the position of particle
$i$ is SPH-smoothed in the form
\begin{eqnarray}
\left(\nabla P_{\rmn{cr}}\right)_{i}=\sum_{j}m_{j}\rho_{i}\left(\frac{P_{{\rmn{cr},i}}}{\rho_{i}^{2}}+\frac{P_{{\rmn{cr},j}}}{\rho_{j}^{2}}\right)\nabla W_{ij}.
\end{eqnarray}

According to \citet{Cleary1999}, taking the harmonic mean of the formal
diffusivities instead of the arithmetic mean forces the flux to be continuous in
implementations of thermal conduction, especially in regions where the
diffusivity changes abruptly. We will see in Appendix~\ref{sec: Numerical Tests}
that this does not work perfectly in our case, but irrespectively we apply the
harmonic mean for the sake of a lower computational cost, as will be discussed
later on, too.

If a particle has no CRs at all, its formal diffusivity will be zero
and therefore also the harmonic means of equations~\eqref{eq:effective
  diffusi1} and \eqref{eq:effective diffusi2}.  Thus, there would be
no exchange between CR pressurized particles and particles without any
CRs, which is unphysical. In such a case we adopt
\begin{eqnarray}
\bar{\kappa}_{\tilde{\varepsilon}}^{ij} & = & \kappa_{\tilde{\varepsilon}}^{i},\\
\bar{\kappa}_{\tilde{n}}^{ij} & = & \kappa_{\tilde{n}}^{i},
\end{eqnarray}
for the formal diffusivities, i.e., we take the diffusivity of the CR
pressurized partner. We also employ this choice if the CR pressure gradient,
$\nabla P_{\rmn{cr}}$, of one partner is equal to zero, since otherwise its
formal diffusivity would be infinite. If both particles have a CR population
and no CR pressure gradient in the current time step, then we switch off
exchange via streaming for this pair.

Let us now return to the issue of kernel derivatives. Indeed, employing
equations \eqref{eq:energy equation 4} and \eqref{eq:number density equation 4} for
the CR streaming implementation will lead to rather noisy results because the
SPH form of the CR pressure gradient already contains one kernel derivative. To
tackle this problem, we follow the suggestions by \citet{Jubelgas2004} and
replace the CR pressure of particle $j$ by its SPH-smoothed equivalent in the
evolution equations, i.e.,
\begin{eqnarray}
P_{\rmn{cr},j}\rightarrow\bar{P}_{\rmn{cr},j}=\sum_{i}\frac{m_{i}}{\rho_{i}}\, P_{\rmn{cr},i}W_{ij},\label{eq:smoothed pressure}
\end{eqnarray}
where the sum extends over all smoothing neighbours $i$ of particle $j$. In
agreement with the findings of \citet{Jubelgas2004}, this choice makes our
scheme less prone to small-scale particle noise, even in the presence of an
interpolated CR pressure gradient. Thus, our evolution equations for the CR
quantities now read
\begin{eqnarray}
\frac{\dd\tilde{\varepsilon}_{\rmn{cr},i}}{\dd t} \!\!& = &\!\! 2\sum_{j}\frac{m_{j}}{\rho_{i}\rho_{j}}\bar{\kappa}_{\tilde{\varepsilon}}^{ij}\left(\frac{\bar{P}_{\rmn{cr},j}-P_{\rmn{cr},i}}{|\vec{x}_{ij}|^{2}}\right)\vec{x}_{ij}\cdot\nabla_{i}\overline{W}_{ij},\label{eq:final energy evol}\\
\frac{\dd\tilde{n}_{\rmn{cr},i}}{\dd t} \!\!& = &\!\! 2\sum_{j}\frac{m_{j}}{\rho_{i}\rho_{j}}\bar{\kappa}_{\tilde{n}}^{ij}\left(\frac{\bar{P}_{\rmn{cr},j}-P_{\rmn{cr},i}}{|\vec{x}_{ij}|^{2}}\right)\vec{x}_{ij}\cdot\nabla_{i}\overline{W}_{ij}.\label{eq:final number evol}
\end{eqnarray}
Note that we still use the proper CR pressure in the formal diffusivity in
equation~\eqref{eq:effective diffusi1}, rather than an interpolant.

Equations \eqref{eq:final number evol} and \eqref{eq:final energy evol} show
that the SPH particle $i$ will gain CR energy and CRs from SPH particle $j$ if
its pressure is less than the (smoothed) pressure of $j$, conforming with our
picture of CRs streaming down their gradient. Here it is important to note that
for particles which have no CRs at all, the intrinsic CR pressure,
$P_{\mathrm{cr}}$, must not be replaced by the smoothed pressure
$\bar{P}_{\mathrm{cr}}$ in the transport equations above. Otherwise these
particles could loose energy that they do not have if there are pressurized
particles in the vicinity that contribute to a non-zero
$\bar{P}_{\mathrm{cr}}$.

Furthermore, we need to discuss the time step that we adopt for our explicit
time integration scheme. We decided to use a criterion of the form
\begin{eqnarray}
\Delta t<\varepsilon\frac{1}{\lambda c_{\mathrm{s}}}\left(\frac{m}{\rho}\right)^{1/3}.\label{eq:timestep}
\end{eqnarray}
Equation \eqref{eq:timestep} is based on the notion that effects of CR
streaming should take place on a time-scale equal to the time that CRs
need to cross the mean inter-particle spacing
$\left(m/\rho\right)^{1/3}$, travelling at the streaming speed
$\lambda c_{\mathrm{s}}$. Here, $\varepsilon$ is a parameter for which
we used $\varepsilon=0.004$ in most of our simulations. This choice
provides us with a good balance between accuracy and computational
cost. We note that this choice would still be fairly expensive for
cosmological simulations, suggesting that a replacement of the
explicit time integration by an implicit scheme
\citep[e.g.,][]{Petkova2009} should be considered in future work.

In order to ensure conservation of CR energy and particle number during the
exchange between SPH particles, we follow \citet{Jubelgas2008} and apply their
manifestly conservative scheme. Instead of equations \eqref{eq:final number evol} and
\eqref{eq:final energy evol}, we use
\begin{eqnarray}
m_{i}\frac{\dd\tilde{\varepsilon}_{\rmn{cr},i}}{\dd t} & = & \frac{\dd E_{\rmn{cr},i}}{\dd t},\label{eq:conserv1}\\
m_{i}\frac{\dd\tilde{n}_{\rmn{cr},i}}{\dd t} & = & \frac{\dd N_{\rmn{cr},i}}{\dd t},
\end{eqnarray}
to calculate the changes of CR energy and particle number for an individual SPH particle
$i$ due to CR streaming. This translates the CR number density and energy
density per unit mass into proper particle number and energy.  Thus, the
exchange terms between the SPH neighbours $i$ and $j$ over which we sum, now
read
\begin{eqnarray}
\frac{\dd E_{ij}}{\dd t} & = & 2\frac{m_{i}m_{j}}{\rho_{i}\rho_{j}}\bar{\kappa}_{\tilde{\varepsilon}}^{ij}\left(\frac{\bar{P}_{\rmn{cr},j}-P_{\rmn{cr},i}}{|\vec{x}_{ij}|^{2}}\right)\vec{x}_{ij}\cdot\nabla_{i}\overline{W}_{ij},\\
\frac{\dd N_{ij}}{\dd t} & = & 2\frac{m_{i}m_{j}}{\rho_{i}\rho_{j}}\bar{\kappa}_{\tilde{n}}^{ij}\left(\frac{\bar{P}_{\rmn{cr},j}-P_{\rmn{cr},i}}{|\vec{x}_{ij}|^{2}}\right)\vec{x}_{ij}\cdot\nabla_{i}\overline{W}_{ij}.
\end{eqnarray}
More importantly, in the SPH loop of particle $i$, we only inject an amount
$0.5\,(\dd E_{ij}/\dd t)\Delta t_{i}$ of energy into its CR population. In
addition, $-0.5(\dd E_{ij}/\dd t)\Delta t_{i}$ is given to particle
$j$ at the same time. After $j$'s loop is complete, the effective change in the
CR energy (and correspondingly in the CR particle number) for both particles is
then given by
\begin{eqnarray}
\Delta E_{i} & = & \frac{1}{2}\left(\frac{{\rm d}E_{ij}}{{\rm d}t}\Delta
  t_{i}-\frac{{\rm d}E_{ji}}{{\rm d}t}\Delta t_{j}\right),\label{eq:total energy change}\\
\Delta E_{j} & = & \frac{1}{2}\left(\frac{{\rm d}E_{ji}}{{\rm
      d}t}\Delta t_{j}-\frac{{\rm d}E_{ij}}{{\rm d}t}\Delta t_{i}\right).\label{eq:total number change}
\end{eqnarray}
Here we assume for simplicity that $j$ is the only neighbour of $i$ and vice
versa. In total, during each SPH loop, the change in the CR energy and particle
number, respectively, is zero even if each particle has its own time step.
Hence, the total change in CR energy and particle number, equations \eqref{eq:total
  energy change} and \eqref{eq:total number change}, may be seen as injection
with the average time step, $0.5\times\left(\Delta t_{i}+\Delta t_{j}\right)$,
provided that $\dd E_{ij}/\dd t=-\dd E_{ji}/\dd t$.

So far, we have only addressed the streaming part of the CR evolution equations
that leads to an exchange of CRs between different SPH particles and therefore
conserves the total energy and particle number in CRs. However, in
Section~\ref{sec:CR-streaming} we saw that the streaming CRs will excite
hydromagnetic waves that get quickly damped in the plasma. This gives rise to an
additional loss term of CR energy (but not of CR particles) for individual SPH
particles. We account for this wave heating term in each time step $\Delta t$ by
transferring an amount
\begin{eqnarray}
\Gamma_{i}=\frac{\lambda c_{\mathrm{s}}\left|\nabla P_{\mathrm{cr}}\right|_{i}}{\varrho_{i}}\,\Delta t\label{eq:wave heating}
\end{eqnarray}
of energy (per unit mass) from the CR population of each particle $i$ into its
thermal reservoir. This will lead to a continuous heating of the gas as long the
magnitude of the CR pressure gradient at the specific location is non-zero,
i.e., as long as the CRs are streaming through the rest frame of the gas.  The
updated CR spectral cutoff and normalization are derived from CR
streaming-induced changes in $\Delta\varepsilon$ and $\Delta n$ according to the
same scheme as in \citet{Jubelgas2008}.

\section{Numerical Tests}
\label{sec: Tests}

\subsection{Testing the Streaming Implementation}
\label{sec: Numerical Tests}

\begin{figure*}
\begin{tabular}{ccc}
\hspace{-1.0em}\includegraphics[scale=0.32]{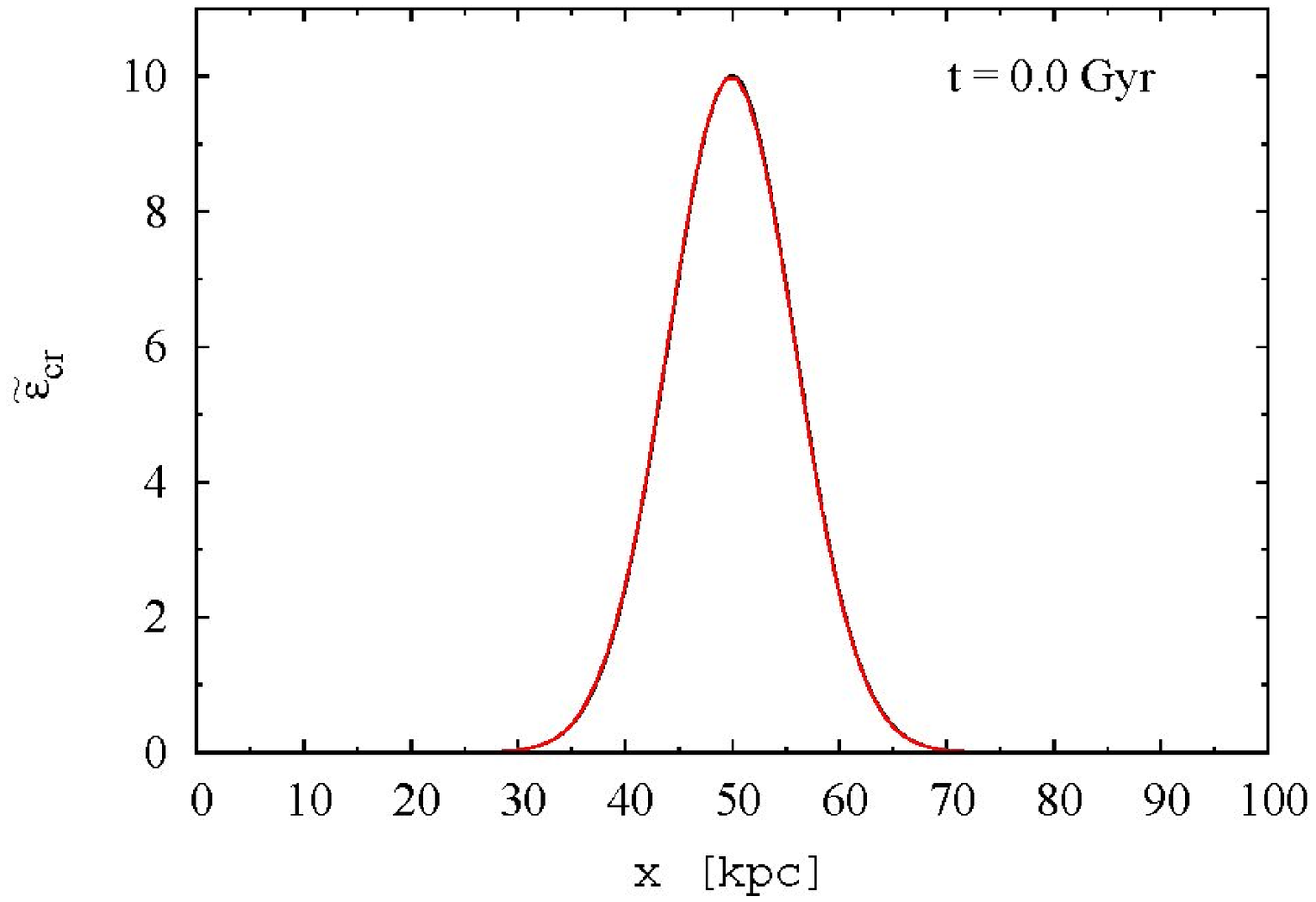} & 
\hspace{-2.5em}\includegraphics[scale=0.32]{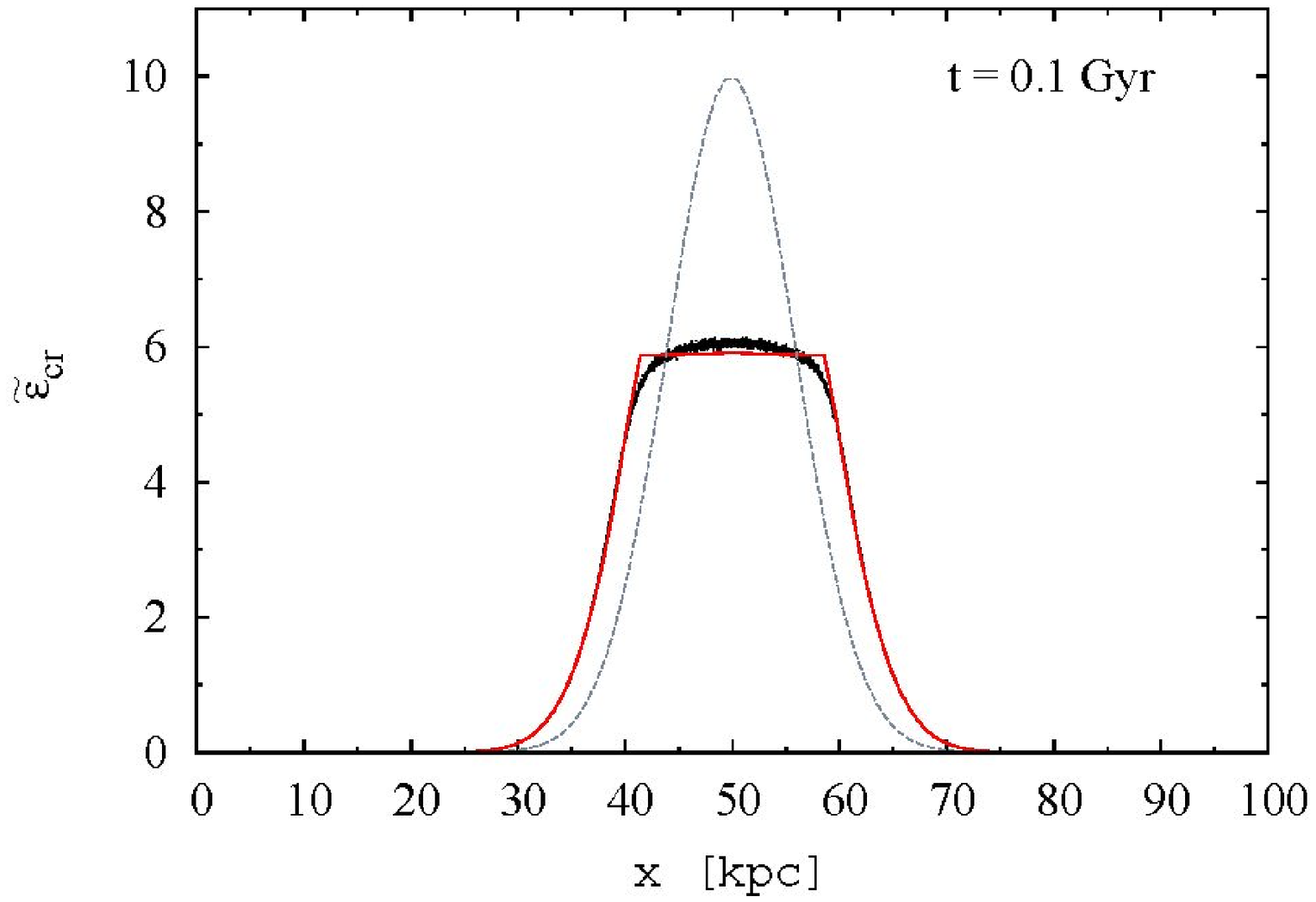} & 
\hspace{-2.5em}\includegraphics[scale=0.32]{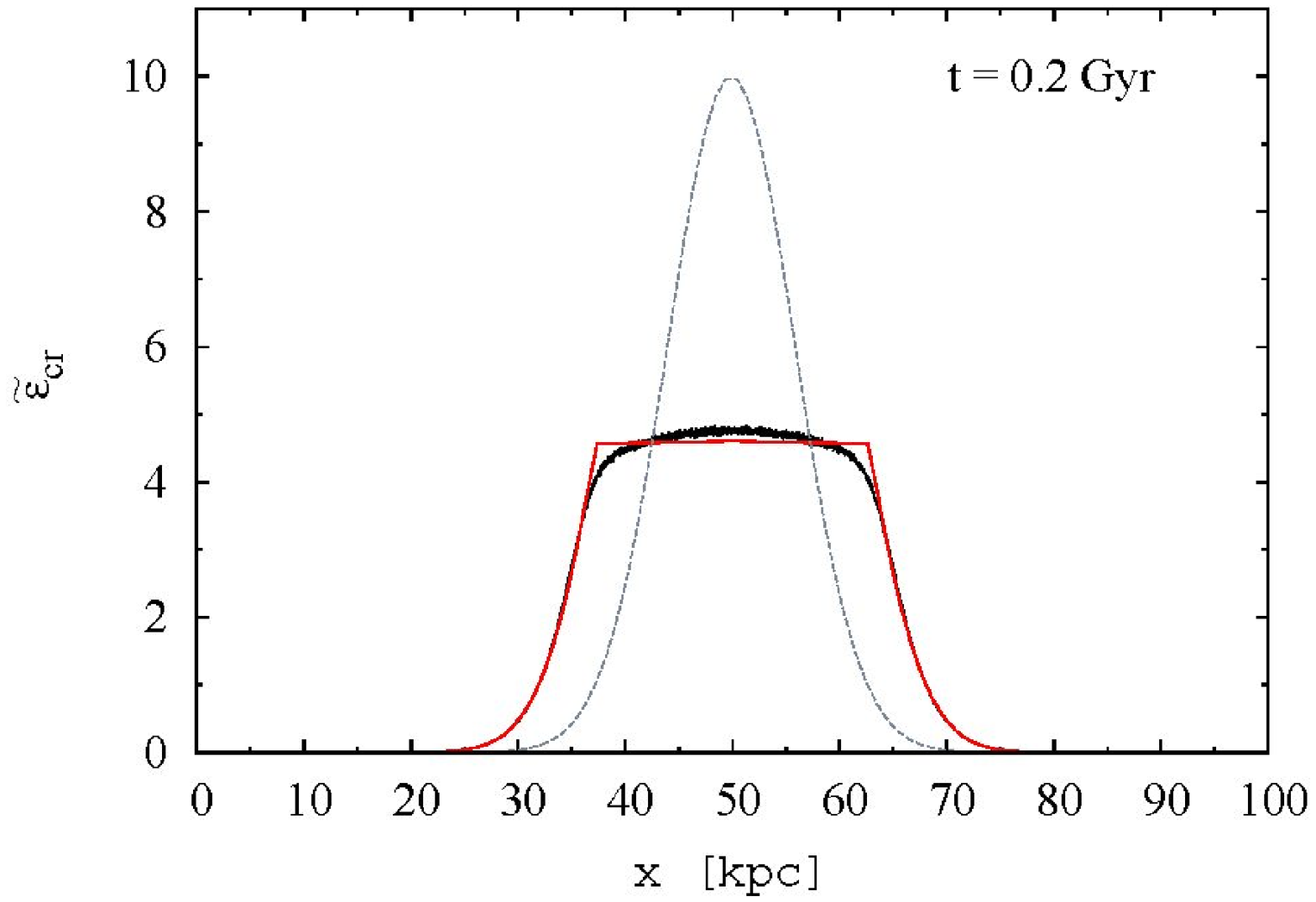} \\
\hspace{-1.0em}\includegraphics[scale=0.32]{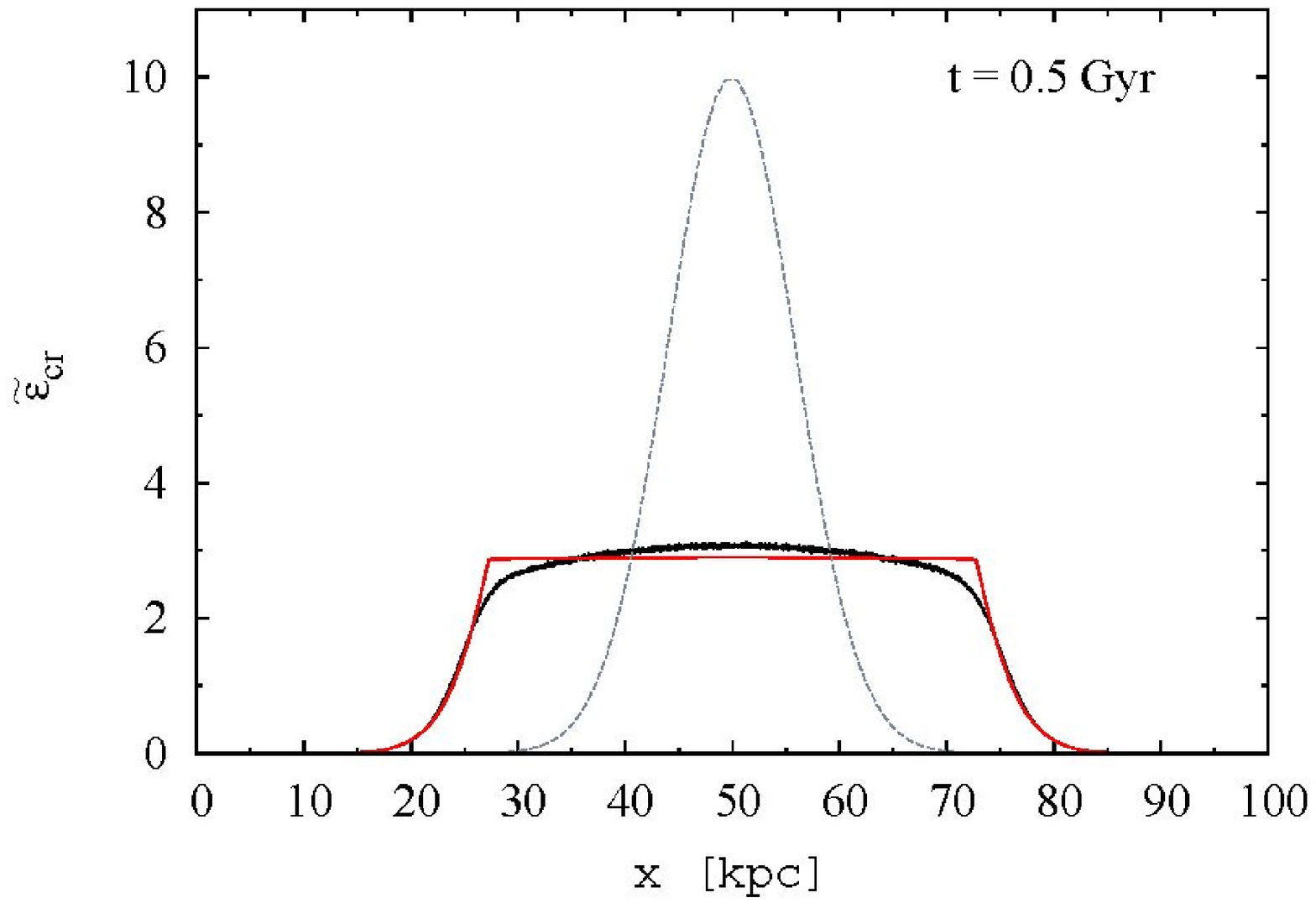} & 
\hspace{-2.5em}\includegraphics[scale=0.32]{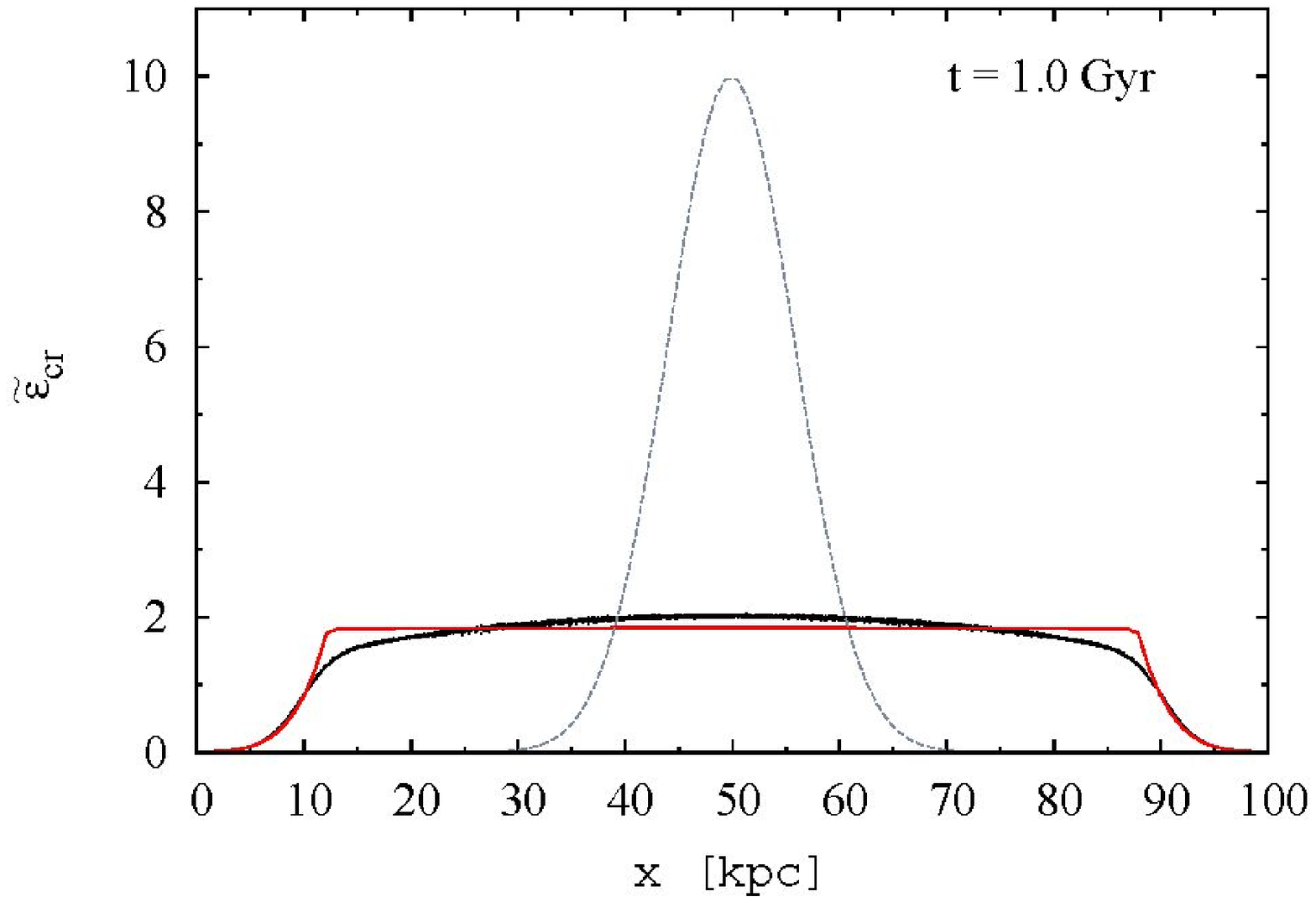} & 
\hspace{-2.5em}\includegraphics[scale=0.32]{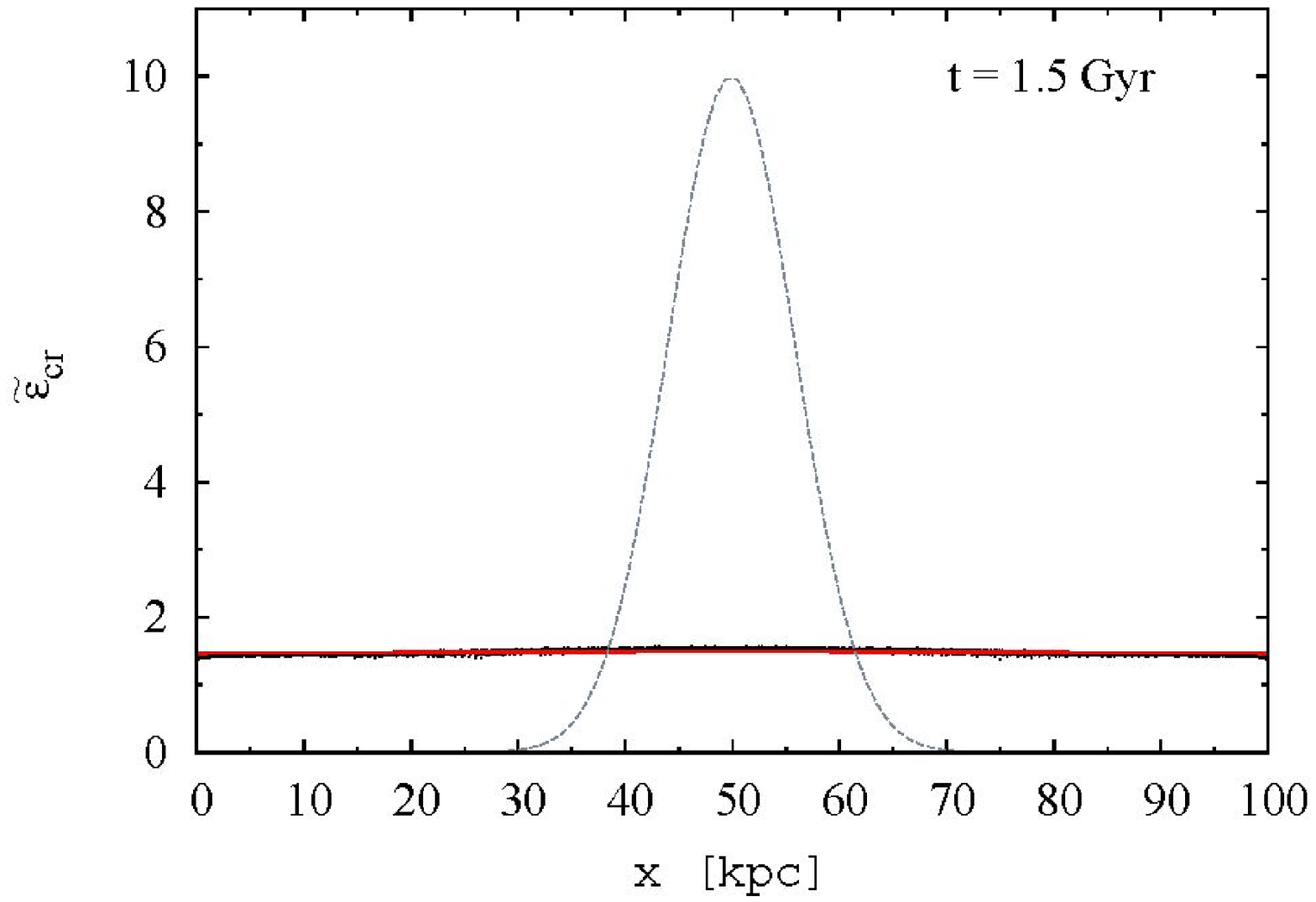}
\end{tabular}
\begin{tabular}{ccc}
\hspace{-1.0em}\includegraphics[scale=0.32]{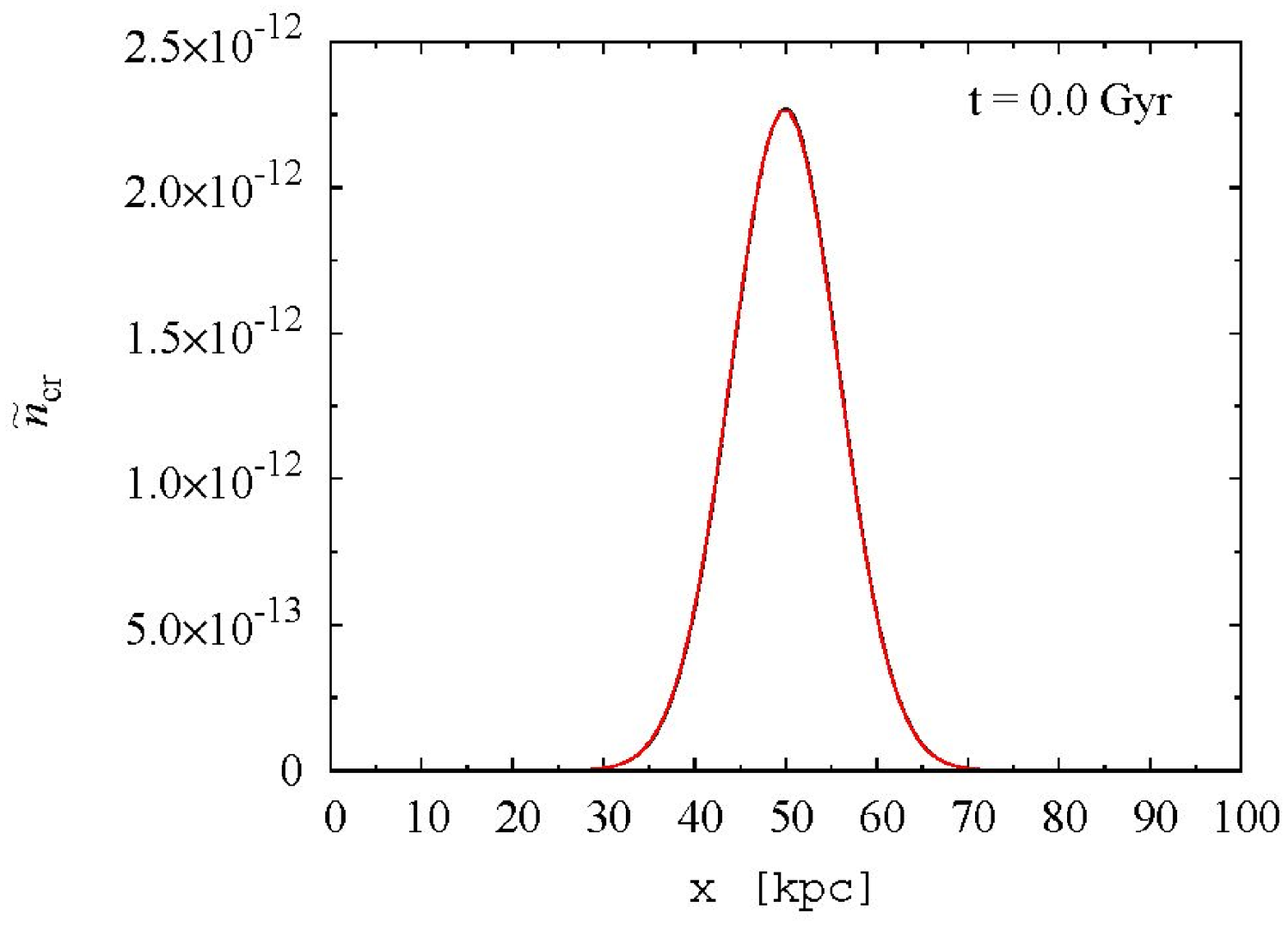} & 
\hspace{-2.5em}\includegraphics[scale=0.32]{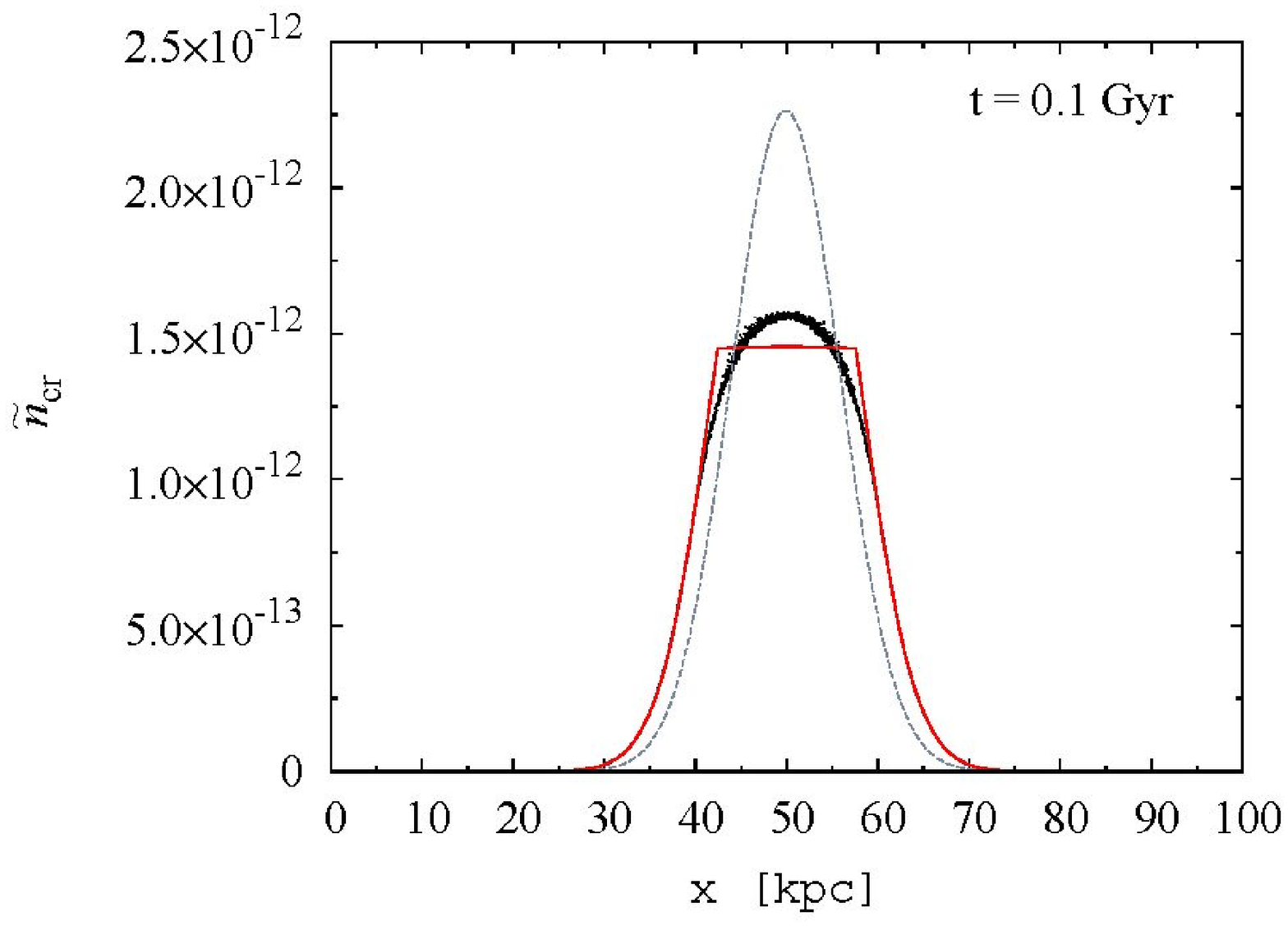} & 
\hspace{-2.5em}\includegraphics[scale=0.32]{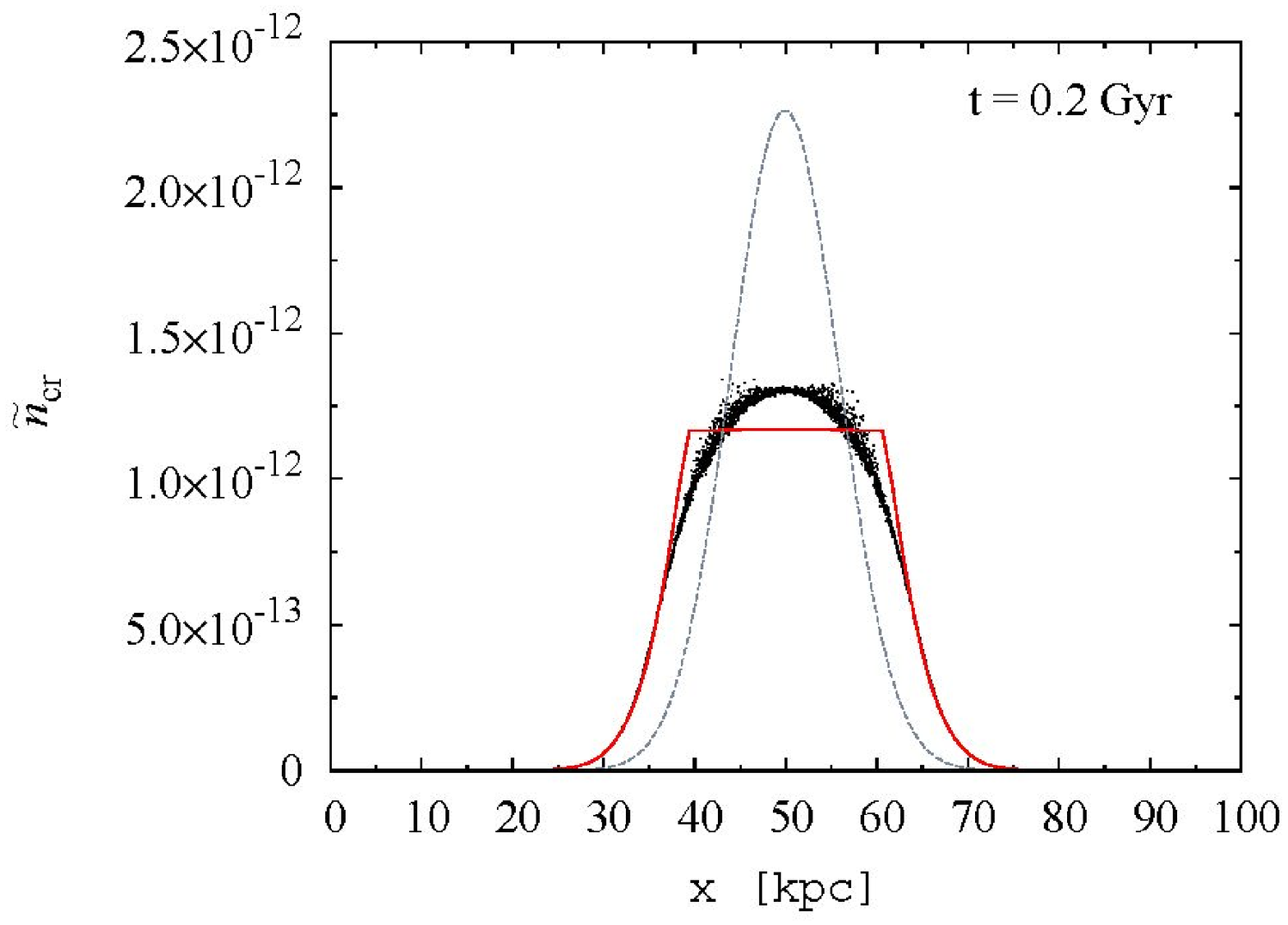} \\
\hspace{-1.0em}\includegraphics[scale=0.32]{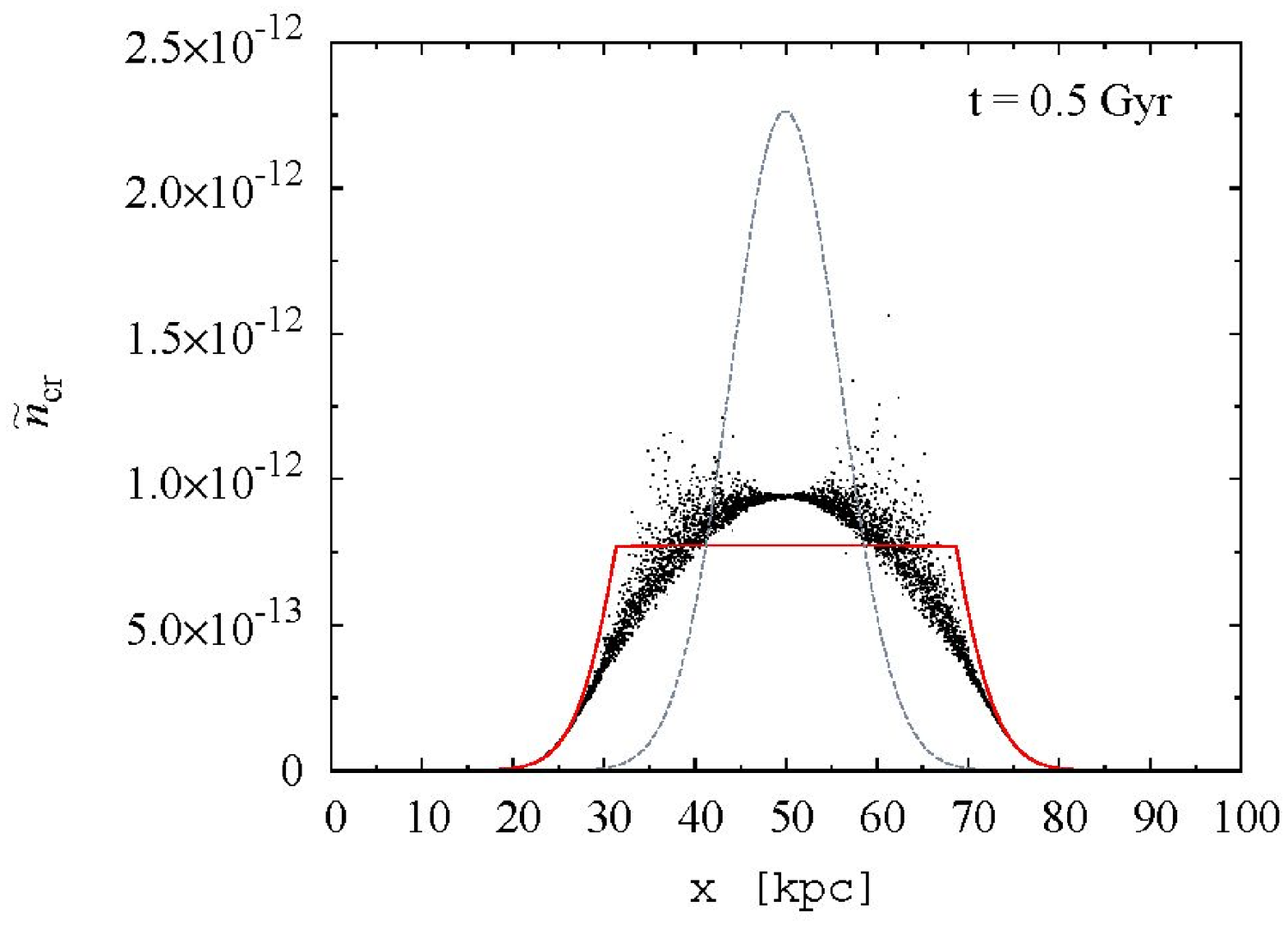} & 
\hspace{-2.5em}\includegraphics[scale=0.32]{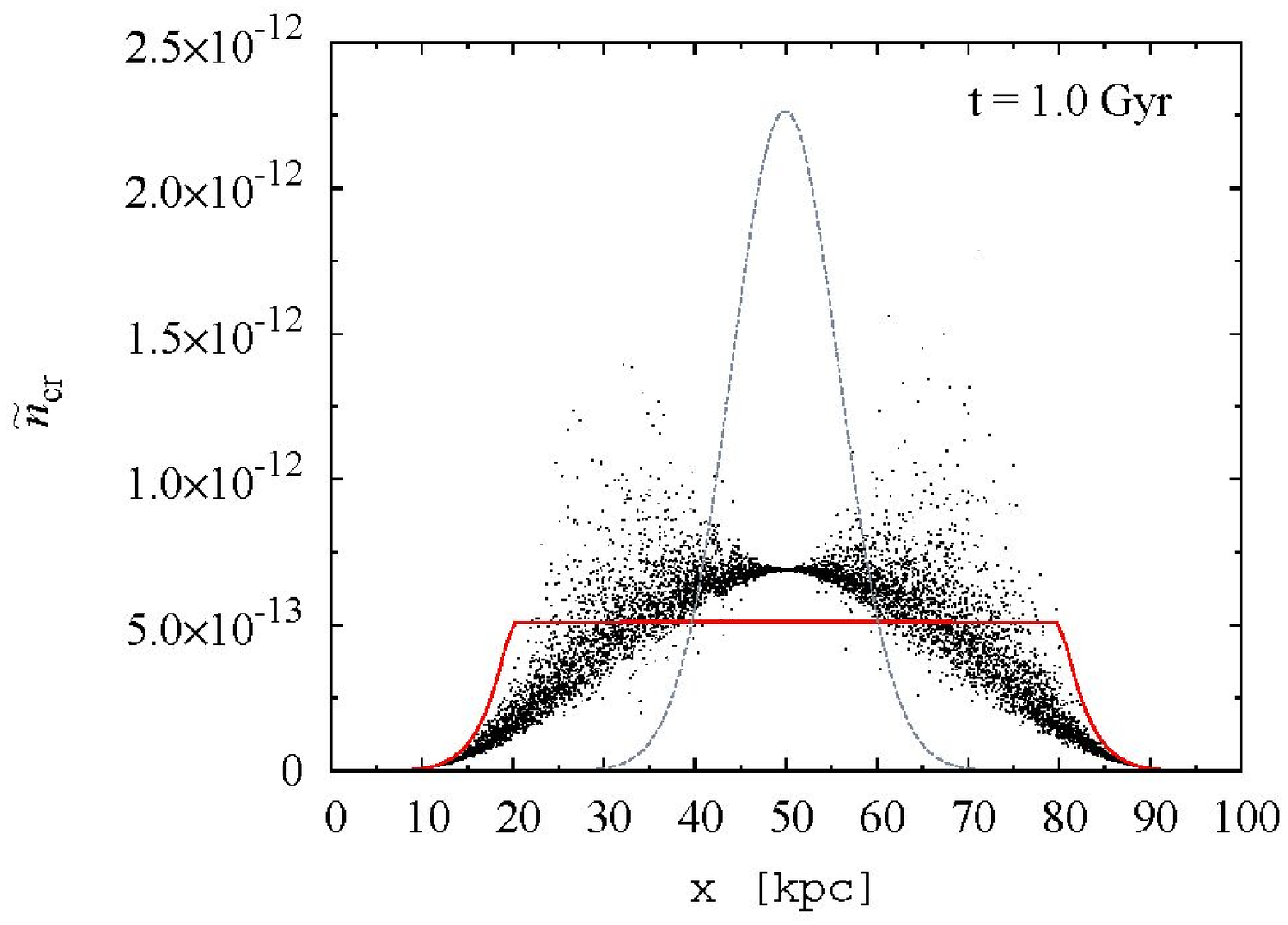} & 
\hspace{-2.5em}\includegraphics[scale=0.32]{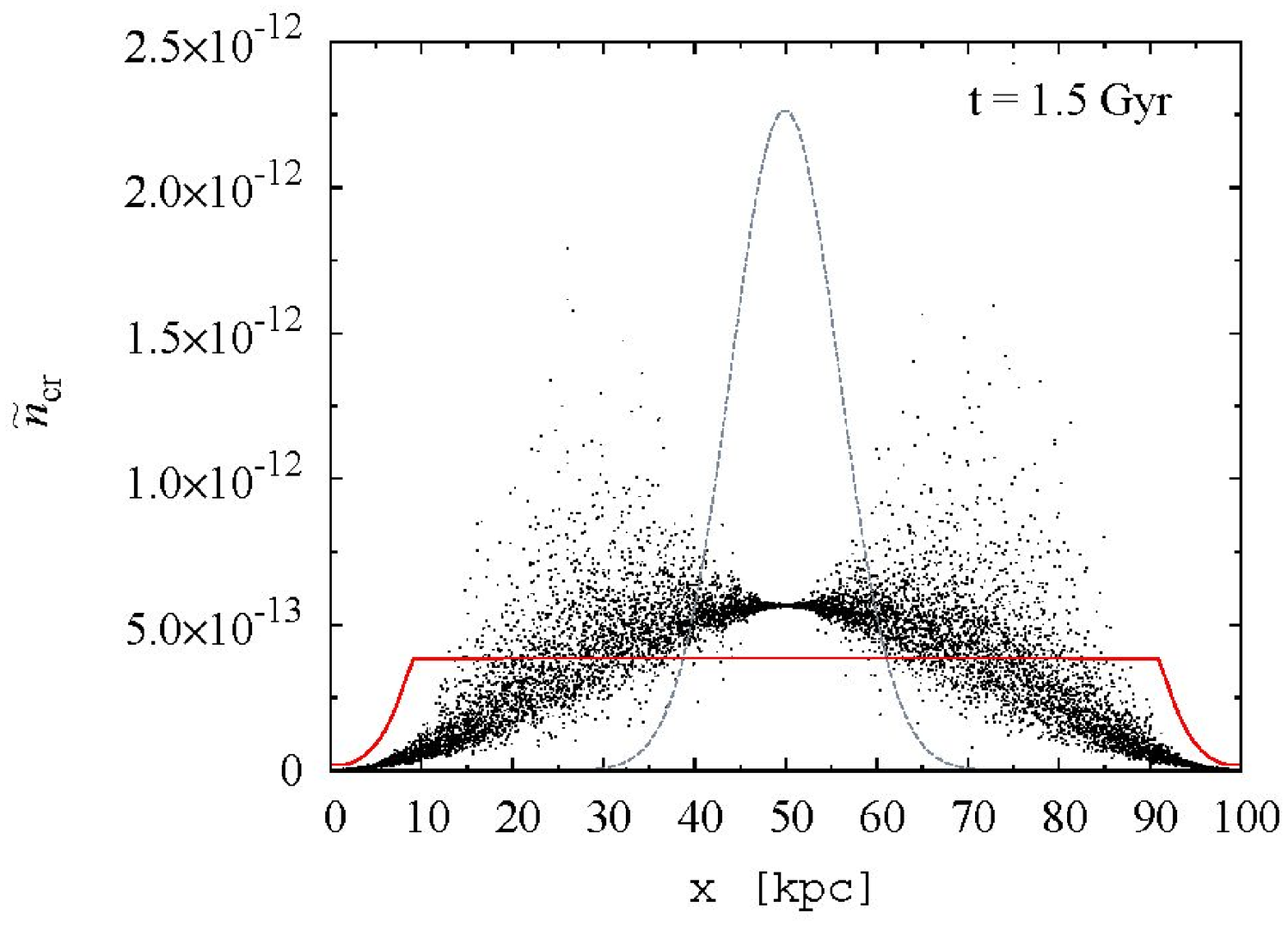}
\end{tabular}
\caption{Numerical results of the CR streaming implementation in {\sc Gadget-2}
  for the time evolution of an initial Gaussian distribution in CR energy (first
  two rows) and number density (last two rows), respectively. From the top left to the bottom right panel in the first two
  rows, the simulation times increase as $t=0.0,\,0.1,\,0.2,\,0.5,\,1.0,\,1.5$
  Gyr, and correspondingly for the last two rows. The black dots indicate the SPH
  particle values, the red line represents the results obtained via Fromm's
  method, and the dashed line indicates the initial distribution. The sound speed
  is $c_\rmn{s} = 20\,\rmn{km\, s}^{-1}$ throughout the volume, but the
  propagation speed is larger than that, because of the appearance of CR
  pressure in the formal diffusivities. The SPH particle positions resemble a
  ``glass''-like distribution with a mean inter-particle spacing of 1 kpc. For
  this test, we switched off the gravitational and hydrodynamical accelerations.}
\label{fig: 1D tests including pressure in diffusivity-1}
\end{figure*}

In this section we report results for our CR streaming implementation for a
number of one-dimensional test cases. We face the difficulty here that in
general no analytical solutions of equations \eqref{eq:energy equation 3} and
\eqref{eq:number density equation 3}, respectively, exist. This is due to the
fact that the advection speed (the streaming velocity, equation \eqref{eq:streaming
  velocity}) changes sign at extrema in the CR pressure distribution. Only in
some trivial cases, where no extrema are present, it is possible to give an
analytical solution. Therefore, the only way to verify our scheme in a more
realistic setting is to compare it to results obtained with other numerical
recipes. There exist several techniques to numerically solve advection-type
equations with a spatially varying transport velocity. We decided to use Fromm's
method which belongs to the class of finite volume methods and is described in
detail in \citet{LeVe}.  Also, a recent discussion of the properties of the
streaming equation as well as different ways to numerically solve it on a
one-dimensional grid was given by \citet{Sharma2010}.

As far as the initial conditions for our implementation in {\sc
  Gadget-2} are concerned, we use an ensemble of $10^4$ SPH particles
at rest (i.e.  $\left|\vvel\right|=0$) in a periodic slab of
matter. Both, the gravitational and the hydrodynamical accelerations
are switched off to ensure that the changes in the CR properties are
solely caused by CR streaming.  For the same reason, we neglect
Coulomb, hadronic and wave damping losses. The slab has a volume of
$10\times10\times100\,\mathrm{kpc^{3}}$, corresponding to a mean
inter-particle spacing of $1\,\mathrm{kpc}$, and the particle mass is
chosen such that the matter density is
$10^{10}\,\mathrm{M_{\odot}kpc^{-3}}$.  Since we use periodic boundary
conditions for the short dimensions of our simulation box, boundary
effects are avoided and the setting becomes effectively
one-dimensional. The temperature throughout the simulation box is
assumed to be constant and chosen such that the sound speed is
$c_\rmn{s} = 20\,\mathrm{km\, s^{-1}}$.  Furthermore, in order to test
our scheme in the presence of numerical noise in the gas density, as
encountered in cosmological simulations, the initial particle
positions resemble an irregular, glass-like distribution. The
simulations are carried out using $50$ smoothing neighbours. Regarding
the parameters of the CR feedback model, we adopt $\lambda=1$ and
$\varepsilon=0.004$ for the streaming part and the CR populations are
initialized with a fixed spectral slope $\alpha=2.25$ and a fixed
low-momentum cut-off $q=10.0$, throughout the simulation volume. Note
that the choice for $\varepsilon$ given above leads to a comparatively
small time step, which is necessary to prevent overshooting in our
explicit time integration scheme, but results in a high computational
cost. The expensiveness of numerically solving the streaming equation
was also encountered in the one-dimensional tests conducted by
\citet{Sharma2010}, even when implicit methods were used.  The grid
for our one-dimensional reference simulations adopting Fromm's method
is made up of $1000$ mesh points and are based on a time step of size
$\Delta t=10^{-4}\, h/\vel_{\rmn{st}}$.

\begin{figure*}
\begin{tabular}{cc}
\includegraphics[width=0.48\linewidth]{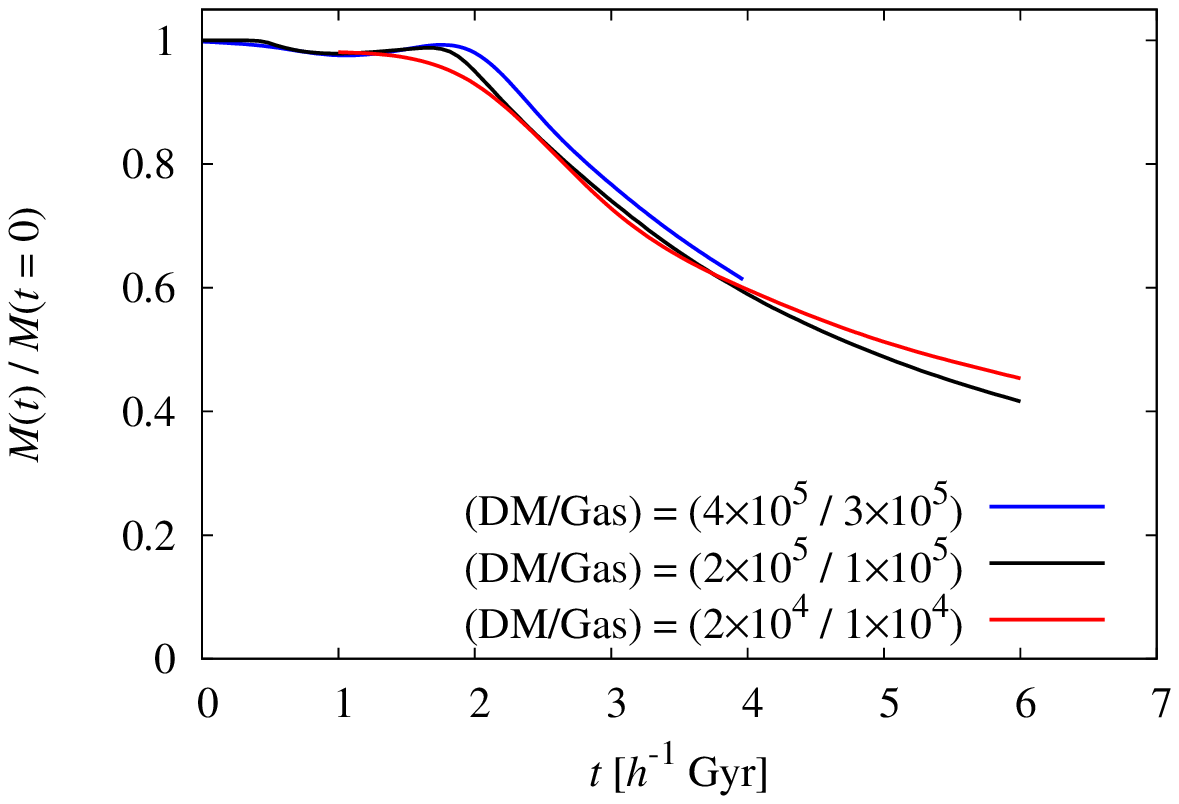}&
\includegraphics[width=0.48\linewidth]{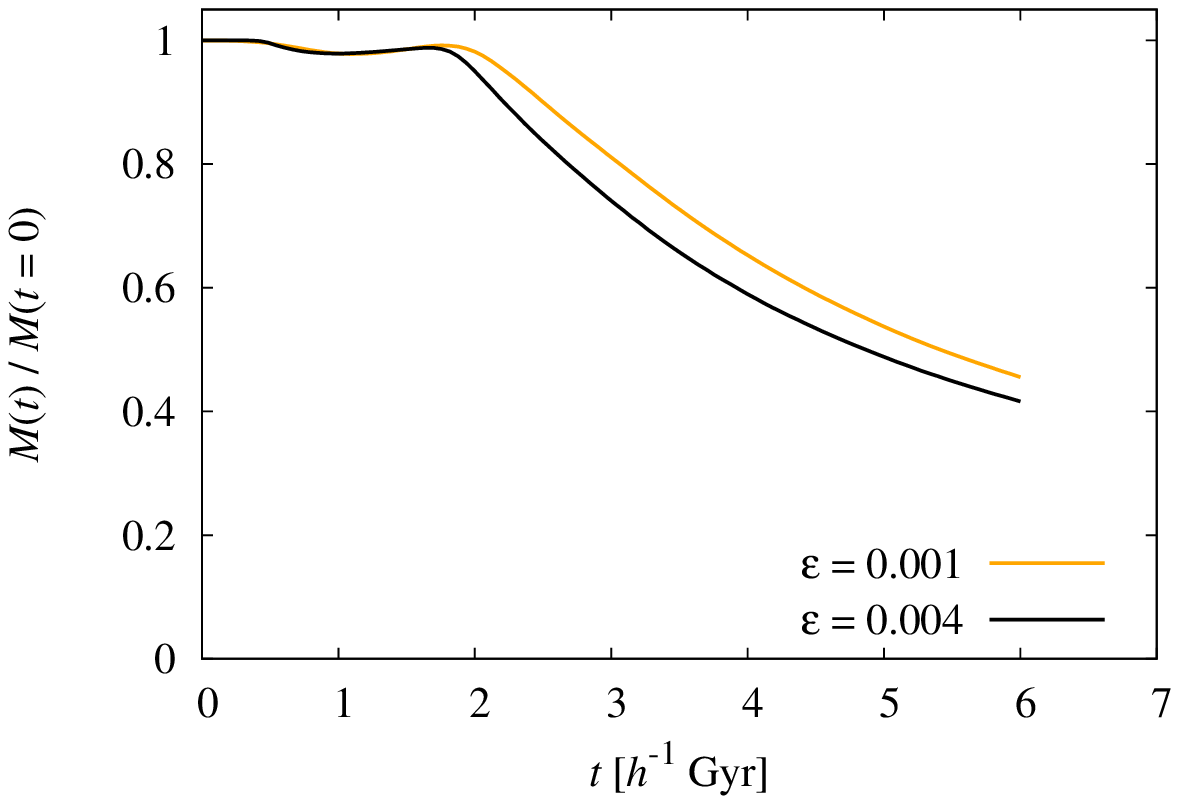}
\end{tabular}
\caption{Resolution study of mass loss history, $M(t)$, in the $10^9\,\hm$
  halo. Left panel: despite varying the number of SPH and DM particles used to
  sample dark matter and baryonic gas by a factor of 20, $M(t)$ is
  converged. Right panel: we vary the parameter $\varepsilon$, which regulates
  the size of the streaming time step of equation~\eqref{eq:timestep}, and
  demonstrate approximate convergence of $M(t)$.}
\label{fig: resolution}
\end{figure*}

The streaming equations that we solve read
\begin{eqnarray}
\frac{\dd\tilde{\varepsilon}_{\mathrm{cr}}}{\dd t} & = & -\frac{1}{\rho}\frac{\partial}{\partial x}\left[\vel_{\mathrm{st}}\,\left(\rho\,\tilde{\varepsilon}_{\mathrm{cr}}+P_{\mathrm{cr}}\right)\right],\label{eq:energy density (test case) 2}\\
\frac{\dd \tilde{n}_{\mathrm{cr}}}{\dd t} & = & -\frac{1}{\rho}\frac{\partial}{\partial x}\left(\vel_{\mathrm{st}}\, \rho\,\tilde{n}_{\mathrm{cr}}\right),\label{eq:number density (test case) 2}
\end{eqnarray}
with the magnitude of the streaming speed given by the sound speed and the
direction determined by the local CR pressure gradient,
\begin{equation}
\vel_{\mathrm{st}}=\begin{cases}
-c_{\mathrm{s}} & \mbox{for}\quad\frac{\partial P_{_{\mathrm{cr}}}}{\partial x}>0,\\
+c_{\mathrm{s}} & \mbox{for}\quad\frac{\partial P_{_{\mathrm{cr}}}}{\partial x}<0.
\end{cases}
\end{equation}
While the CR description in {\sc Gadget-2} allows for a varying effective
adiabatic index $\gamma_{\rm cr}$ of CRs, we do not account for this possibility in
our simplified Fromm scheme. Instead, we adopt a fixed adiabatic index of
$\gamma_\rmn{cr} \simeq 1.34$, suitable for the spectral slope and cut-off given
above.  Using the relation $P_\rmn{cr} =
(\gamma_\rmn{cr}-1)\,\rho\tilde{\varepsilon}_\rmn{\mathrm{cr}}$, the CR pressure is then
simply proportional to the energy density and can be replaced by the latter in
the streaming equation, equation~\eqref{eq:energy density (test case) 2}, evolved by
the Fromm scheme.

Figure \ref{fig: 1D tests including pressure in diffusivity-1} compares the
results obtained with our CR streaming implementation and those obtained using
Fromm's method for the time evolution of the streaming equations given
above. Here, we simulated the spatial and temporal evolution of the CR
population, starting from an initial Gaussian distribution in both, CR energy
density and number density, over a time interval of $1.5$ Gyr.

While the overall agreement of both schemes is good, there exist significant
differences in the transition region between the flat inner part of the CR
energy density and the outer flanks, with Fromm's method showing a sharper
transition. This discrepancy is most likely due to the fact that SPH is not
able to accurately describe sharp discontinuities but rather tends to smooth
them out. Fromm's method in contrast should give a numerical solution quite
close to reality, since the flat inner part of the distribution arises owing to
the diffusive nature of the CR streaming equation around extrema, as shown by
\citet{Sharma2010}.  However, in both cases the flanks seem to travel at almost
the same speed, which indicates a good representation of the advective aspect
of the CR streaming equations in our implementation.

Basically all the aforementioned considerations also hold true for the
evolution of CR number density, but additional complications
arise. Starting at about $t=0.5$ Gyr, the results become quite noisy.
However, this is not problematic, since the bulk of the SPH particles
shows no extreme scatter. More importantly, our main requirement is to
obtain a good representation of the dynamically relevant CR
quantities, i.e., the pressure and energy density. Since this is
fulfilled, the noise in number density should not significantly
influence our results concerning the dynamical impact of CRs on the
evolution of galactic winds.  Note also that the propagation of CR
number density is slower than that of CR energy density. Hence
streaming is not able to smooth out the former until $t=1.5$ Gyr but
this is because of the disappearance of the CR pressure gradient (and
correspondingly $\vel_{\mathrm{st}}=0$) rather than due to
inaccuracies in our scheme.

Apart from an initial Gaussian distribution, we also tested a step
function with each side of the step set to a different sound speed
(and thus $\vel_{\mathrm{st}}$). Also in this case we observed a good
overall agreement of CR pressure and energy density evolution in both
schemes. However, for large steps in the sound speed (temperature) the
CR energy flux became discontinuous for several SPH particles located
at the step. This is not an issue for our isolated haloes but could be
problematic in simulations of cosmological structure formation.  We
have found a slightly modified SPH formulation of CR streaming, that
behaves better in the aforementioned test case. By taking the
geometric mean of the formal diffusivities in the streaming equations,
rather than the harmonic one, i.e.
\begin{equation}
2\frac{\kappa^{i}\cdot\kappa^{j}}{\kappa^{i}+\kappa^{j}}\quad\rightarrow\quad\sqrt{\kappa^{i}\cdot\kappa^{j}}, \label{eq: from harmonic to geom-1}
\end{equation}
the CR energy flux is well-behaved even at large jumps in the temperature field. However, due to the geometric mean being bigger than the harmonic one, this modified scheme needs a more conservative time step and was therefore
not used for our isolated halo simulations.
In order to reduce the computational cost as well as to make the scheme more stable for cosmological simulations, we intend to implement
an implicit time integration scheme in the future.

\subsection{Resolution Tests}
\label{sec: resolution}

We perform a number of resolution tests to study numerical convergence of the
mass loss history and CR Alfv\'en-wave heating. In Fig.~\ref{fig: resolution},
we demonstrate that we obtain converged results for our standard choice of the
parameter $\varepsilon=0.004$, which regulates the size of the streaming time
step of equation~\eqref{eq:timestep}, and the number of SPH and DM particles of
$1\times10^5$ and $2\times10^5$, respectively.

\bsp

\label{lastpage}

\end{document}